\documentclass[11pt]{article}
\usepackage{setspace}
\usepackage{amsmath, amssymb, bm} 

\usepackage{epsfig,epstopdf}
\usepackage{graphicx,amssymb}
\usepackage{subfig}
\usepackage{multirow}
\usepackage{multicol}
\usepackage{amsmath}
\usepackage{color}
\usepackage{booktabs}
\usepackage{algorithmic}
\usepackage{algorithm}
\usepackage{natbib}
\usepackage{bm}
\usepackage{enumitem}
\usepackage{titling}
\allowdisplaybreaks
\usepackage{setspace}
\usepackage[left=1in,top=1in,right=1in,bottom=1in, nohead]{geometry}
\usepackage{authblk}

\usepackage[colorlinks=true, linkcolor=blue,citecolor=blue,urlcolor=blue,filecolor=blue]{hyperref}
\usepackage{color}
\usepackage{natbib}\bibpunct{(}{)}{;}{a}{}{,}

\usepackage{titlesec}
\titlespacing*{\section}
{0pt}      
{0.9ex plus 0.2ex minus 0.2ex}  
{0.6ex plus 0.2ex}              

\titlespacing*{\subsection}
{0pt}
{0.7ex plus 0.2ex minus 0.2ex}
{0.4ex plus 0.2ex}

\titlespacing*{\subsubsection}
{0pt}
{0.5ex plus 0.2ex minus 0.2ex}
{0.3ex plus 0.2ex}

\newtheorem{theorem}{Theorem}
\newtheorem{lemma}{Lemma}

\newenvironment{definition}[1][Definition]{\begin{trivlist}
		\item[\hskip \labelsep {\bfseries #1}]}{\end{trivlist}}

\newenvironment{remark}[1][Remark]{\begin{trivlist}
		\item[\hskip \labelsep {\bfseries #1}]}{\end{trivlist}}

\newcommand{\qed}{\nobreak \ifvmode \relax \else \ifdim\lastskip<1.5em \hskip-\lastskip \hskip1.5em plus0em minus0.5em \fi \nobreak \vrule height0.75em width0.5em depth0.25em\fi}

\newcommand{\bmx}{\boldsymbol{x}}
\newcommand{\bmX}{\boldsymbol{X}}
\newcommand{\bmZ}{\boldsymbol{Z}}
\newcommand{\bmV}{\boldsymbol{V}}
\newcommand{\bmU}{\boldsymbol{U}}
\newcommand{\G}{\mathcal{G}}

 \newcommand{\diag}{\text{diag}}

\newcommand{\Ga}{\text{Ga}}

\usepackage{tikz}
\usepackage{appendix}
\usepackage{float}

\providecommand{\bq}{\mathbf q}
\providecommand{\bmmu}{\boldsymbol{\mu}}
\providecommand{\Var}{\text{Var}}
\providecommand{\Cov}{\text{Cov}}
\providecommand{\E}{\text{E}}

\begin{document}
	
	\title{Finite-horizon maintenance under geometric-process models: Quantifying model and decision uncertainty}
	
	\vspace{-1em}
	\author[1]{\small Wenhui~Wu}
	\author[2]{\small Bing~Xing~Wang }
	\author[2]{\small Wei~Wang }
	\author[1]{\small Piao~Chen \footnote{Corresponding author: piaochen@intl.zju.edu.cn}}
	\affil[1]{ZJUI Institute, Zhejiang University, China}
	\affil[2]{School of Statistics and Data Science, Zhejiang Gongshang University, China}
	
	\date{}
	\maketitle

	\setlength{\abovedisplayskip}{2pt plus 1.1pt minus 1.1pt}
	\setlength{\belowdisplayskip}{2pt plus 1.1pt minus 1.1pt}
	\setlength{\abovedisplayshortskip}{2pt plus 1pt minus 1pt}
	\setlength{\belowdisplayshortskip}{2pt plus 1pt minus 1pt}

	\vspace{-2cm}
	

	\begin{abstract}
		\noindent
		\vspace{-0.8cm}

		Geometric process (GP) models are widely used for repairable systems whose inter-failure times evolve over successive failure-repair cycles. Maintenance decisions under GP models are often made with limited failure data and are subject to uncertainty in baseline selection and the resulting maintenance decision. Existing approaches typically rely on model ranking and plug-in estimates, which can understate uncertainty surrounding next-cycle replacement decisions.
		This paper develops a unified generalized-inference framework for finite-horizon, event-driven maintenance decision making under GP models. The proposed framework quantifies uncertainty in GP model parameters, cycle-specific reliability, hazard rate, one-cycle cost rate, and the associated optimal replacement time. For exponential and gamma baselines, the framework yields generalized confidence intervals through generalized pivotal quantities. For the Weibull baseline, it is implemented through a generalized fiducial approach. It also extends to a finite-sample discrimination procedure for quantifying baseline-selection uncertainty. Theoretical results establish the asymptotic validity of the proposed procedures and their interval estimates. Simulation study shows reliable finite-sample coverage for model parameters and nonlinear maintenance-relevant quantities, and satisfactory discrimination performance. A case study on aircraft air-conditioning failures shows that, in early cycles, replacement schedules with similar expected cost can imply materially different risk profiles, so a single plug-in recommendation may be misleading.
		
		
	\end{abstract}

	\section{Introduction}
	
	Maintenance decision making for repairable systems is inherently sequential and often must rely on limited unit-specific operational records, particularly in the early cycles. In many industrial settings, equipment experiences repeated failure-repair cycles, and the associated inter-failure times are typically not identically distributed across cycles because of aging and imperfect maintenance, leading to systematic cycle-to-cycle distributional drift. The geometric process (GP), introduced by \citet{Lam1988a} (see also \citealp{Lam1988b}), provides a parsimonious way to model such nonstationary behavior by imposing a geometric ratio relationship across cycles, so that the failure-time distribution evolves in a structured manner from one cycle to the next. Subsequent extensions accommodated deterioration (or improvement) trends and broader repairable-system configurations \citep{Lam1992OptimalGP, LamZhang1996TwoComponent, LamZhangZheng2002Multistate, PerezOconMontoro2004Transient}. Building on these developments, substantial maintenance literature has adopted GP-based models to design policies under imperfect repair, including preventive replacement strategies \citep{zhang2002, WangZhang2006Preventive, LAM2007b, WangZhangYam2017GGP} and delayed-repair settings \citep{Zhang2016}.
	
	Despite this modeling appeal, GP-based maintenance optimization does not always align with practical decision making, where actions are updated sequentially as failures, inspections, or condition information arrive. Many maintenance studies rely on long-run average criteria, typically formulated through an expected cost rate and justified by renewal-reward arguments. Such criteria are useful for evaluating the long-run performance of a maintenance rule under approximately stable operating conditions and long planning horizons \citep{ARTS2025}. In practice, however, maintenance decisions are often revised after each new observation using currently available information, which has motivated finite-horizon rules such as one-cycle optimality \citep{CoolenSchrijner2007, Hamidi2016} and condition-based maintenance policies \citep{CAI2023, ESPOSITO2026}. This practical setting raises a central question for GP-based maintenance: how should uncertainty be quantified and incorporated when decisions are made cycle by cycle from limited data?
	
	Two layers of uncertainty are especially important in this context. The first is baseline-distribution uncertainty. Maintenance recommendations under a GP model can be sensitive to the choice of baseline distribution because the baseline family determines the implied reliability pattern and therefore affects the recommended policy. This remains true even under flexible extensions such as the generalized GP \citep{WangYam2017GGP} and the doubly GP \citep{Wu2018DGP}. In practice, the baseline is often selected using Akaike information criterion (AIC) or Bayesian information criterion (BIC). These criteria provide a best-fitting model or a ranking among candidate baselines, but they do not quantify whether the observed improvement in fit is large enough to be regarded as statistically meaningful, particularly when samples are small or competing baselines fit similarly well. The second layer is decision uncertainty. In most implementations, model parameters are first estimated and then treated as known when evaluating a policy. Estimation uncertainty is therefore typically ignored in both policy evaluation and cycle-by-cycle decisions, even though incorporating it can materially affect the recommended maintenance intensity \citep{DEJONGE2015}.
	
	These two layers of uncertainty become particularly pronounced in the small-sample settings that often arise in early operating cycles. Failure histories are then short, candidate baselines can be difficult to distinguish, and uncertainty in the baseline parameters or the geometric ratio may propagate across cycles and induce substantial uncertainty in decision quantities such as reliability, hazard rate, and the next-cycle optimal replacement time. Existing studies have developed analytical results for GP models, including distributional properties \citep{BraunLiZhao2005Properties} and moment evaluation for the associated counting process \citep{Aydogdu2014WeibullGP, PekalpAydogdu2018SecondMoment}. However, these studies have mainly concentrated on parameter inference, leaving uncertainty in decision-oriented quantities and the resulting policy implementation much less explored. Consequently, maintenance policies based solely on point estimates can be fragile in early cycles, because even a modest change in the estimated replacement time may shift the decision across operational constraints such as maintenance windows or risk {bounds}. A few studies have begun to view parameter uncertainty as a decision risk in policy selection \citep{Wu2017}, and resampling-based evaluations have reported confidence levels for reliability measures under periodic maintenance \citep{Muller2017}, but a unified treatment of both baseline-selection uncertainty and decision uncertainty under GP models is still lacking.
	
	A natural way to address this gap is to quantify uncertainty directly at the decision level by constructing interval estimates for maintenance-relevant quantities such as reliability, cost rates, and optimal replacement times or thresholds. Related ideas have been explored in preventive maintenance. Large-sample and graphical procedures have been used to derive confidence {bounds} for optimal maintenance intervals \citep{HalimTang2009Graphical}; bootstrap methods have been applied to obtain intervals for optimal maintenance times or reliability measures \citep{GilardoniColosimoDallaValle2013, Muller2017}; and Bayesian decision-analytic formulations have translated posterior uncertainty into actions through utility- or cost-based criteria \citep{FangHuang2008}. However, these approaches can be difficult to rely on when only a few failures are observed. Wald-type and delta-type intervals depend on large-sample normal approximations and, for complex decision functions, may also require derivatives that are tedious to derive and implement. Bayesian credible intervals offer an alternative, but under short failure histories they may be sensitive to prior specification. Bootstrap procedures reduce reliance on analytic approximations, yet they can remain unstable for highly nonlinear decision quantities in sparse-data settings, so maintenance recommendations may still vary substantially once estimation uncertainty is taken into account \citep{SANCHEZ2009, LAGGOUNE2010}.

	To handle these difficulties, this paper develops a generalized-inference framework for finite-horizon maintenance decision making under GP models. Its main appeal is not merely that it avoids large-sample normal approximations or prior specification, but that it can be used to infer directly on decision-relevant quantities that are complicated functions of the model parameters. In GP-based maintenance, quantities such as cycle-specific reliability, hazard, one-cycle cost rates, and the next-cycle optimal replacement time are all nonlinear functions of the underlying parameters, and some are defined implicitly through optimization. Generalized inference provides a convenient way to transfer estimation uncertainty to these decision objects without requiring case-by-case asymptotic linearization or explicit derivative calculations for each target quantity. This feature is particularly valuable in early cycles, where only sparse failure data are available and the induced uncertainty at the decision level can be substantial. The same framework also supports formal discrimination among candidate baseline distributions, and hence complements simple AIC or BIC ranking when explicit control of baseline-selection risk is needed.

	The primary contributions of this study are threefold. First, we formulate a finite-horizon maintenance framework for repairable systems under the GP, tailored to cycle-by-cycle, event-driven decision making, and evaluate maintenance-relevant functions, including the one-cycle cost rate and the resulting replacement time, with interval estimates rather than point values. Second, we develop generalized inference procedures for GP models to construct small-sample confidence intervals for model parameters and decision-relevant functions, without relying on asymptotic normal approximations or prior specifications. This enables uncertainty in parameter estimation to be carried through to the maintenance decision, improving the robustness of finite-horizon actions in early cycles. Third, we propose a discrimination test for baseline identification among candidate distributions. In contrast to AIC or BIC ranking, the proposed test controls the risk of incorrect baseline selection at a specified significance level, thereby providing an explicit measure of statistical support for the selected model.
	
	The remainder of the paper is organized as follows. Section~\ref{sec:model} introduces the GP model and the finite-horizon reliability measures considered in this study. Section~\ref{sec:exp} develops the generalized inference procedure for the exponential baseline distribution, and Section~\ref{sec:wei_gam} extends the framework to the gamma and Weibull baseline distributions. Section~\ref{sec:test} presents the proposed discrimination test for baseline selection. Section~\ref{sec:simulation} provides simulation results, and Section~\ref{sec:case} illustrates the proposed methods using an industrial case study. Section~\ref{sec:conclusion} concludes the paper.

	\section{GP model and reliability metrics}\label{sec:model}

	Consider a repairable system that experiences successive failures over time. Let $\{X_i,\, i=1,2,\dots,n\}$ denote the inter-failure times, where $X_i$ is the elapsed time between the $(i-1)$-th and $i$-th failures. Repeated failures and repairs may induce a systematic trend in $X_i$. To capture such event-driven evolution, we model $\{X_i,\, i=1,2,\dots,n\}$ using a GP.
	\begin{definition}\label{def:gp}
		A sequence of random variables $\{X_i,\, i=1,2,\dots,n\}$ is said to be a GP if there exists a constant $a>0$ such that $X_1,X_2,\dots,X_n$ are independent and, for each $i=1,2,\dots,n$, the distribution function of $X_i$ satisfies
		\begin{equation*}
			F_i(x)=F\!\left(a^{\,i-1}x\right), \qquad x\ge 0,
		\end{equation*}
		where $F(\cdot)$ denotes the cumulative distribution function (CDF) of $X_1$. 
	\end{definition}
	The constant $a$ is called the geometric ratio, which governs the stochastic trend across failure cycles. When $a>1$, the inter-failure times tend to shorten with $i$, reflecting deterioration (e.g., imperfect repairs). When $a=1$, the GP reduces to a renewal process. When $0<a<1$, the inter-failure times tend to lengthen, indicating performance improvement.
	A convenient equivalent representation is obtained by standardization $Y_i=a^{\,i-1}X_i$. Under the GP assumption, $\{Y_i,\, i=1,2,\ldots,n\}$ are independent and identically distributed  (i.i.d.) with a common distribution function $F(\cdot)$ and density $f(\cdot)$, where $f(\cdot)$ is the probability density function (PDF) of $X_1$.
	
	Standardization $Y_i$ highlights the event-driven nature of the GP, namely, each operating cycle has its own inter-failure time distribution through the ratio parameter $a$. This motivates the cycle-specific reliability measures defined after each failure, which are closely related to the survival and hazard functions commonly used in survival analysis and modeling of the repairable system \citep{Nakagawa2005, KleinbaumKlein2012}.
	For $\tau>0$, the reliability  after the $(i-1)$-th failure can be written as 
	\begin{equation*}
		R_i(\tau) = \Pr\!\left(X_i>\tau \,\middle|\, X_1,\ldots,X_{i-1}\right) = \Pr(X_i>\tau)=1-F(a^{i-1}\tau).
	\end{equation*}
	To characterize the instantaneous failure risk after the $(i-1)$-th repair, we define the  hazard rate in the failure index $i$ as 
	\begin{equation*}
		h_i(\tau)=\frac{f_i(\tau)}{R_i(\tau)},
	\end{equation*}
	where $f_i(\cdot)$ denotes the PDF of $X_i$.

	From a maintenance decision perspective, $R_i(\tau)$ serves as a short-term performance measure after the $(i-1)$-th repair. Given the observed failure history up to that point, it quantifies the probability that the system will continue operating for at least $\tau$ time units without failure, and is therefore directly relevant to sequential maintenance or inspection decisions that are updated after each failure rather than fixed in advance on a long-run basis. In contrast, $h_i(\tau)$ characterizes the instantaneous failure risk at time $\tau$ since the last repair, conditional on survival up to $\tau$, and thus captures the marginal risk of continuing operation for a little longer. This makes it particularly useful for assessing the timing and urgency of preventive actions. In the GP framework, both $R_i(\tau)$ and $h_i(\tau)$ are cycle-specific quantities that vary with the failure index $i$, reflecting how past failures and repairs affect the risk profile in each operating cycle. They therefore provide an event-driven description of system reliability and complement long-run performance criteria in maintenance planning. Consequently, in the following we focus on statistical inference for these cycle-specific reliability measures, with particular attention to constructing confidence intervals that quantify their uncertainty under the GP model.

	\section{Generalized inference for the exponential GP}\label{sec:exp}

	Suppose that $\{X_i,\, i=1,2,\ldots,n\}$ follows a GP with geometric ratio parameter $a$, and that the first inter-failure time $X_1$ has an exponential distribution with rate parameter $\lambda$. 
	Under the standardization $Y_i=a^{\,i-1}X_i$, we have $Y_i\sim \mathrm{Exp}(\lambda)$, so the exponential GP is characterized by the two-dimensional parameter vector $\boldsymbol{\theta}_{\mathrm E}=(a,\lambda)^{T}$. We begin with this baseline because it provides the simplest setting for introducing the proposed generalized-inference framework. Although the underlying lifetime model is elementary, the GP structure already makes the relevant reliability quantities cycle-specific through their joint dependence on $a$ and $\lambda$. This allows the main ideas behind generalized inference to be clearly presented  before the framework is extended to more complex baselines in Section~\ref{sec:wei_gam}.

	\subsection{Likelihood and point estimation}
	
	We first introduce the likelihood-based estimators, which play two roles in what follows: they provide natural point estimates, and they also serve as the observed-data inputs from which the confidence intervals are constructed. {Given $\bmX = \{X_i,\, i=1,2,\ldots,n\}$, the log-likelihood function for $\boldsymbol{\theta}_{\mathrm E}$ is}
	\begin{equation*}
		\ell(\boldsymbol{\theta}_{\mathrm E}; \bmX)
		= n\log \lambda + \frac{n(n-1)}{2}\log a - \lambda \sum_{i=1}^n a^{i-1} X_i .
		\label{eq:loglike}
	\end{equation*}
	By differentiating $\ell(\boldsymbol{\theta}_{\mathrm E};\bmX)$ with respect to $\boldsymbol{\theta}_{\mathrm E}$ and setting the derivatives equal to zero, the maximum likelihood (ML) estimator $\hat{\boldsymbol{\theta}}_{\mathrm E}=(\hat{a},\hat{\lambda})^{T}$ is obtained, where $\hat a$ is the solution to
	\begin{equation}
		\sum_{i=1}^n (n-2i+1)\, a^{\,i-1} X_i = 0,
		\label{eq:a}
	\end{equation}
	and the corresponding ML estimator of $\lambda$ is
	\begin{equation}
		\hat{\lambda}=\frac{n}{\sum_{i=1}^n \hat{a}^{\,i-1} X_i}.
		\label{eq:lambda}
	\end{equation}
	The estimation equations already reveal the basic inferential challenge: the GP ratio $a$ and the baseline parameter $\lambda$ are coupled through the transformed inter-failure times $a^{i-1}X_i$. As a result, uncertainty in $a$ propagates directly to the inferred reliability profile of each cycle.

	The usual regular theory for ML estimators is not directly applicable here, because $X_i$'s are independent but not identically distributed, and the score component for $a$ contains the increasing weight $i-1$. As a result, the estimator of $a$ has a faster convergence rate than that of $\lambda$, leading to a nonstandard asymptotic distribution. 
	We therefore establish the asymptotic distribution of the ML estimator $\hat{\boldsymbol{\theta}}_{\mathrm E}$ in the following theorem. The proof is given in Section~\ref{proof.mle_exp} of the Supplementary Material.
	\begin{theorem}\label{thm:mle_exp}
		Let $D_{n,\mathrm E}=\diag(n^{3/2},n^{1/2}),$ and $\hat{\boldsymbol{\theta}}_{\mathrm E}$ be defined by Equations \eqref{eq:a}-\eqref{eq:lambda}. Then 
		\[
		D_{n,\mathrm E}(\hat{\boldsymbol{\theta}}_{\mathrm E}-\boldsymbol{\theta}_{\mathrm E})
		\xrightarrow{d}
		N(\bm0,\Sigma_{\mathrm E}^{-1}),
		\]
		where
		\[
		\Sigma_{\mathrm E}^{-1}=
		\begin{pmatrix}
			12a^2 & -6a\lambda\\
			-6a\lambda & 4\lambda^2
		\end{pmatrix}.
		\]
	\end{theorem}

	\subsection{Generalized pivotal quantities for the model parameters}

	For the present problem, the main difficulty lies not only in inference on the model parameters $a$ and $\lambda$, but also in inference on maintenance-relevant quantities such as $R_i(\tau)$ and $h_i(\tau)$, which are nonlinear functions of these parameters. In small samples, large-sample linearization may be unreliable, while exact classical pivots are difficult to obtain because the GP structure induces a nonlinear coupling between $a$ and $\lambda$. Generalized inference addresses this difficulty by constructing generalized pivotal quantities (GPQs) from transformed variables with fully known, parameter-free distributions. Specifically, the estimation equations are first rewritten in terms of auxiliary random variables whose distributions do not involve unknown parameters, and the resulting relationships are then solved for suitable functions of the estimators and the parameters. When the resulting random quantity has a distribution free of unknown parameters, it can play the role of a pivot, and inverting this relationship yields a GPQ for the parameter of interest. Once GPQs for $a$ and $\lambda$ are obtained, GPQs for derived reliability measures can be constructed by substitution,  allowing uncertainty to propagate directly from the fitted GP model to the cycle-specific quantities used in subsequent maintenance decisions.

	{To implement this idea, let $Z_i = \lambda a^{i-1} X_i, \ i=1,2,\ldots,n.$}
	{Under the exponential GP model, $Z_1,\ldots,Z_n$ are i.i.d. with $\mathrm{Exp}(1)$.}
	This transformation is crucial because it removes the unknown parameters from the distributional part of the problem and leaves only algebraic relationships involving the estimators, the parameters, and the auxiliary variable $Z_i$.
	Substituting $X_i=Z_i/(\lambda a^{i-1})$ into Equation~\eqref{eq:a}, we obtain
	\begin{equation}\label{eq:g1exp}
		\sum_{i=1}^n (n - 2i + 1)\left(\frac{\hat a}{a}\right)^{i-1} Z_i = 0 .
	\end{equation}
	{Let $\phi_{a,\mathrm E}(\bmZ)$ denote the root of Equation~\eqref{eq:g1exp} corresponding to $\hat a/a$, where $\bmZ=(Z_1,\ldots,Z_n)^T$. Since $\bmZ$ has a fully known distribution free of $a$ and $\lambda$, the distribution induced by this root is also free of unknown parameters. We therefore define the GPQ for $a$ by
		\begin{equation}
			\G_{a,\mathrm E}
			= \frac{\hat a}{\phi_{a,\mathrm E}(\bmZ)},
			\label{eq:a_gpqexp}
		\end{equation}
		where $\hat a$ is the ML estimate obtained from the observed sample $\bmx=\{x_i,\,i=1,2,\ldots,n\}$.
	}
	This step captures the core logic of generalized inference: rather than searching for a closed-form pivot for $a$ directly, we construct one indirectly through the parameter-free auxiliary variable {$\bmZ$} and then invert the relation.
	
	Next, combining the transformation $Z_i$ with Equation~\eqref{eq:lambda} gives
	\begin{equation}
		\frac{\hat\lambda}{\lambda}
		=
		\frac{n}{\sum_{i=1}^n \left(\frac{\hat a}{a}\right)^{i-1}Z_i}.
		\label{eq:glamexp}
	\end{equation}
	Substituting $\phi_{a,\mathrm E}(\bmZ)$ into Equation~\eqref{eq:glamexp}, we define
	\begin{equation}
		\phi_{\lambda,\mathrm E}(\bmZ)
		=
		\frac{n}{\sum_{i=1}^n (\phi_{a,\mathrm E}(\bmZ))^{\,i-1}Z_i}.
		\label{eq:g2exp}
	\end{equation}
	Again, because $\bmZ$ is parameter-free, the distribution of $\phi_{\lambda,\mathrm E}(\bmZ)$ does not involve any unknown parameter. The GPQ for $\lambda$ is therefore defined by
	\begin{equation}
		\G_{\lambda,\mathrm E}
		=
		\frac{\hat\lambda}{\phi_{\lambda,\mathrm E}(\bmZ)},
		\label{eq:lam_gpqexp}
	\end{equation}
	where $\hat \lambda$ is the ML estimate obtained from the observed sample $\bmx$.

	Taken together, $\G_{a,\mathrm E}$ and $\G_{\lambda,\mathrm E}$ form the basis for generalized inference under the exponential GP model. In practice, the generalized confidence intervals (GCIs) for $a$ and $\lambda$ are constructed from the empirical quantiles of the simulated GPQs $\G_{a,\mathrm E}$ and $\G_{\lambda,\mathrm E}$, rather than from a Wald-type normal approximation centered at the ML estimators. To justify this procedure theoretically, the following theorem establishes the joint asymptotic distribution of $\G_{a,\mathrm E}$ and $\G_{\lambda,\mathrm E}$. The proof is given in Section~\ref{proof.gpq_exp} of the Supplementary Material.

	\begin{theorem}\label{thm:gpq_exp}
		Let $\G_{a,\mathrm E}$ and $\G_{\lambda,\mathrm E}$ be the GPQs defined in Equations~\eqref{eq:a_gpqexp} and \eqref{eq:lam_gpqexp}. Then
		\[
		\begin{pmatrix}
			n^{3/2}(\G_{a,\mathrm E}-a) \\
			n^{1/2}(\G_{\lambda,\mathrm E}-\lambda)
		\end{pmatrix}
		\xrightarrow{d}
		N\!\left(
		\mathbf 0,\,
		2 \Sigma_{\mathrm E}^{-1}
		\right),
		\]
		where $\Sigma_{\mathrm E}^{-1}$  is given in Theorem \ref{thm:mle_exp}.
		Consequently, the GPQs $\G_{a,\mathrm E}$ and $\G_{\lambda,\mathrm E}$ are consistent and asymptotically normal.
	\end{theorem}
	
	Theorem~\ref{thm:gpq_exp} describes the unconditional large-sample behavior of the GPQs. Its limiting covariance contains two independent first-order sources of variation, one from the observed-sample ML estimator and the other from the auxiliary sample used in the GPQ construction. This explains why the unconditional covariance matrix is twice the limiting covariance matrix of the ML estimator. In constructing GCIs, however, the observed data are held fixed and the interval quantiles are computed from the auxiliary parameter-free simulation. Therefore, the Monte Carlo distribution used for interval construction reflects the auxiliary component of the GPQ variation, rather than the doubled unconditional variation.

	\subsection{GCIs for reliability metrics}
	
	The ultimate quantities of interest in maintenance planning are not only the model parameters, but also the cycle-specific reliability and hazard functions. {Under the exponential GP, $R_{i,\mathrm E}(\tau)=\exp\{-\lambda a^{i-1}\tau\},\ h_{i,\mathrm E}(\tau)=\lambda a^{i-1}.$}
	Because these quantities are explicit functions of $(a,\lambda)$, their GPQs can be obtained using the substitution method proposed by \cite{weerahandi_2004}, namely by replacing $(a,\lambda)$ with ${(\G_{a,\mathrm E},\G_{\lambda,\mathrm E})}$, which is given by
	\begin{equation}\label{eq:rihi_expgp}
		\G_{R_{i,\mathrm E}(\tau)} = \exp\{- \G_{\lambda,\mathrm E}(\G_{a,\mathrm E})^{i-1} \tau\}, \qquad \G_{h_{i,\mathrm E}(\tau)} =\G_{\lambda,\mathrm E}(\G_{a,\mathrm E})^{i-1} .
	\end{equation}
	
	This substitution step is important conceptually: once the uncertainty in the basic GP parameters has been represented through GPQs, the same uncertainty is transmitted automatically to the maintenance-relevant functions derived from them. In this way, generalized inference moves from parameter estimation to decision-oriented reliability assessment without requiring separate asymptotic derivations for each target quantity.
	The $100(1-\gamma)\%$ GCI for $R_{i,\mathrm E}(\tau)$ is obtained from the $\gamma/2$ and $1-\gamma/2$ quantiles of $\G_{R_{i,\mathrm E}(\tau)}$, and the GCI for $h_{i,\mathrm E}(\tau)$ is obtained in the same way from $\G_{h_{i,\mathrm E}(\tau)}$. {In practice, these quantiles can be evaluated by Monte Carlo simulation. The resulting procedure is summarized in Section \ref{app.alg:gci_exp} of the Supplementary Material. }

	\section{Extension to other baseline distributions}\label{sec:wei_gam}
	
	The exponential GP clarifies the basic mechanism of generalized inference, but it is often too restrictive for maintenance data whose failure behavior departs from a constant hazard pattern. Two of the most important alternatives in reliability and maintenance are the gamma and Weibull distributions. The gamma baseline accommodates greater skewness and dispersion, and includes the exponential model as a special case, while the Weibull baseline is the standard choice for modeling monotone aging effects through increasing or decreasing hazards. However, extending the generalized-inference framework to these baselines is not a routine substitution. For the gamma GP, the GPQ strategy can still be retained, but the main obstacle is the construction of a suitable GPQ for the shape parameter. For the Weibull GP, by contrast, a tractable GPQ construction is no longer available, so the same generalized-inference philosophy is implemented through a generalized fiducial distribution instead. This section develops these two extensions.
	
	\subsection{Generalized inference for the gamma GP}
	
	Suppose that $\{X_i,i=1,2,\dots,n\}$ is a GP with ratio parameter $a$ and that $X_1$ follows a gamma distribution $\mathrm{Ga}(\alpha,\beta)$, where $\alpha$ and $\beta$ are the shape and rate parameters, respectively. Under the standardization $Y_i=a^{i-1}X_i$, we have $Y_i \sim \mathrm{Ga}(\alpha,\beta)$, so the gamma GP is characterized by $\boldsymbol{\theta}_{\mathrm G}=(a,\alpha,\beta)^T$. {Given $\bmX = \{X_i,i=1,2,\dots,n\}$,  the log-likelihood function for $\boldsymbol{\theta}_{\mathrm G}$ is given by}
	\[
	\ell(\boldsymbol{\theta}_{\mathrm G}; \bmX)
	= n\alpha\log\beta
	- n\log\Gamma(\alpha)
	+ \frac{n(n-1)\alpha}{2}\log a
	+ (\alpha-1)\sum_{i=1}^n \log X_i
	- \beta \sum_{i=1}^n a^{i-1}X_i.
	\]
	The ML estimator {$\hat{\boldsymbol{\theta}}_{\mathrm G} =(\hat a,\hat  \alpha,\hat \beta)^T$} is obtained by solving the following equations
	\begin{align}
		&\sum_{i=1}^n (n - 2i + 1)Y_i = 0, \label{eq:amle}\\
		&\log \alpha - \psi(\alpha) - \log(\bar{Y}/\tilde{Y}) = 0, \label{eq:alpmle}\\
		&\beta \sum_{i=1}^n Y_i - n\alpha = 0, \label{eq:lammle}
	\end{align}
	where $\bar{Y} = \dfrac{1}{n}\sum_{i=1}^n Y_i$, $\tilde{Y} = \left(\prod_{i=1}^n  Y_i \right)^{1/n}$ and $\psi(\cdot)$ is the digamma function.

	For the gamma GP, the nonstandard asymptotic distribution of the ML estimator $\hat{\boldsymbol{\theta}}_{\mathrm G}$ is stated in the following theorem, with the proof given in Section~\ref{proof.mle_gamgp} of the Supplementary Material.
	\begin{theorem}\label{thm:mle_gamgp}
		Let $D_{n,\mathrm G}=\diag(n^{3/2},n^{1/2},n^{1/2})$ and $\hat{\boldsymbol{\theta}}_{\mathrm G}$ be defined by Equations \eqref{eq:amle}-\eqref{eq:lammle}. Then 
		\[
		D_{n,\mathrm G}(\hat{\boldsymbol{\theta}}_{\mathrm G}-\boldsymbol{\theta}_{\mathrm G})
		\xrightarrow{d}
		N(\bm0,\Sigma_{\mathrm G}^{-1}),
		\]
		where
		\[
		\Sigma_{\mathrm G}^{-1}=
		\begin{pmatrix}
			\dfrac{12a^2}{\alpha} & 0 & -\dfrac{6a\beta}{\alpha}\\[8pt]
			0 & \dfrac{\alpha}{\alpha\psi_1(\alpha)-1} & \dfrac{\beta}{\alpha\psi_1(\alpha)-1}\\[10pt]
			-\dfrac{6a\beta}{\alpha} & \dfrac{\beta}{\alpha\psi_1(\alpha)-1}
			& \dfrac{\beta^2\{4\alpha\psi_1(\alpha)-3\}}{\alpha\{\alpha\psi_1(\alpha)-1\}}
		\end{pmatrix},
		\]
		and $\psi_1(\cdot)$ is the trigamma function.
	\end{theorem}

	As in Section~\ref{sec:exp}, our goal is not only to obtain a point estimation of $\boldsymbol{\theta}_{\mathrm G}$, but also to construct interval inference for the cycle-specific reliability and hazard measures. The main difficulty now lies in the shape parameter $\alpha$. In the exponential GP, a simple rescaling yields parameter-free auxiliary variables immediately. However, for the gamma GP, $\alpha$ enters the model through the shape of the baseline distribution and cannot be isolated by the same argument. Existing GPQ constructions for gamma models often depend on auxiliary approximations, such as moment-matching $\chi^2$ arguments or Cornish--Fisher expansions \citep{chen2017gamma,wang2018gamma}. By contrast, our construction is based on the exact gamma quantile representation and therefore avoids these additional approximation layers altogether. This yields a new GPQ for $\alpha$ that is fully consistent with the generalized-inference principle of building inference from parameter-free auxiliary randomness with known distribution, after which $a$ and $\beta$ can be recovered conditionally.

	To construct a GPQ for $\alpha$, we exploit the quantile representation of the gamma distribution. Since $\beta Y_i \sim \mathrm{Ga}(\alpha,1)$ for $i=1,\ldots,n$, let $F_G(\cdot;\alpha)$ denote the CDF of $\mathrm{Ga}(\alpha,1)$, and let $Q(u;\alpha)=F_G^{-1}(u;\alpha)$ for $0<u<1$. If $\bmU=(U_1,\ldots,U_n)$ is a random sample from $\mathrm{Unif}(0,1)$, then the probability integral transform implies that {$
		\beta Y_i \overset{d}{=} Q(U_i;\alpha), \ i=1,\ldots,n.$}
	This representation is especially useful here because the likelihood equation for $\alpha$ in Equation~\eqref{eq:alpmle} depends only on the sample through $\log(\bar Y/\tilde Y)$. We therefore define the auxiliary statistic
	\begin{equation}\label{eq:T_alpha}
		T_n(\alpha;\bmU)
		=
		\log\!\left(
		\frac{\dfrac{1}{n}\sum_{i=1}^n Q(U_i;\alpha)}
		{\left\{\prod_{i=1}^n Q(U_i;\alpha)\right\}^{1/n}}
		\right).
	\end{equation}
	By construction, $T_n(\alpha;\bmU)$ depends on $\alpha$ but is free of $a$ and $\beta$. Its large-sample behavior is summarized in the following lemma; the proof is given in Section~\ref{proof.T_alpha_gam} of the Supplementary Material.

	\begin{lemma}\label{lem:T_alpha_limit_gam}
		Define $g(\alpha)=\log(\alpha)-\psi(\alpha)$ for $\alpha>0$. 
		Then,
		\begin{enumerate}
			\item {For each fixed $\alpha>0$, $T_n(\alpha;\bmU) \xrightarrow{\mathrm{a.s.}} g(\alpha)$;}
			\item $\sqrt{n}\{T_n(\alpha;\bmU)-g(\alpha)\} \xrightarrow{d}  N\!\bigl(0,\ \psi_1(\alpha)-1/\alpha\bigr)$;
			\item $g(\alpha)$ is strictly decreasing on $(0,\infty)$.
		\end{enumerate}
	\end{lemma}

	To connect the auxiliary statistic with the observed sample, we first form the observed counterpart of the quantity appearing in the estimation equation for $\alpha$. Replacing the unknown $a$ with its ML estimate $\hat a$, define
	\begin{equation}\label{eq:tobs_gamgp}
		t_{\mathrm{obs}}
		=
		\log\!\left(
		\frac{\sum_{i=1}^n \hat a^{\,i-1}x_i / n}
		{\left(\prod_{i=1}^n \hat a^{\,i-1}x_i\right)^{1/n}}
		\right).
	\end{equation}
	We then define the GPQ for $\alpha$, denoted by $\G_{\alpha,\mathrm G}$, as the solution to
	\begin{equation}\label{eq:alpha_star_def}
		T_n(\alpha;\bmU) = t_{\mathrm{obs}}.
	\end{equation}
	This construction matches the observed estimation equation with its auxiliary counterpart built from the exact gamma quantile representation, and therefore provides the key new step in the gamma extension. In this way, the GPQ for $\alpha$ is obtained without introducing the additional approximation layers used in existing methods. In general, $\G_{\alpha,\mathrm G}$ is computed numerically. The next theorem shows that this solution exists uniquely when $t_{\mathrm{obs}}>0$; the proof is given in Section~\ref{proof.alpha_star_unique_gam} of the Supplementary Material.

	\begin{theorem}\label{thm:alpha_star_unique_gam}
		If $t_{\mathrm{obs}}>0$, then
		the solution to
		Equation~\eqref{eq:alpha_star_def} exists and is unique.
	\end{theorem}

	Once {$\G_{\alpha,\mathrm G}$} has been obtained, the remaining parameters can be handled using $V_i=2\beta Y_i \sim \chi^2(2\alpha)$ for $i=1,\ldots,n$.
	To construct the GPQ for $a$, we replace the unknown shape parameter $\alpha$ by its GPQ $\G_{\alpha,\mathrm G}$ and generate an auxiliary sample $V_i^* \sim \chi^2(2 \G_{\alpha,\mathrm G})$. The corresponding auxiliary version of Equation~\eqref{eq:amle} is
	\begin{equation}\label{eq:g3_gamgp}
		\sum_{i=1}^n (n-2i+1)\left(\frac{\hat a}{a}\right)^{i-1}V_i^* = 0.
	\end{equation}
	Let $\phi_{a,\mathrm G}(\bmV^*)$ denote the root of Equation~\eqref{eq:g3_gamgp} corresponding to $\hat a/a$, where $\bmV^*=(V_1^*,\ldots,V_n^*)^T$. By construction, the distribution of $ \phi_{a,\mathrm G}(\bmV^*)$ is free of unknown parameters, so the GPQ for $a$ is
	\begin{equation}\label{eq:a_gamgp}
		\G_{a,\mathrm G}
		=\frac{\hat a}{\phi_{a,\mathrm G}(\bmV^*)},
	\end{equation}
	where $\hat a$ is the ML estimate obtained from the observed sample $\bmx$.
	Moreover, since $\sum_{i=1}^n V_i \sim \chi^2(2n\alpha)$, replacing $a$ and $\alpha$ by $\G_{a,\mathrm G}$ and $\G_{\alpha,\mathrm G}$ leads to the GPQ for $\beta$,
	\begin{equation}\label{eq:beta_gamgp}
		\G_{\beta,\mathrm G}=\frac{\sum_{i=1}^n V_i^*}{2 \sum_{i=1}^n (\G_{a,\mathrm G})^{\,i-1}x_i},
	\end{equation}
	where $\sum_{i=1}^n V_i^* \sim \chi^2(2n{\G_{\alpha,\mathrm G}})$.

	The above construction yields {$\G_{\boldsymbol{\theta}_{\mathrm G}}=(\G_{a,\mathrm G},\G_{\alpha,\mathrm G},\G_{\beta,\mathrm G})^{T}$}, a complete GPQ for the gamma GP. Compared with the moment-matching $\chi^2$ approximation in \citet{chen2017gamma} and the Cornish-Fisher expansion in \citet{wang2018gamma}, the present construction uses the exact gamma quantile function and auxiliary uniform variables to handle the shape parameter. It therefore provides a new GPQ route for the gamma GP that avoids those additional approximation layers. 
	
	We next summarize the asymptotic behavior of $\G_{\boldsymbol{\theta}_{\mathrm G}}$. The proof is given in Section~\ref{proof.AN_gamgp} of the Supplementary Material.
	
	\begin{theorem}\label{thm:AN_gamgp}
		Let ${\G_{\alpha,\mathrm G}}$, $\G_{a,\mathrm G}$, and $\G_{\beta,\mathrm G}$ be the GPQs defined in Equations~\eqref{eq:alpha_star_def}, \eqref{eq:a_gamgp}, and \eqref{eq:beta_gamgp}, respectively. Then
		\[
		\begin{pmatrix}
			n^{3/2}(\G_{a,\mathrm G}-a)\\[2pt]
			n^{1/2}({\G_{\alpha,\mathrm G}}-\alpha)\\[2pt]
			n^{1/2}(\G_{\beta,\mathrm G}-\beta)
		\end{pmatrix}
		\xrightarrow{d}
		N\!\left(
		\mathbf 0,\,
		2 \Sigma_{\mathrm G}^{-1}
		\right),
		\]
		where $\Sigma_{\mathrm G}^{-1}$ is given in Theorem \ref{thm:mle_gamgp}.
		Consequently, the GPQs $\G_{a,\mathrm G}$, ${\G_{\alpha,\mathrm G}}$, and $\G_{\beta,\mathrm G}$ are consistent and asymptotically normal.
	\end{theorem}

	This variance-doubling phenomenon is not specific to the exponential case; it also occurs in the gamma case. Although the gamma GP involves an additional shape parameter, its GPQs admit the same first-order representation as the difference between the observed-sample ML estimator and an independent auxiliary-sample counterpart. Since these two terms are asymptotically independent and have the same limiting covariance, the unconditional limiting covariance matrix of the GPQs is twice that of the ML estimator.

	The maintenance quantities of interest are again obtained by substitution. Let $F_G(\cdot;\alpha,\beta)$ and $f_G(\cdot;\alpha,\beta)$ denote the CDF and PDF of $\mathrm{Ga}(\alpha,\beta)$, respectively. Plugging ${\G_{\boldsymbol{\theta}_{\mathrm G}}=(\G_{a,\mathrm G},\G_{\alpha,\mathrm G},\G_{\beta,\mathrm G})^T}$ into the closed-form expressions of the cycle-specific reliability $R_{i,\mathrm G}(\tau)$ and hazard  $h_{i,\mathrm G}(\tau)$ yields
	\begin{equation}\label{eq:rihi_gamgp}
		\G_{R_{i,\mathrm G}(\tau)}
		=
		1-F_G\!\left((\G_{a,\mathrm G})^{\,i-1}\tau;\G_{\alpha,\mathrm G},\G_{\beta,\mathrm G}\right),
		\quad
		\G_{h_{i,\mathrm G}(\tau)}
		=
		\frac{(\G_{a,\mathrm G})^{\,i-1}f_G\!\left((\G_{a,\mathrm G})^{\,i-1}\tau;\G_{\alpha,\mathrm G},\G_{\beta,\mathrm G}\right)}
		{\G_{R_{i,\mathrm G}(\tau)}}.
	\end{equation}
	Hence the uncertainty encoded in $\G_{\boldsymbol{\theta}_{\mathrm G}}$ is transferred directly to the cycle-specific reliability profile relevant for subsequent maintenance decisions. By Theorem~\ref{thm:AN_gamgp}, percentile GCIs built from these GPQs are asymptotically valid. {In practice, the required quantiles are evaluated by Monte Carlo simulation. The resulting procedure is summarized in Section \ref{app.alg:gci_gamma} of the Supplementary Material.}

	\subsection{Generalized inference for the Weibull GP}
	
	The Weibull baseline is important in maintenance applications because it is the standard lifetime model for monotone aging behavior. Through its shape parameter, it can represent decreasing, constant, or increasing hazards, so it is often the natural benchmark when deterioration effects are of primary concern. However, the extension from the exponential and gamma cases is no longer straightforward. The GPQ strategy used above depends on the availability of analytically manageable pivotal relations. For the Weibull GP, such a construction becomes difficult because the shape parameter enters both power and scale-type terms, and its coupling with the geometric ratio prevents a convenient parameter-free transformation analogous to those used earlier.
	
	Rather than leaving the generalized-inference framework, we continue along the same basic principle: uncertainty should still be propagated from known auxiliary randomness to the unknown parameters and then to the maintenance quantities of interest. For the Weibull GP, this is achieved through generalized fiducial inference \citep{Hannig2009GFI}, which starts from a data-generating structural equation instead of an explicit GPQ. In this sense, the Weibull analysis is a continuation of the same inferential logic, but implemented through a different generalized-inference device.
	
	Suppose that $\{X_i,i=1,2,\dots,n\}$ is a GP with ratio parameter $a$, and that $X_1$ follows a Weibull distribution $\mathrm{WB}(m,\eta)$, where $m$ is the shape parameter and $\eta$ is the scale parameter. Let $\boldsymbol{\theta}_{\mathrm W} = (a,m,\eta)^T$. {Based on $\bmX=\{X_i, i=1,2,\ldots,n\}$, the likelihood function for $\boldsymbol{\theta}_{\mathrm W}$ is}
	\begin{equation}\label{eq:likelihood_weigp}
		L(\boldsymbol{\theta}_{\mathrm W}; \bmX) = 
		\frac{m^n}{\eta^{mn}}
		a^{\frac{mn(n-1)}{2}}
		\prod_{i=1}^n X_i^{m-1}
		\exp\!\left[-\sum_{i=1}^n\left(\frac{a^{i-1}X_i}{\eta}\right)^m\right].
	\end{equation}
	Let $\hat{\boldsymbol{\theta}}_{\mathrm W}=(\hat a,\hat m,\hat \eta)^T$ denote the corresponding ML estimator. Details of its numerical computation can be found in \citet{Usta2022Bayesian2022}. As in the exponential and gamma GP cases, the GP ratio estimator has a nonstandard convergence rate due to the increasing weight $i-1$ in the score function. The asymptotic distribution of $\hat{\boldsymbol{\theta}}_{\mathrm W}$ is given in the following theorem, with the proof provided in Section~\ref{proof.mle_weigp} of the Supplementary Material. 
	\begin{theorem}\label{thm:mle_weigp}
		Let $D_{n,\mathrm W}=\diag(n^{3/2},n^{1/2},n^{1/2})$ and $\hat{\boldsymbol{\theta}}_{\mathrm W}$ be the ML estimator based on the likelihood function in Equation \eqref{eq:likelihood_weigp}. Then 
		\[
		D_{n,\mathrm W}(\hat{\boldsymbol{\theta}}_{\mathrm W}-\boldsymbol{\theta}_{\mathrm W})
		\xrightarrow{d}
		N(\bm0,\Sigma_{\mathrm W}^{-1}),
		\]
		where
		\[
		\Sigma_{\mathrm W}^{-1}
		=
		\begin{pmatrix}
			\dfrac{12a^2}{m^2} & 0 & \dfrac{6a\eta}{m^2}\\[8pt]
			0 & \dfrac{6m^2}{\pi^2} & \dfrac{6(1-c_E)\eta}{\pi^2}\\[10pt]
			\dfrac{6a\eta}{m^2} & \dfrac{6(1-c_E)\eta}{\pi^2}
			& \dfrac{\eta^2}{m^2}\!\left(4+\dfrac{6(1-c_E)^2}{\pi^2}\right)
		\end{pmatrix},
		\]
		and $c_E$ is the Euler--Mascheroni constant.
	\end{theorem}

	Let $F_W(\cdot;m,\eta)$ denote the CDF of $\mathrm{WB}(m,\eta)$, and define
	\[
	U_i = F_W(a^{i-1}X_i;m,\eta), \qquad i=1,\ldots,n.
	\]
	Then $U_1,\ldots,U_n$ are i.i.d. with $\mathrm{Unif}(0,1)$, and the observed inter-failure times satisfy the structural equation
	\begin{equation}\label{eq:dg_weigp}
		X_i = H_{\mathrm W}(U_i;\boldsymbol{\theta}_{\mathrm W})
		= \frac{\eta}{a^{\,i-1}}\Bigl[-\log(1-U_i)\Bigr]^{1/m},\qquad i=1,\ldots,n.
	\end{equation}
	This structural equation serves as the starting point for generalized fiducial inference under the Weibull GP model.
	For the observed sample $\bmx=\{x_i,\ i=1,2,\ldots,n\}$, let
	$
	u_i = F_W(a^{i-1}x_i;m,\eta).
	$
	According to \citet{Hannig2009GFI}, the associated Jacobian is
	\begin{equation}\label{eq:J_weigp_general}
		J(\bmx;\boldsymbol{\theta}_{\mathrm W})
		=
		\sum_{1\le i_1<i_2<i_3\le n}
		\left|
		\det\!\left(
		\frac{\partial}{\partial\boldsymbol{\theta}_{\mathrm W}}
		\bigl(H_{\mathrm W}(u_{i_1};\boldsymbol{\theta}_{\mathrm W}),H_{\mathrm W}(u_{i_2};\boldsymbol{\theta}_{\mathrm W}),H_{\mathrm W}(u_{i_3};\boldsymbol{\theta}_{\mathrm W})\bigr)^T
		\right)
		\right|.
	\end{equation}
	For the Weibull GP, this expression simplifies to
	\begin{equation}\label{eq:J_weigp_explicit_short}
		J(\bmx;\boldsymbol{\theta}_{\mathrm W})
		=
		\frac{1}{a m \eta}
		\sum_{1\le i<j<k\le n}
		x_i x_j x_k
		\Bigl|
		(k-j)\log x_i+(i-k)\log x_j+(j-i)\log x_k
		\Bigr|.
	\end{equation}
	As shown by \citet{hannig2013}, the generalized fiducial distribution of $\boldsymbol{\theta}_{\mathrm W}$ has an absolutely continuous limiting density of the form
	\begin{equation}\label{eq:gfi_density_general_weigp}
		f_{F}(\boldsymbol{\theta}_{\mathrm W} ; \bmx)=
		\frac{
			L(\boldsymbol{\theta}_{\mathrm W} ;\bmx)\,J(\bmx;\boldsymbol{\theta}_{\mathrm W})
		}{
			\int_{0}^{\infty}\!\!\int_{0}^{\infty}\!\!\int_{0}^{\infty}
			L(\boldsymbol{\theta}_{\mathrm W};\bmx)\,J(\bmx;\boldsymbol{\theta}_{\mathrm W})\,da\,dm\,d\eta
		},
	\end{equation}
	where $L(\boldsymbol{\theta}_{\mathrm W};\bmx)$ is the likelihood function in Equation~\eqref{eq:likelihood_weigp} evaluated at the observed data $\bmx$. Substituting Equations~\eqref{eq:likelihood_weigp} and \eqref{eq:J_weigp_explicit_short} into Equation~\eqref{eq:gfi_density_general_weigp} yields the induced generalized fiducial density
	\begin{equation}\label{eq:gfi_density_weigp}
		f_{F}(\boldsymbol{\theta}_{\mathrm W} ; \bmx)
		\propto
		\frac{m^{\,n-1}}{\eta^{mn+1}}\, a^{\frac{mn(n-1)}{2}-1}\prod_{i=1}^n x_i^{m-1}
		\exp\!\left[-\sum_{i=1}^n\left(\frac{a^{i-1}x_i}{\eta}\right)^m\right].
	\end{equation}
	
	Equation~\eqref{eq:gfi_density_weigp} is the Weibull counterpart of the GPQs developed in the previous sections, in the sense that it provides a generalized-inference distribution on the parameter space from which interval estimates can be obtained. The following theorem establishes the large-sample validity of the resulting GCIs; the proof is given in Section~\ref{proof.AN_weigp} of the Supplementary Material.

	\begin{theorem}\label{thm:AN_weigp}
		Let $\G_{\boldsymbol{\theta}_{\mathrm W}}=(\G_{a,\mathrm W},\G_{m,\mathrm W},\G_{\eta,\mathrm W})^T$ be the generalized fiducial draw induced by the generalized fiducial density $f_F(\boldsymbol{\theta}_{\mathrm W};\bmx)$ in Equation~\eqref{eq:gfi_density_weigp}. Then
		\[
		\begin{pmatrix}
			n^{3/2}(\G_{a,\mathrm W}-a)\\[2pt]
			n^{1/2}(\G_{m,\mathrm W}-m)\\[2pt]
			n^{1/2}(\G_{\eta,\mathrm W}-\eta)
		\end{pmatrix}
		\xrightarrow{d}
		N\!\left(
		\mathbf 0,\,
		2 \Sigma_{\mathrm W}^{-1}
		\right),
		\]
		where $\Sigma_{\mathrm W}^{-1}$ is given in Theorem \ref{thm:mle_weigp}. Consequently, the generalized fiducial draw $\G_{\boldsymbol{\theta}_{\mathrm W}}$ is consistent and asymptotically normal.
	\end{theorem}

	The same covariance-doubling result holds for the Weibull case. The local expansion of the fiducial density shows that the fiducial draw has the same first-order fluctuation around the observed ML estimator as the ML estimator has around the true parameter. These two asymptotically independent fluctuations add when the draw is centered at the true parameter, leading to an unconditional limiting covariance matrix twice that of the ML estimator.

	Then GCIs for each component of $\boldsymbol{\theta}_{\mathrm W}$, as well as for any smooth function of $\boldsymbol{\theta}_{\mathrm W}$, can be constructed from the empirical quantiles of the fiducial draws. This includes the cycle-specific reliability $R_{i,\mathrm W}(\tau)$ and hazard function $h_{i,\mathrm W}(\tau)$ under the Weibull GP, namely
	\begin{equation}\label{eq:rihi_weigp}
		\G_{R_{i,\mathrm W}(\tau)}
		=\exp\!\left\{
		-\left(
		\frac{(\G_{a,\mathrm W})^{i-1}\tau}{\G_{\eta,\mathrm W}}
		\right)^{\G_{m,\mathrm W}}
		\right\},
		\
		\G_{h_{i,\mathrm W}(\tau)}
		=
		\frac{\G_{m,\mathrm W}(\G_{a,\mathrm W})^{i-1}}{\G_{\eta,\mathrm W}}
		\left(
		\frac{(\G_{a,\mathrm W})^{i-1}\tau}{\G_{\eta,\mathrm W}}
		\right)^{\G_{m,\mathrm W}-1}.
	\end{equation}
	Thus, even though an explicit GPQ is unavailable, the generalized-inference objective remains the same: uncertainty is pushed from the parameter model to the maintenance-relevant quantities used in subsequent decisions.
	
	Because direct sampling from $f_F(\boldsymbol{\theta}_{\mathrm W};\bmx)$ is not available in closed form, we generate fiducial draws by a Gibbs sampler with Metropolis--Hastings updates. Specifically, given $(a,m,\bmx)$, the conditional fiducial distribution of $\eta^m$ is the inverse gamma distribution with shape parameter $n$ and scale parameter $\sum_{i=1}^n (a^{\,i-1}x_i)^m$, denoted by $\mathrm{IGa}\!\left(n,\sum_{i=1}^n (a^{\,i-1}x_i)^m\right)$. Moreover, the conditional fiducial density of $a$, given $(m,\eta,\bmx)$, is proportional to
	\[
	\pi(a\mid m,\eta,\bmx)\propto
	a^{\frac{mn(n-1)}{2}-1}
	\exp\!\left\{-\frac{\sum_{i=1}^n (a^{\,i-1}x_i)^m}{\eta^m}\right\},
	\]
	and the conditional fiducial density of $m$, given $(a,\eta,\bmx)$, is proportional to
	\[
	\pi(m\mid a,\eta,\bmx)\propto
	m^{\,n-1}a^{\frac{mn(n-1)}{2}-1}\prod_{i=1}^n x_i^{m-1}\eta^{-(mn+1)}
	\exp\!\left\{-\frac{\sum_{i=1}^n (a^{\,i-1}x_i)^m}{\eta^m}\right\}.
	\]
	Since $\pi(a\mid m,\eta,\bmx)$ and $\pi(m\mid a,\eta,\bmx)$ do not belong to standard parametric families, $a$ and $m$ are updated by Metropolis--Hastings steps with normal proposals. After discarding burn-in draws, the retained fiducial samples are used to construct GCIs for $R_{i, \mathrm W}(\tau)$ and $h_{i, \mathrm W}(\tau)$ through their empirical quantiles. {The detailed algorithm is summarized in Section~\ref{app.alg:gfd_weibull_gp} of the Supplementary Material.}

	\section{Discrimination of baseline distributions for GP models}\label{sec:test}

	The preceding sections show that generalized inference for GP-based maintenance depends on the assumed baseline distribution. The baseline family determines the standardization $Y_i=a^{\,i-1}X_i$, and therefore affects parameter estimation, the cycle-specific reliability and hazard functions, and ultimately the maintenance decisions derived from them. Baseline discrimination is thus not a separate model-selection exercise, but an integral part of the uncertainty quantification problem studied in this paper. In practice, competing baselines are often compared using information criteria such as AIC or BIC. Although these criteria provide a convenient ranking, they do not quantify how uncertain that ranking is in finite samples, which is precisely the setting of interest here because GP applications often involve only a short sequence of failures.
	
	To address this issue, we formulate baseline discrimination as a hypothesis-testing problem based on the maximized log-likelihood difference. This choice is natural for two reasons. First, it is directly tied to the likelihood framework used throughout Sections~\ref{sec:exp} and \ref{sec:wei_gam}. Second, it admits an interpretation in terms of the Kullback--Leibler information criterion (KLIC): under misspecification, log-likelihood differences compare the relative closeness of competing models to the true data-generating mechanism at their pseudo-true parameters \citep{White1982, Schennach2017}. Hence the resulting procedure applies not only to nested candidates but also to nonnested baseline families such as the exponential, gamma, and Weibull distributions considered here.

	Suppose that two candidate baseline distributions with densities $f_0(\cdot;\boldsymbol\vartheta_0)$ and $f_1(\cdot;\boldsymbol\vartheta_1)$ are under consideration. The discrimination problem for the GP $\bmX=\{X_i,\ i=1,2,\dots,n\}$ is formulated as
	\begin{equation}
		H_0:\ \bm X \sim GP(a_0,f_0(\cdot;\boldsymbol\vartheta_0))
		\quad \text{versus} \quad
		H_1:\ \bm X \sim GP(a_1,f_1(\cdot;\boldsymbol\vartheta_1)),
		\label{eq:H0H1}
	\end{equation}
	where $\bm X \sim GP(a_k,f_k)$ for $k=0,1$ means that $\bm X$ is generated from a GP with ratio parameter $a_k$ and baseline density $f_k(\cdot;\boldsymbol\vartheta_k)$.
	Based on the observed data $\bmx=\{x_i,\ i=1,2,\ldots,n\}$, the ratio of maximized likelihoods is defined as
	\begin{equation}
		\mathrm{RML}=\frac{\max_{a_0,\boldsymbol\vartheta_0}
			\prod_{i=1}^n a_0^{\,i-1}
			f_0(a_0^{\,i-1}x_i;\boldsymbol\vartheta_0)
		}{\max_{a_1,\boldsymbol\vartheta_1}
			\prod_{i=1}^n a_1^{\,i-1}
			f_1(a_1^{\,i-1}x_i;\boldsymbol\vartheta_1)}.
		\label{eq:RML}
	\end{equation}
	Equivalently, we work with the log-likelihood difference
	\begin{equation}
		\mathcal S(\bmx)=
		\ell_0(\hat a_0,\hat{\boldsymbol\vartheta}_0;\bmx)-
		\ell_1(\hat a_1,\hat{\boldsymbol\vartheta}_1;\bmx),
		\label{eq:Tstat}
	\end{equation}
	where $\ell_k(\hat a_k,\hat{\boldsymbol\vartheta}_k;\bmx)$ is the maximized log-likelihood under $H_k$, $k=0,1$. 
	A negative value of $\mathcal S(\bmx)$ favors $H_1$, whereas a positive value favors $H_0$.

	The remaining issue is how to calibrate $\mathcal S(\bmx)$ in the small-sample GP setting. As emphasized earlier, asymptotic approximations alone may be unreliable when only a few failures are available. We therefore exploit the generalized-inference machinery developed in Sections~\ref{sec:exp} and \ref{sec:wei_gam}. Under the null model $H_0$, let $(\G_{a_0},\G_{\boldsymbol\vartheta_0})$ denote a draw from the inferential distribution of $(a_0,\boldsymbol\vartheta_0)$, namely the distribution induced by the GPQs for the exponential or gamma GP, or by the generalized fiducial distribution for the Weibull GP. Conditional on $(\G_{a_0},\G_{\boldsymbol\vartheta_0})$, one can generate pseudo baseline observations from $f_0(\cdot;\G_{\boldsymbol\vartheta_0})$, transform them back to the GP scale using $\G_{a_0}$, and recompute the statistic $\mathcal S(\bmx)$. 
	Repeating this procedure yields an empirical approximation to the finite-sample null distribution of $\mathcal S(\bmx)$ while preserving the model-specific uncertainty quantified in the previous sections. In this sense, the discrimination procedure is fully integrated with the generalized-inference framework rather than appended to it. {The resulting procedure is in Section~\ref{app.alg:gpq_discrimination} of the Supplementary Material.
	}

	\vspace{-1em}
	
	\begin{remark}
		When more than two baseline families are entertained, the above procedure can be applied pairwise. In particular, for the exponential, gamma, and Weibull GP models considered in this paper, the test offers a formal complement to AIC or BIC ranking by quantifying whether the observed likelihood advantage of one baseline over another is large relative to the inferential uncertainty induced by the short GP record. This is precisely the type of baseline-selection uncertainty that the introduction identified as one of the two major sources of uncertainty in GP-based maintenance analysis.
	\end{remark}

	\section{Simulation study}\label{sec:simulation}
	
	This section reports Monte Carlo evidence on the finite-sample performance of the proposed methods. The simulation study has two objectives. The first is to assess the proposed GCIs for model parameters and cycle-specific reliability measures, which constitute the main inferential output of the paper. The second is to evaluate the proposed discrimination procedure for selecting among candidate GP baselines. Throughout, all results are based on $10{,}000$ Monte Carlo replications, and $B=10{,}000$ is used in the corresponding algorithms.
	
	\subsection{Interval estimation}\label{subsec:sim_intervals}

\begin{figure}[H]
		\centering
		\includegraphics[width=0.9\linewidth]{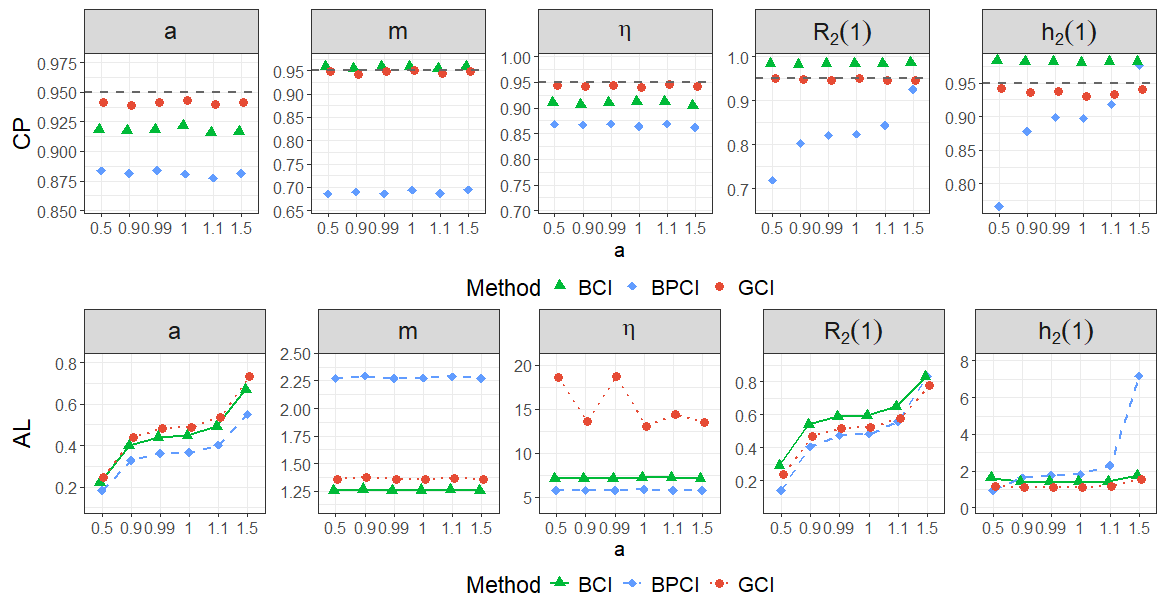}
		\caption{Coverage probabilities and average interval lengths  of 95\% confidence intervals for the Weibull GP with $n=10$.}
		\label{fig:wei_n10}
	\end{figure}

\begin{figure}[H]
		\centering
		\includegraphics[width=0.9\linewidth]{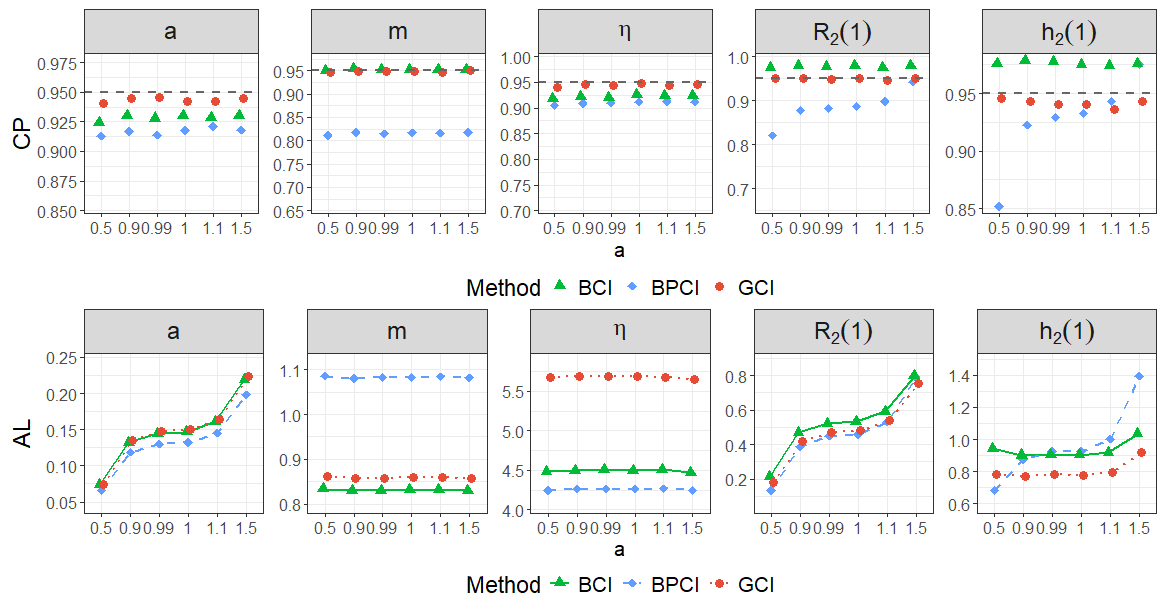}
		\caption{Coverage probabilities and average interval lengths  of 95\% confidence intervals for the Weibull GP with $n=20$.}
		\label{fig:wei_n20}
	\end{figure}

\begin{figure}[htbp]
		\centering
		\includegraphics[width=0.9\linewidth]{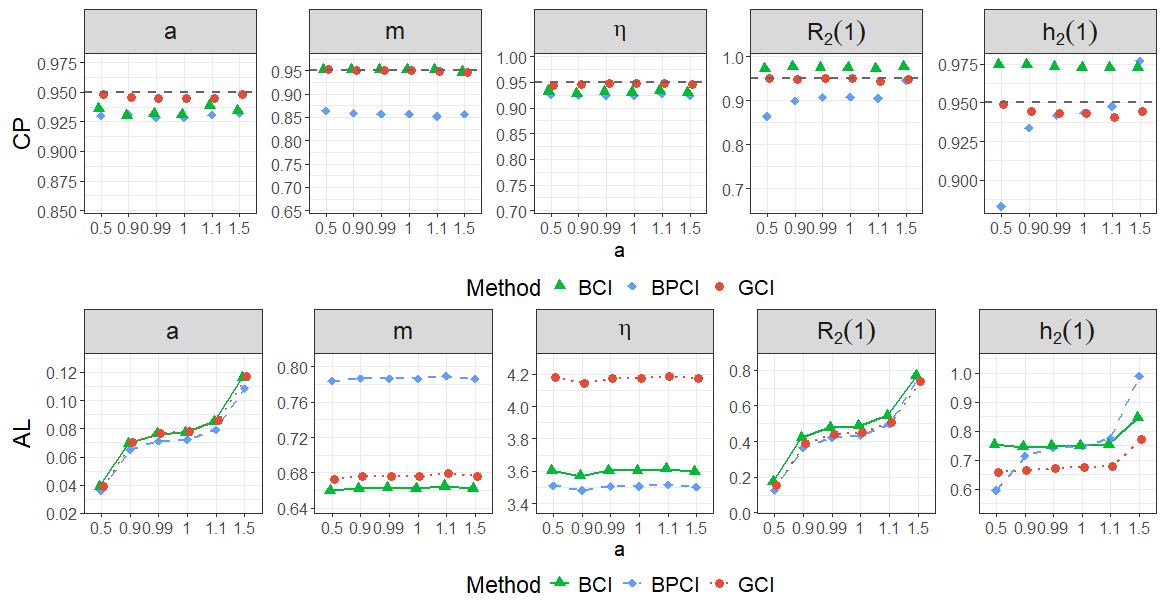}
		\caption{Coverage probabilities and average interval lengths  of 95\% confidence intervals for the Weibull GP with $n=30$.}
		\label{fig:wei_n30}
	\end{figure}

	We examine the interval estimation performance under the Weibull GP, taking $m=1.13$, $\eta=2.75$, $a\in\{0.5,0.9,{0.99},1,1.1,1.5\}$, and $n\in\{10,20,30\}$. For the cycle-specific reliability measures, we use one-sided confidence bounds, namely the lower bound for $R_i$ and the upper bound for $h_i$, because maintenance decisions are usually based on conservative assessments of the minimum reliable operating level and the maximum plausible failure risk. We also report the bootstrap confidence intervals (BPCIs) and the Bayesian credible intervals (BCIs) proposed by \cite{ChenLiLam2010} for comparison. The prior distributions used for the BCIs are $a\sim \mathrm{Unif}(0.001,3)$, $m\sim \mathrm{Ga}(1,1)$, and $\eta\sim \mathrm{Ga}(1,0.0001)$. The performance is evaluated in terms of coverage probability (CP) and average interval length (AL). For the one-sided bounds, the corresponding length is defined as $1$ minus the lower bound for $R_i$ and as the upper bound for $h_i$. The results are presented in Figures~\ref{fig:wei_n10}--\ref{fig:wei_n30}.

	The main finding is that the proposed GCIs achieve CPs close to the nominal level $0.95$ across all sample sizes for both the model parameters $(a,m,\eta)$ and the cycle-specific reliability measures $R_2(1)$ and $h_2(1)$. By comparison, the BCIs exhibit mild undercoverage for $a$ and $\eta$, and overcoverage for $R_2(1)$ and $h_2(1)$ in several settings. The BPCIs show more pronounced undercoverage, especially for $m$ and $h_2(1)$. These results suggest that the proposed GCIs provide the most reliable coverage performance under the Weibull GP.
	In terms of AL, the ALs for the model parameters generally decrease as $n$ increases. The BPCIs often yield wider intervals for $m$ than either the BCIs or GCIs, but this extra width does not translate into better coverage. For $R_2(1)$ and $h_2(1)$, the lengths tend to increase with $a$. This is because, for $a>1$, a larger GP ratio parameter corresponds to faster deterioration, so that $R_i(\tau)$ decreases while $h_i(\tau)$ increases at a fixed cycle $i$ and time $\tau$.
	Overall, the proposed generalized-inference method continues to provide the most stable near-nominal coverage across both model parameters and derived reliability measures under the Weibull GP, further supporting the use of generalized inference for uncertainty quantification in maintenance-relevant functions when failure data are limited.

	\vspace{-1em}
	
	\subsection{Discrimination test}\label{subsec:sim_test}
	
\begin{figure}[H]
		\centering
		\includegraphics[width=0.9\linewidth]{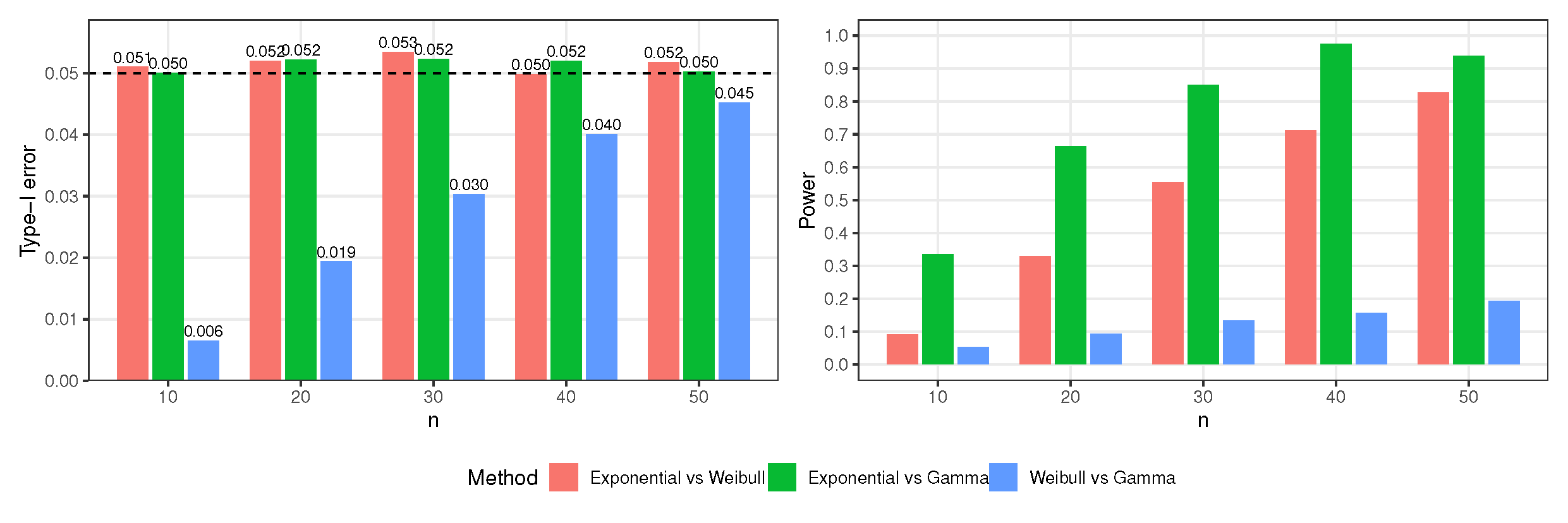}
		\caption{Empirical Type-I error and power for the discrimination test among the exponential, Weibull, and gamma baselines when $n = 10,20,30,40,50$.}
		\label{fig:test}
	\end{figure}

	We next examine the finite-sample behavior of the proposed generalized-inference discrimination procedure. At the nominal level $0.05$, Figure~\ref{fig:test} reports the empirical Type-I error and power for $n=10,20,30,40,50$ in the three pairwise comparisons among the exponential, Weibull, and gamma baselines.

	A first important result is that the empirical Type-I error is well controlled across all three tests. The exponential--Weibull and exponential--gamma comparisons remain close to the nominal level throughout, while the Weibull--gamma comparison is slightly conservative for small $n$ but approaches $0.05$ as $n$ increases. This indicates that the proposed procedure provides satisfactory size control even when the baseline families are compared under limited sample sizes.
	
	The second key result concerns power. As expected, power generally increases with $n$ in all three pairwise comparisons, but the rate of increase differs across model pairs. The exponential--gamma comparison is the easiest to distinguish and consistently achieves the highest power. The exponential--Weibull comparison shows moderate power, whereas the Weibull--gamma comparison is the most difficult under the settings considered. This ranking is intuitively consistent with the relative similarity of the candidate baseline families and shows that the proposed discrimination test is informative in finite samples, while still reflecting the genuine difficulty of distinguishing flexible models such as Weibull and gamma from one another.
	
	Taken together, the simulation results suggest that the proposed discrimination procedure maintains appropriate size and offers increasing power as more failure cycles are observed. This complements the interval-estimation results above: the proposed framework performs well both in quantifying uncertainty conditional on a chosen baseline and in formally assessing uncertainty about the baseline itself.

	\section{ Case study}\label{sec:case}

\begin{figure}[H]
		\centering
		\includegraphics[width=0.8\linewidth]{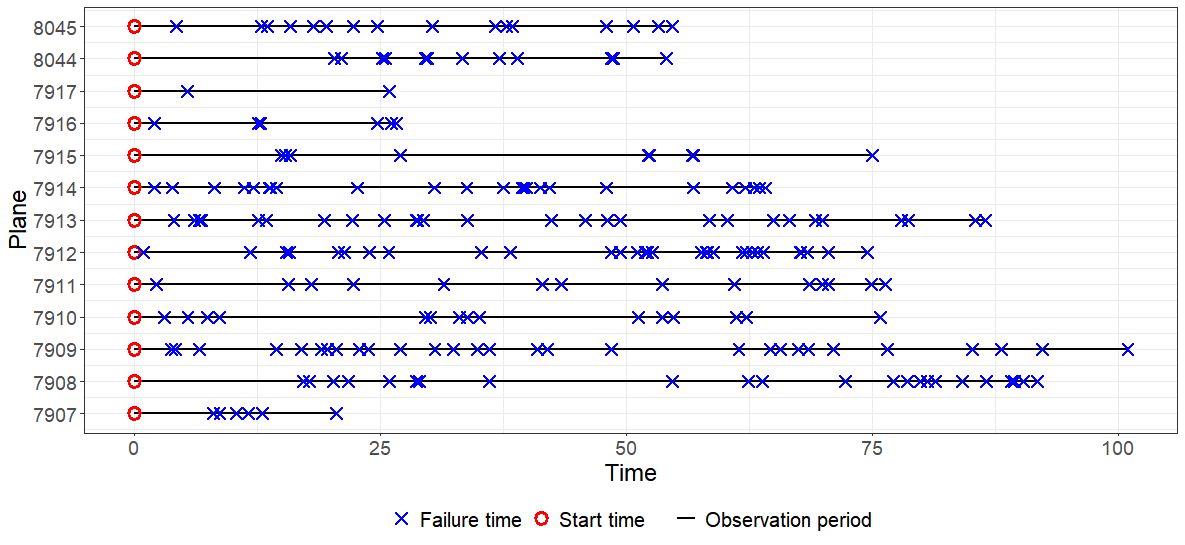}
		\caption{Failure times in operating days of air-conditioning equipment for Boeing 720 jet aircraft.}
		\label{fig:aircraft_all}
	\end{figure}

	This case study uses the successive-failure data on the air-conditioning systems of Boeing 720 jet aircraft reported by \citet{Proschan1963}. The full dataset contains 13 aircraft, each with recorded operating times between successive failures. To provide an overall view of the data, Figure~\ref{fig:aircraft_all} plots the cumulative failure-time histories for all 13 aircraft after converting the original times from hours to days. The figure shows substantial variation in failure patterns across aircraft. For instance, the failure events of Plane 7912 become increasingly clustered over time, whereas those of Plane 7913 become more widely spaced, suggesting different trends in the inter-failure times. Such cross-aircraft variation is consistent with differences in operating conditions, maintenance history, and usage intensity, and it motivates the use of the GP model to capture systematic changes in successive inter-failure times. In the following, we focus on Plane 7913, which contains 27 inter-failure times and was previously analyzed by \citet{Lam2007}, who showed that a GP provides an adequate fit to this aircraft.

	\subsection{Baseline selection and model fitting}\label{subsec:baseline_mle}

	To choose a baseline distribution for the GP model, we first use information criteria for an initial comparison. Table~\ref{tab:aic} reports the AIC, BIC, and log-likelihood under the exponential, gamma, and Weibull baselines. The exponential baseline is favored by AIC and BIC, while the Weibull baseline gives the largest log-likelihood. The gamma baseline is close to the Weibull baseline in terms of log-likelihood, whereas its AIC and BIC are slightly larger.
	
	\begin{table}[H]
		\centering
		\begin{spacing}{1} 
			\caption{Information criteria for different baselines of Plane 7913.}
			\label{tab:aic}
			\begin{tabular}{lcccc}
				\toprule
				Baseline & AIC  & BIC & Log-likelihood \\
				\midrule
				Exponential & \textbf{120.509} & \textbf{123.100} & $-58.254$ \\
				Gamma & 122.210 & 126.098 & $-58.105$ \\
				Weibull &  121.898 & 125.785 & $\mathbf{-57.949}$ \\
				\bottomrule
			\end{tabular}
		\end{spacing}
	\end{table}
	
	Information criteria provide only a post-sample comparison and do not directly quantify the strength of evidence in a finite sample, especially when competing models fit similarly well. We therefore  apply the proposed  discrimination test at $\gamma=0.05$, rejecting $H_0$ when $\mathcal S_0<\mathcal S_c$. Since the Weibull baseline attains the largest log-likelihood and is widely used in  reliability analysis, we take it as the null model and examine whether there is sufficient evidence to reject it in favor of alternative baselines. The results are summarized in Table~\ref{tab:test}. The test does not reject the Weibull baseline against either the exponential or gamma alternative. Therefore, although the exponential model is preferred by AIC and BIC, the discrimination results do not provide sufficient evidence against the Weibull baseline. Taking both pieces of evidence into account, we use the Weibull distribution as the working baseline for the subsequent analysis.

	\begin{table}[H]
		\centering
		\begin{spacing}{1} 
			\caption{Discrimination results for baselines of Plane 7913 at significance level $\gamma=0.05$.}
			\label{tab:test}
			\begin{tabular}{lllll}
				\toprule
				Hypothesis   & $\mathcal S_0$  & $\mathcal S_c$ & Decision \\
				\midrule
				$H_0$: Weibull  $H_1$: Exponential
				& 0.305  & 0.006   & Fail to reject $H_0$ \\
				$H_0$: Weibull $H_1$: Gamma & 0.156 & -0.459 & Fail to reject $H_0$ \\
				
				\bottomrule
			\end{tabular}
		\end{spacing}
	\end{table}

	Table~\ref{tab:ex_pa} reports the ML estimates and the corresponding 95\% BPCIs, BCIs, and GCIs for the Weibull GP parameters based on the full dataset. These results summarize the overall failure pattern of Plane 7913 under the fitted Weibull GP model and provide a reference for the subsequent sequential reliability and maintenance analyses. The estimate of the GP ratio parameter $a$ is slightly less than 1, suggesting that the inter-failure times tend to become longer over successive cycles. Across the three methods, the interval estimates are broadly similar, which indicates that inference is fairly stable when the full dataset is used.

	\begin{table}[H]
		\centering
		\begin{spacing}{1} 
			\caption{ML and interval estimates for model parameters of Plane 7913.}
			
			\label{tab:ex_pa}
			\begin{tabular}{ccccccccccccc}
				\toprule
				& Parameter & ML estimate  & BPCI & BCI & GCI \\
				\midrule
				
				& $a$ 	&0.986   &(0.942, 1.031) &(0.930, 1.026) & (0.944, 1.031)\\[0.6ex]
				& $m$	& 1.135	&(0.899,  1.693) 	&(0.785, 1.470) 	&(0.778,  1.452)  \\[0.6ex]
				& $\eta$ &2.750  &(1.321, 5.088)     &(1.172, 5.179)  &(1.441, 5.751) \\
				\bottomrule
			\end{tabular}
		\end{spacing}
	\end{table}

	\subsection{Reliability assessment under sequential updating}\label{subsec:example_CI}

\begin{figure}[H]
		\centering
		\includegraphics[width=0.9 \linewidth]{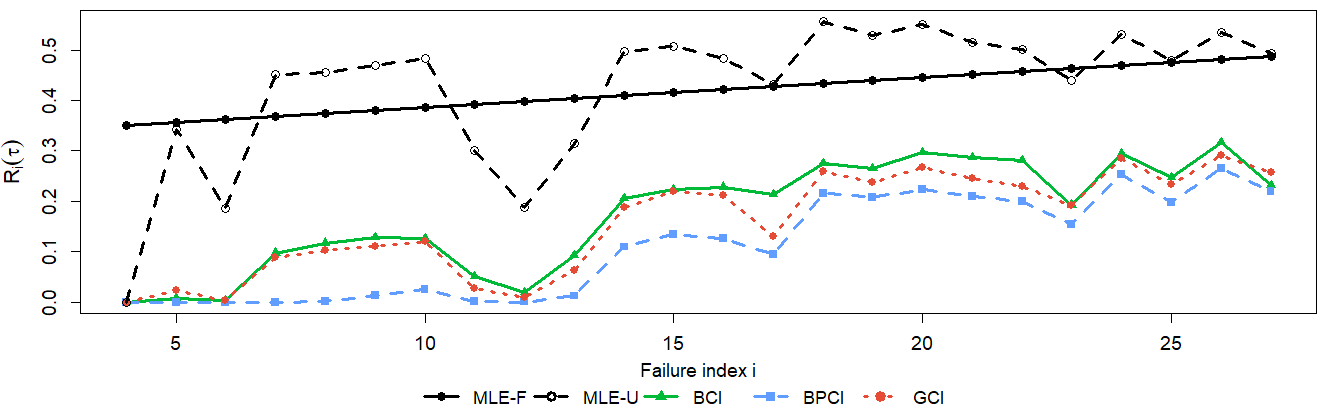}
		\caption{The 95\% lower bounds of $R_{i}(3)$ across failure index $i$ of Plane 7913.}
		\label{fig:Ri_low}
	\end{figure}

	Figure~\ref{fig:Ri_low} reports $R_i(3)$ across the failure index $i$, together with the corresponding 95\% lower bounds obtained from the bootstrap, Bayesian, and generalized inference methods. Here, MLE-F denotes the ML estimates based on the full dataset, while MLE-U denotes the updating estimates obtained sequentially using the failure data available up to cycle $i$. This distinction is important because the full-data estimates summarize the overall reliability pattern after all failures have been observed, whereas the sequential estimates reflect the information actually available when maintenance decisions are updated over time.
	
	As can be seen, the full-data ML estimates follow a smooth increasing trend, while the sequential ML estimates are much more variable in the early cycles. The increasing pattern of the full-data estimates is consistent with the fitted Weibull GP model, for which the estimate of the GP ratio parameter $a$ is slightly less than 1, implying that the inter-failure times tend to become longer over successive cycles. By contrast, the sequential estimates are based only on the limited failure history available at each stage and are therefore more sensitive to local observations. In particular, the third and fourth inter-failure times are much shorter than the preceding observations, so the sequential estimates give a very low assessment of short-term reliability in the early stage.
	
	This feature is also reflected in the lower bounds. At the beginning of the sequence, all three methods produce 95\% lower bounds that are close to zero, and the lower bounds are essentially zero at $i=4$. This indicates that, given the limited early-cycle information, the available evidence provides very weak support for a reliable short-term operating condition. As additional failures are observed, the lower bounds increase and become more stable, showing that the uncertainty in $R_i(3)$ is gradually reduced. From the middle cycles onward, the three lower-bound curves are close to each other, which suggests that, once more failure information has accumulated, the differences among the inferential methods become much less important. 
	In addition, the simulation results in Section~\ref{subsec:sim_intervals} show that, under the Weibull GP model with parameter values matching those in this example, the proposed GCIs yield lower bounds whose CPs are closer to the nominal level than the corresponding BPCIs and BCIs, especially for small samples.
	
	From the viewpoint of maintenance decision-making, the figure suggests that early-cycle decisions based on sequential reliability assessment may be much more conservative than those suggested by the full-data estimates. In particular, when the lower bound of $R_i(3)$ is close to zero, the available information does not support delaying replacement, so a risk-based policy would tend to favor earlier action. Therefore, the difference between MLE-U and MLE-F is not only statistical, but can also lead to different replacement decisions in the early cycles.

	\subsection{Maintenance implications and replacement decisions}
	
	In this subsection, we concentrate on the maintenance implications of the sequentially updated analysis. Because the three methods give similar results in the previous reliability assessment, and the simulation study under the same parameter setting has already demonstrated the superior interval performance of the proposed GCI method, only the GCI-based maintenance results are presented here. 
	
	\subsubsection{Cycle-by-cycle updating of the replacement decision}
	
	We translate the uncertainty in parameter estimation into uncertainty in the next-cycle replacement decision. Consider an age-replacement policy with threshold $\tau$: in the $i$-th operating cycle, the unit undergoes preventive replacement at age $\tau$ if no failure occurs before $\tau$, and corrective replacement upon failure if $X_i<\tau$. Let $C_f$ and $C_p$ denote the costs of corrective and preventive replacement, respectively, where typically $C_f>C_p$.
	To allow adaptive decisions across cycles, we adopt the one-cycle optimality criterion, which minimizes the expected cost per unit time within a single cycle \citep{CoolenSchrijner2007}.
	Consequently, the one-cycle expected cost rate for the replacement strategy $\tau$ in the $i$-th cycle is
	\begin{equation}
		C_{i}(\tau)
		= C_f\int_{0}^{\tau}\frac{1}{x}f_i(x)\,\mathrm{d}x
		+\frac{C_p}{\tau}R_i(\tau),
		\label{eq:one_cycle_expected_cost_rate}
	\end{equation}
	where $f_i(\cdot)$ is the PDF of $X_i$. 
	
	Under the Weibull GP assumption, the one-cycle optimal replacement time in the $i$-th cycle  admits the closed-form expression
	\begin{equation}
		\tau_i^\star=
		\frac{\eta}{a^{\,i-1}}
		\left(\frac{C_p}{m(C_f-C_p)}\right)^{1/m}.
		\label{eq:tau_star_weibull_gp_onecycle}
	\end{equation}
	Moreover, the cost rate $C_{i}(\tau)$ is given by
	\begin{equation}
		C_{i}(\tau)=
		C_f\,\frac{a^{\,i-1}}{\eta}\,
		\gamma\!\left(1-\frac{1}{m},\, \left(\frac{a^{\,i-1}\tau}{\eta}\right)^m\right)
		+\frac{C_p}{\tau}\exp\!\left[-\left(\frac{a^{\,i-1}\tau}{\eta}\right)^m\right].
		\label{eq:one_cycle_expected_cost_rate_weibull_gp_closed}
	\end{equation}
	where $\gamma(\cdot,\cdot)$ is the lower incomplete gamma function.

\begin{figure}[h]
		\centering
		\includegraphics[width=0.75\linewidth]{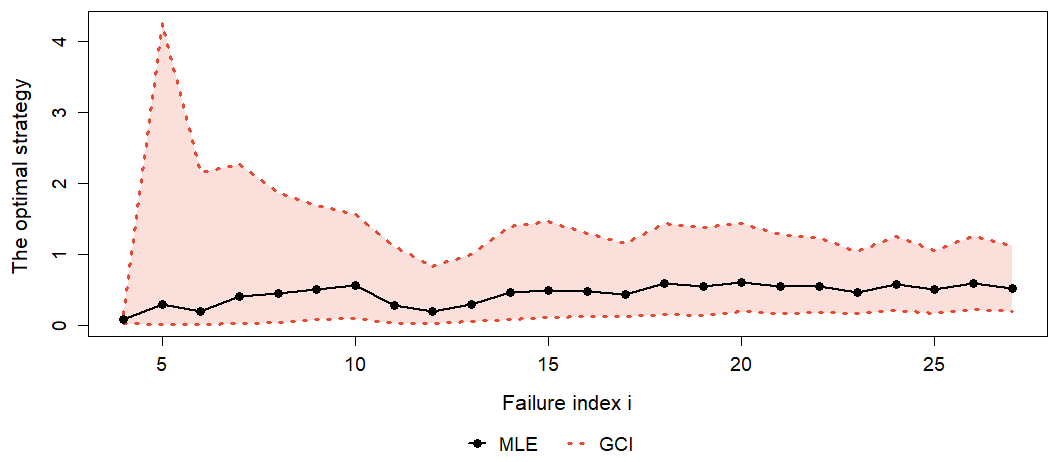}
		\caption{ {One-cycle optimal replacement time $\tau_i^*$ across failure index $i$ under the Weibull GP.}}
		\label{fig:tau_twoside}
	\end{figure}

	Figure~\ref{fig:tau_twoside} reports the one-cycle optimal replacement time under the Weibull GP when the policy is re-optimized after each newly observed failure. The plug-in ML estimates yield a single value of $\tau_i^\star$ in each cycle, and the resulting sequence varies only modestly with $i$. 
	In contrast, the 95\% GCIs for $\tau_i^\star$ are very wide in the early cycles and contract as the failure history accumulates.
	In the early stage, the width of the interval indicates that the data do not locate the optimum sharply. Over a broad range of $\tau_i$, the one-cycle cost rates are hard to distinguish, so many thresholds remain plausible at the 95\% level. A single point recommendation, therefore, conveys more certainty than the data support. As $i$ increases, the interval narrows, suggesting that the information in the observed failures becomes sufficient to stabilize the next-cycle recommendation.
	
	Table~\ref{tab:tau_Ri_hi_cost_bounds} further connects the decision intervals for $\tau_i^\star$ to action-relevant quantities by reporting the plug-in values of $R_i(\tau_i^\star)$, $h_i(\tau_i^\star)$, and $C_i(\tau_i^\star)$ at the point estimates and at the lower and upper bounds. Overall, the lower-bound strategy corresponds to an earlier preventive replacement with higher reliability at the planned time, whereas the upper-bound strategy corresponds to a later action with lower reliability. The table also shows that substantially different reliability levels may be associated with only modest differences in cost rate. For example, at $i=9$, the point and upper-bound strategies yield cost rates of 12.705 and 13.767, although the corresponding reliabilities are 0.912 and 0.670, respectively. At $i=13$, the three cost rates are all close to 360, again indicating that the one-cycle cost objective is fairly flat around its optimum. These results show that reporting an interval for $\tau_i^\star$ is useful in practice, since it reveals a range of nearly cost-equivalent decisions with materially different risk implications.
	
	\begin{table}[H]
		\centering
		\begin{spacing}{1}
			\caption{{Plug-in estimates of $R_i(\tau_i^*)$, $h_i(\tau_i^*)$, and $C_i(\tau_i^*)$ at failure indices $i=4,9,13$, based on the point estimates and the lower and upper confidence bounds of $\tau_i^*$.}}
			\label{tab:tau_Ri_hi_cost_bounds}
			\renewcommand{\arraystretch}{1.15}
			\begin{tabular}{c c c c c c}
				\toprule[0.5mm]
				$i$ & Strategy & $\tau_i^*$ & $R_i(\tau_i^*)$ & $h_i(\tau_i^*)$ & $C_i(\tau_i^*)$ \\
				\hline
				4
				& point & 0.084 & 0.980 & 0.039 & 14.441 \\
				& lower & 0.016 & 1.000 & 0.000 & 64.426 \\
				& upper & 0.143 & 0.690 & 0.427 & 31.796 \\
				9 
				& point & 0.498 & 0.912 & 0.334 & 12.705 \\
				& lower & 0.076 & 0.990 & 0.228 & 20.494 \\
				& upper & 1.684 & 0.670 & 0.427 & 13.767 \\
				13 
				& point & 0.292 & 0.896 & 0.376 & 360.280 \\
				& lower & 0.043 & 0.984 & 0.368 & 373.181 \\
				& upper & 0.990 & 0.686 & 0.381 & 361.675 \\
				\bottomrule[0.5mm]
			\end{tabular}
		\end{spacing}
	\end{table}

	\subsubsection{Sensitivity to cost structure and risk preference}
	
\begin{figure}[h]
		\centering
		\includegraphics[width=0.75\linewidth]{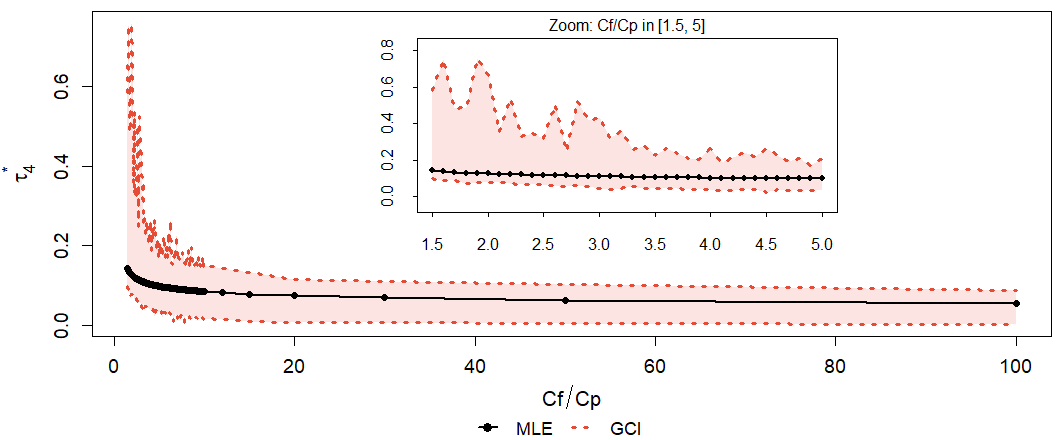}
		\caption{ The optimal replacement time $\tau_4^*$ versus cost ratio $C_f/C_p$ for Plane 7913 under the Weibull GP model at failure index $i=4$.}
		\label{fig:tau_policy}
	\end{figure}

	Figure~\ref{fig:tau_policy} links the one-cycle replacement decision at  failure index $i=4$ to the cost structure. As the ratio $C_f/C_p$ increases, the plug-in policy recommends earlier replacement, as higher failure costs naturally lead to a more preventive replacement strategy. The figure also suggests that the degree of decision uncertainty depends on the cost ratio. 
	When $C_f/C_p$ is small, the GCI is relatively wide, indicating that the data support only a broad range of plausible replacement times. In this setting, reporting only a point-optimal $\tau_4^\star$ may overstate precision. As $C_f/C_p$ becomes larger, the plug-in estimate shifts toward smaller $\tau_4^\star$, and the GCI becomes narrower, indicating a more stable preventive replacement decision.
	Overall, the figure shows a clear shift from greater flexibility at low cost ratios to a more concentrated replacement window as the failure cost becomes more dominant.
	
	\section{Conclusion}
	\label{sec:conclusion}
	
	This paper studies finite-horizon maintenance decision making for repairable systems under the GP model. Motivated by event-driven settings in which maintenance decisions are updated after each newly observed failure, we move beyond long-run average criteria and focus on cycle-specific quantities that are directly relevant to practice. The proposed generalized-inference framework quantifies uncertainty not only in the GP model parameters, but also in maintenance-relevant quantities such as cycle-specific reliability, hazard rate, and the one-cycle optimal replacement time. It also incorporates a finite-sample baseline discrimination procedure, thereby addressing both decision uncertainty and baseline-selection uncertainty under limited failure information.
	
	The numerical studies and aircraft case study demonstrate the value of reporting uncertainty at the decision level. The proposed GCIs provide stable coverage for model parameters and nonlinear maintenance-relevant quantities, while the discrimination procedure complements information-criterion ranking by formally quantifying baseline-selection uncertainty. In the case study, early-cycle failure histories lead to substantial uncertainty in the optimal replacement time. Replacement times within the estimated interval may yield similar expected cost, but correspond to markedly different reliability and risk profiles, showing that a single plug-in recommendation can overstate the precision supported by the data. As more failures are observed, and as failure costs become larger relative to preventive replacement costs, the decision interval narrows and the recommended maintenance schedule becomes easier to justify. Overall, the proposed framework provides a data-supported range of replacement times rather than a single point recommendation, offering more transparent and practically useful decision support for maintenance planning with limited failure data.
	
	\section*{Disclosure statement}
	No potential conflict of interest was reported by the authors.
	\section*{Data availability statement}
	The authors confirm that the data supporting the findings of this study are available within the supplementary materials.
	
	\clearpage
	\setcounter{section}{0}
	\setcounter{subsection}{0}
	\setcounter{subsubsection}{0}
	\setcounter{figure}{0}
	\setcounter{table}{0}
	\setcounter{equation}{0}
	\setcounter{algorithm}{0}
	\renewcommand{\thesection}{S\arabic{section}}
	\renewcommand{\thefigure}{S\arabic{figure}}
	\renewcommand{\thetable}{S\arabic{table}}
	\renewcommand{\theequation}{S\arabic{equation}}
	\renewcommand{\thealgorithm}{S\arabic{algorithm}}
	\renewcommand{\theHsection}{supp.\arabic{section}}
	\renewcommand{\theHfigure}{supp.\arabic{figure}}
	\renewcommand{\theHtable}{supp.\arabic{table}}
	\renewcommand{\theHequation}{supp.\arabic{equation}}
	\renewcommand{\theHalgorithm}{supp.\arabic{algorithm}}
	\setlength\parindent{0.7cm}
	\setstretch{1.5}
	\title{\bf{Supplementary Material for} \\
		Finite-horizon maintenance under geometric-process models: Quantifying model and decision uncertainty}
	\author{}
	\date{}

	\maketitle
	
	In this supplementary material, we use the following notation. The symbol ``$\xrightarrow{\mathrm{a.s.}}$'' denotes almost sure convergence, ``$\xrightarrow{P}$'' denotes convergence in probability, and ``$\xrightarrow{d}$'' denotes convergence in distribution. The symbol ``$\to$'' denotes the standard limit as $n\to\infty$.

	\section{Proof of Theorem \ref{thm:mle_exp}}\label{proof.mle_exp}
	

	To establish the consistency and asymptotic distribution of the maximum likelihood (ML) estimator $(\hat a, \hat\lambda)$, we analyze the estimation equation $\Phi_n(\hat a)=0$ at the true value $a$, where
	\[
	\Phi_n(u)=\sum_{i=1}^n (n-2i+1)u^{\,i-1}X_i, \,\, u>0.
	\]
	Let $Y_i=a^{i-1}X_i$, $i=1,2,\ldots,n$. 
	Then, $Y_1,Y_2,\ldots,Y_n$ are independent and identically distributed exponential random variables with rate parameter $\lambda$. Write $\varepsilon_i=\lambda Y_i-1$, where $E(\varepsilon_i)=0$ and $\Var(\varepsilon_i)=1$. Since
	$\sum_{i=1}^n(n-2i+1)=0$, we have
	\[
	\Phi_n(a)=\sum_{i=1}^n (n-2i+1)Y_i
	=\frac1\lambda\sum_{i=1}^n (n-2i+1)\varepsilon_i
	=O_p(n^{3/2}).
	\]
	Moreover,
	\[
	\Phi_n'(a)=\frac1a\sum_{i=1}^n (n-2i+1)(i-1)Y_i.
	\]
	{
		Let $c_{i,n}=(n-2i+1)(i-1)$. Then
		\[
		n^{-3}\Phi_n'(a)
		=\frac{1}{a\lambda n^3}\sum_{i=1}^n c_{i,n}
		+\frac{1}{a \lambda n^3}\sum_{i=1}^n c_{i,n} \varepsilon_i.
		\]
		Since $\sum_{i=1}^n c_{i,n}=-{n(n^2-1)}/{6},$ the first term of $n^{-3}\Phi_n'(a)$ converges to $-1/(6a\lambda)$. 
		For the second term, note that $\sum_{i=1}^n c_{i,n}^2=O(n^5)$, then we obtain
		\[
		\Var\!\left(\frac{1}{a \lambda n^3}\sum_{i=1}^n c_{i,n} \varepsilon_i \right)
		=\frac{1}{a^2\lambda^2 n^6}\sum_{i=1}^n c_{i,n}^2
		=O(n^{-1})\to 0,
		\]
		Hence, the second term of $n^{-3}\Phi_n'(a)$ is $o_p(1)$, and therefore $n^{-3}\Phi_n'(a)\xrightarrow{P}-{1}/{(6a\lambda)}.$

		We now prove the local uniform convergence of  $n^{-3}\Phi_n'(u)$ in a shrinking neighborhood of $a$. For fixed $M>0$, define
		$
		I_{n,M}=\{u:|u-a|\le Mn^{-3/2}\}.
		$
		Uniformly for $u\in I_{n,M}$, we can obtain $\max_{1\le i\le n}
		\left|\left({u}/{a}\right)^{i-1}-1\right|=O(n^{-1/2}),
		\,\, 1/u-1/a=O(n^{-3/2}).$
		Hence,
		\[
		\max_{1\le i\le n}\sup_{u\in I_{n,M}}
		\left|
		\frac1u\left(\frac{u}{a}\right)^{i-1}
		-\frac1a
		\right|
		=O(n^{-1/2}),
		\]
		and consequently,
		\[
		\begin{aligned}
			\sup_{u\in I_{n,M}}|\Phi_n'(u)-\Phi_n'(a)|
			&\le
			\sum_{i=1}^n |c_{i,n}|Y_i
			\sup_{u\in I_{n,M}}
			\left|
			\frac1u\left(\frac{u}{a}\right)^{i-1}
			-\frac1a
			\right|  \\
			&\le
			\left[
			\max_{1\le i\le n}\sup_{u\in I_{n,M}}
			\left|
			\frac1u\left(\frac{u}{a}\right)^{i-1}
			-\frac1a
			\right|
			\right]
			\sum_{i=1}^n |c_{i,n}|Y_i  \\
			&=
			O(n^{-1/2})\sum_{i=1}^n |c_{i,n}|Y_i .
		\end{aligned}
		\]
		Since $E\left(\sum_{i=1}^n |c_{i,n}|Y_i\right)
		=\lambda^{-1}\sum_{i=1}^n |c_{i,n}|=O(n^3),$
		we have $\sum_{i=1}^n |c_{i,n}|Y_i=O_p(n^3).$
		It follows that
		$
		\sup_{u\in I_{n,M}}
		n^{-3}|\Phi_n'(u)-\Phi_n'(a)|
		=o_p(1).
		$
		Together with the convergence of $n^{-3}\Phi_n'(a)$, we obtain
		\[
		\begin{aligned}
			\sup_{u\in I_{n,M}}
			\left|n^{-3}\Phi_n'(u)+\frac{1}{6a\lambda}\right|
			&\le\sup_{u\in I_{n,M}}n^{-3}|\Phi_n'(u)-\Phi_n'(a)| +\left|n^{-3}\Phi_n'(a)+\frac{1}{6a\lambda}\right|.
		\end{aligned}
		\]
		The first term is $o_p(1)$, and the second term is also $o_p(1)$ because
		$n^{-3}\Phi_n'(a)\xrightarrow{P}-1/(6a\lambda)$. Hence,
		\[
		\sup_{u\in I_{n,M}}
		\left|
		n^{-3}\Phi_n'(u)+\frac{1}{6a\lambda}
		\right|
		\xrightarrow{P}0.
		\]
		
		We now use this local uniform result to expand $\Phi_n(u)$ around $a$ on the $n^{-3/2}$ scale. Let $u_n(t)=a+tn^{-3/2}$ for fixed $t$. By the mean value theorem, $\Phi_n(u_n(t))-\Phi_n(a)=tn^{-3/2}\Phi_n'(\tilde a_n),$
		where $\tilde a_n$ lies between $a$ and $u_n(t)$. Hence
		$\tilde a_n\in I_{n,|t|}$. Therefore, by the preceding local uniform result,
		$n^{-3}\Phi_n'(\tilde a_n)=-{1}/{(6a\lambda)}+o_p(1).$
		It follows that
		\[
		n^{-3/2}\{\Phi_n(u_n(t))-\Phi_n(a)\}
		=
		t n^{-3}\Phi_n'(\tilde a_n)
		=
		-\frac{t}{6a\lambda}+o_p(1).
		\]
		Then, we have
		\[
		n^{-3/2}\Phi_n(a\mp Mn^{-3/2})
		=
		n^{-3/2}\Phi_n(a)\pm \frac{M}{6a\lambda}+o_p(1).
		\]
		Hence, by the tightness of $n^{-3/2}\Phi_n(a)$, for any $\varepsilon>0$, there exists
		$M>0$ such that
		\[
		\liminf_{n\to\infty}
		\Pr\left\{
		\Phi_n(a-Mn^{-3/2})>0,\ 
		\Phi_n(a+Mn^{-3/2})<0
		\right\}
		\ge 1-\varepsilon .
		\]
		By the continuity of $\Phi_n(u)$ and the intermediate value theorem,
		there exists $u_n^*\in[a-Mn^{-3/2},a+Mn^{-3/2}]$ such that $\Phi_n(u_n^*)=0.$ Therefore, for any $\varepsilon>0$, one can choose $M>0$ such that
		$
		\liminf_{n\to\infty}
		\Pr\left\{
		\text{there exists a root of $\Phi_n(u)=0$ in }
		[a-Mn^{-3/2},a+Mn^{-3/2}]
		\right\}
		\ge 1-\varepsilon .
		$
		Let $\hat a$ denote this local root corresponding to the ML estimator. Then, on the above event, $|\hat a-a|\le Mn^{-3/2}.$
		Hence, for any $\varepsilon>0$, one can choose $M>0$ such that
		$
		\liminf_{n\to\infty}
		\Pr\left\{n^{3/2}|\hat a-a|\le M\right\}
		\ge 1-\varepsilon .
		$
		This implies $n^{3/2}(\hat a-a)=O_p(1)$, or equivalently $\hat a-a=O_p(n^{-3/2}).$ In particular, $\hat a-a=o_p(1)$, and hence $\hat a\xrightarrow{P}a$.
		
		We next show the consistency of $\hat\lambda$.
		Write ${\hat a}/{a}=1+\Delta_n,$ where $\Delta_n={(\hat a-a)}/{a}=O_p(n^{-3/2}).$
		Then $n|\Delta_n|=O_p(n^{-1/2})=o_p(1)$.
		Hence, by the mean value theorem, for some $\xi_{i,n}$ between $0$ and $\Delta_n$, $(1+\Delta_n)^{i-1}-1=(i-1)(1+\xi_{i,n})^{i-2}\Delta_n.$
		Since $|\xi_{i,n}|\le |\Delta_n|$, we have $1+\xi_{i,n}\le 1+|\Delta_n|$, and thus
		\[
		\bigl|(1+\Delta_n)^{i-1}-1\bigr|
		\le (i-1)(1+|\Delta_n|)^{i-2}|\Delta_n|
		\le n(1+|\Delta_n|)^n|\Delta_n|
		\le n e^{n|\Delta_n|}|\Delta_n|.
		\]
		Since $n|\Delta_n|=o_p(1)$, we have $e^{n|\Delta_n|}=O_p(1)$, and hence
		\[
		\max_{1\le i\le n}\left|\left(\frac{\hat a}{a}\right)^{i-1}-1\right|
		=O_p(n^{-1/2}).
		\]
		Note that
		\[
		\hat\lambda^{-1}
		=\frac1n\sum_{i=1}^n \left(\frac{\hat a}{a}\right)^{i-1}Y_i
		=\frac1n\sum_{i=1}^n Y_i
		+\frac1n\sum_{i=1}^n \left[\left(\frac{\hat a}{a}\right)^{i-1}-1\right]Y_i.
		\]
		Since $\max_{1\le i\le n}\left|\left({\hat a}/{a}\right)^{i-1}-1\right|=O_p(n^{-1/2})$, and  $n^{-1}\sum_{i=1}^n Y_i=O_p(1),$ the second term of $\hat\lambda^{-1}$ satisfies
		\[
		\left|
		\frac1n\sum_{i=1}^n \left[\left(\frac{\hat a}{a}\right)^{i-1}-1\right]Y_i
		\right|
		\le
		\max_{1\le i\le n}\left|\left(\frac{\hat a}{a}\right)^{i-1}-1\right|
		\frac1n\sum_{i=1}^n Y_i
		=
		O_p(n^{-1/2}).
		\]
		Thus $\hat\lambda^{-1} =n^{-1}\sum_{i=1}^n Y_i+O_p(n^{-1/2}),$
		which implies $\hat\lambda^{-1}-\lambda^{-1}=O_p(n^{-1/2}).$
		Hence $\hat\lambda-\lambda=O_p(n^{-1/2})$, and in particular $\hat\lambda\xrightarrow{P}\lambda$.
	}

	Having established the consistency and convergence rates of the ML estimator $\hat{\boldsymbol{\theta}}_{\mathrm E}=(a,\lambda)^T$, we next derive its limiting distribution.
	The two components of the ML estimator converge at different rates: the estimation equation for $a$ involves weights of order $i$, which accumulate to order $n^3$, whereas the score for $\lambda$ has the usual order $n$. This leads to nonstandard normalization $D_n=\diag(n^{3/2},n^{1/2}).$ Now let $S_n(\boldsymbol{\theta}_{\mathrm E})$ and $H_n(\boldsymbol{\theta}_{\mathrm E})$ denote the score vector and Hessian matrix of $\boldsymbol{\theta}_{\mathrm E}$. 
	
	Recall that $\varepsilon_i=\lambda Y_i-1$, then at the true parameter $\boldsymbol{\theta}_{\mathrm E}$, we have  
	\[
	S_n(\boldsymbol{\theta}_{\mathrm E})
	=
	-\sum_{i=1}^n
	\begin{pmatrix}
		\dfrac{i-1}{a}, 
		\dfrac{1}{\lambda}
	\end{pmatrix}^T\varepsilon_i,
	\]
	and hence
	\[
	D_n^{-1}S_n(\boldsymbol{\theta}_{\mathrm E})
	=
	-\sum_{i=1}^n
	\begin{pmatrix}
		\dfrac{i-1}{a\,n^{3/2}}, 
		\dfrac{1}{\lambda\,n^{1/2}}
	\end{pmatrix}^T \varepsilon_i.
	\]
	To apply the Lindeberg--Feller central limit theorem \cite[Proposition~2.27]{vdV1998}, we first verify two conditions. Define
	\[
	\bm\xi_{n,i}
	=-
		\left(\dfrac{i-1}{a n^{3/2}}, 
		\dfrac{1}{\lambda n^{1/2}}\right)^T
	\varepsilon_i,
	\qquad i=1,\ldots,n.
	\]
	Then $\bm\xi_{n,1},\ldots,\bm\xi_{n,n}$ are independent mean-zero random
	vectors, and $D_n^{-1}S_n(\boldsymbol{\theta}_{\mathrm E})=\sum_{i=1}^n \bm\xi_{n,i}.$
	Let
	$q_{n,i}={(i-1)^2}/{(a^2n^3)}+{1}/{(\lambda^2 n)}.$
	Then $\|\bm\xi_{n,i}\|^2=q_{n,i}\varepsilon_i^2$. Since
	$
	\max_{1\le i\le n}q_{n,i}
	\le
	{1}/{(a^2n)}+{1}/{(\lambda^2n)}
	\le {C^2}/{n}
	$
	for some constant $C>0$, we have $\|\bm\xi_{n,i}\| \le {C n^{-1/2}}|\varepsilon_i|.$
	Thus, for any $\eta>0$, $\{\|\bm\xi_{n,i}\|>\eta\} \subset \left\{ |\varepsilon_i|>{\eta n^{1/2}}/{C} \right\}. $
	Therefore,
	\begin{align*}
		\sum_{i=1}^n
		E\left[
		\|\bm\xi_{n,i}\|^2
		I\{\|\bm\xi_{n,i}\|>\eta\}
		\right]
		& \le
		\sum_{i=1}^n
		q_{n,i}
		E\left[
		\varepsilon_i^2
		I\left\{
		|\varepsilon_i|>\frac{\eta n^{1/2}}{C}
		\right\}
		\right]=
		\left(\sum_{i=1}^n q_{n,i}\right)
		E\left[
		\varepsilon_1^2
		I\left\{
		|\varepsilon_1|>\frac{\eta n^{1/2}}{C}
		\right\}
		\right].
	\end{align*}
	Moreover,
	$\sum_{i=1}^n q_{n,i}
	=\sum_{i=1}^n(i-1)^2/(a^2n^3)+{\lambda^{-2}}
	=O(1).$
	Since $E(\varepsilon_1^2)<\infty$ and $\eta n^{1/2}/C\to\infty$, it follows that
	$E\left[
	\varepsilon_1^2 I\left\{ |\varepsilon_1|>{\eta n^{1/2}}/{C} \right\} \right]\to 0. $
	Hence,
	$
	\sum_{i=1}^n
	E\left[
	\|\bm\xi_{n,i}\|^2
	I\{\|\bm\xi_{n,i}\|>\eta\}
	\right]\to0.
	$
	The covariance matrix  of $\bm\xi_{n,i}$ is given by
	\[
	\sum_{i=1}^n \operatorname{Cov}(\bm\xi_{n,i})
	=
	\sum_{i=1}^n
	\begin{pmatrix}
		\dfrac{(i-1)^2}{a^2n^3} & \dfrac{i-1}{a\lambda n^2}\\[6pt]
		\dfrac{i-1}{a\lambda n^2} & \dfrac{1}{\lambda^2 n}
	\end{pmatrix}.
	\]
	Since $\sum_{i=1}^n(i-1)^2 / n^3\to 1/3$, and $\sum_{i=1}^n(i-1)/n^2 \to 1/2$, we obtain
	\[
	\sum_{i=1}^n \operatorname{Cov}(\bm\xi_{n,i})
	\to
	\Sigma_{\mathrm E}
	=
	\begin{pmatrix}
		\dfrac{1}{3a^2} & \dfrac{1}{2a\lambda}\\[6pt]
		\dfrac{1}{2a\lambda} & \dfrac{1}{\lambda^2}
	\end{pmatrix}.
	\]
	Therefore, by the Lindeberg--Feller central limit theorem, $D_n^{-1}S_n(\boldsymbol{\theta}_{\mathrm E}) \xrightarrow{d} N(\bm0,\Sigma_{\mathrm E}). $

	Next, consider the Hessian matrix $H_n(\boldsymbol{\theta}_{\mathrm E})$ as  
	\[
	H_n(\boldsymbol{\theta}_{\mathrm E})
	=
	\begin{pmatrix}
		H_{aa}(\boldsymbol{\theta}_{\mathrm E}) & H_{a\lambda}(\boldsymbol{\theta}_{\mathrm E}) \\[8pt]
		H_{a \lambda}(\boldsymbol{\theta}_{\mathrm E}) &H_{\lambda \lambda}(\boldsymbol{\theta}_{\mathrm E})   \\[8pt]
	\end{pmatrix},
	\]
	where
	\begin{align*}
		& H_{aa}(\boldsymbol{\theta}_{\mathrm E})=-
		\frac1{a^2}\sum_{i=1}^n(i-1)^2
		-\frac1{a^2}\sum_{i=1}^n(i-1)(i-2)\varepsilon_i,\\
		&H_{a\lambda}(\boldsymbol{\theta}_{\mathrm E})
		=-
		\frac1{a\lambda}\sum_{i=1}^n(i-1)
		-\frac1{a\lambda}\sum_{i=1}^n(i-1)\varepsilon_i,\\
		&
		H_{\lambda \lambda}(\boldsymbol{\theta}_{\mathrm E})=-\frac{n}{\lambda^2}.
	\end{align*}
	After normalization by $D_n^{-1}$ on both sides, each entry of the negative Hessian has a deterministic leading term and a centered random remainder. The deterministic leading terms converge to the corresponding entries of $\Sigma_{\mathrm E}$, while the random remainders are $o_p(1)$.  Hence, $-D_n^{-1}H_n(\boldsymbol{\theta}_{\mathrm E})D_n^{-1}\xrightarrow{P}\Sigma_{\mathrm E}.$
	Since $\hat{\boldsymbol{\theta}}_{\mathrm E}\xrightarrow{P}\boldsymbol{\theta}_{\mathrm E}$, the same limit also holds for $H_n(\tilde{\boldsymbol{\theta}}_n)$, where  $\tilde{\boldsymbol{\theta}}_n$ lies between $\hat{\boldsymbol{\theta}}_{\mathrm E}$ and $\boldsymbol{\theta}_{\mathrm E}$.
	Finally, by the first-order Taylor expansion of the score function,
	$
	\bm0=S_n(\hat{\boldsymbol{\theta}}_{\mathrm E})
	=S_n(\boldsymbol{\theta}_{\mathrm E})+H_n(\tilde{\boldsymbol{\theta}}_n)
	(\hat{\boldsymbol{\theta}}_{\mathrm E}-\boldsymbol{\theta}_{\mathrm E}),
	$
	which implies
	$
	D_n(\hat{\boldsymbol{\theta}}_{\mathrm E}-\boldsymbol{\theta}_{\mathrm E})
	=
	\Bigl[-D_n^{-1}H_n(\tilde{\boldsymbol{\theta}}_n)D_n^{-1}\Bigr]^{-1}
	D_n^{-1}S_n(\boldsymbol{\theta}_{\mathrm E}).
	$
	Therefore, by Slutsky's theorem,
	\[
	D_n(\hat{\boldsymbol{\theta}}_{\mathrm E}-\boldsymbol{\theta}_{\mathrm E})
	\xrightarrow{d}
	N(\bm0,\Sigma_{\mathrm E}^{-1}),
	\]
	where
	\[
	\Sigma_{\mathrm E}^{-1}=
	\begin{pmatrix}
		12a^2 & -6a\lambda\\
		-6a\lambda & 4\lambda^2
	\end{pmatrix}.
	\]
	
	\section{Proof of Theorem \ref{thm:gpq_exp}}\label{proof.gpq_exp}
	
	To analyze the GPQs, we introduce an auxiliary sample that has the same exponential GP distribution as the observed sample but is generated from the pivotal variables $Z_i \sim \mathrm{Exp(1)}, \ i=1,2,\ldots,n$. This construction produces an independent copy of the ML estimator, allowing the GPQs to be rewritten in terms of the difference between two asymptotically normal ML estimators.
	Let $\hat{\boldsymbol{\theta}}_{\mathrm{aux}}=(\hat a_{\mathrm{aux}},\hat \lambda_{\mathrm{aux}})^T$ be the ML estimator computed from the auxiliary sample $X_{i,\mathrm{aux}}=Z_i/(\lambda a^{i-1})$, $i=1,2,\ldots,n$. Since $X_{i,\mathrm{aux}}$ follows the same exponential GP model as the observed sample, comparing its likelihood equations with Equation \eqref{eq:g1exp} and Equation \eqref{eq:g2exp} gives $\phi_{a,\mathrm E}=\hat a_{\mathrm{aux}}/a$ and $\phi_{\lambda,\mathrm E}=\hat \lambda_{\mathrm{aux}}/\lambda$. Hence $\G_{a,\mathrm E}=\hat a/ \phi_{a,\mathrm E} = a\hat a/\hat a_{\mathrm{aux}}$ and $\G_{\lambda,\mathrm E}=\hat\lambda /\phi_{\lambda,\mathrm E} =\lambda\hat\lambda/\hat \lambda_{\mathrm{aux}}$. By the ML estimator result just established, $\hat a_{\mathrm{aux}}-a=O_p(n^{-3/2})$ and $\hat \lambda_{\mathrm{aux}}-\lambda=O_p(n^{-1/2})$, so $a/\hat a_{\mathrm{aux}}=1+o_p(1)$ and $\lambda/\hat \lambda_{\mathrm{aux}}=1+o_p(1)$. 
	Then, we have
	\[
	\G_{a,\mathrm E}-a
	=a\frac{\hat a}{\hat a_{\mathrm{aux}}}-a
	=a\frac{\hat a-\hat a_{\mathrm{aux}}}{\hat a_{\mathrm{aux}}}
	=\bigl\{(\hat a-a)-(\hat a_{\mathrm{aux}}-a)\bigr\}\frac{a}{\hat a_{\mathrm{aux}}}.
	\]
	Since $\hat a_{\mathrm{aux}}\xrightarrow{P}a$, we have ${a}/{\hat a_{\mathrm{aux}}}=1+o_p(1).$
	Therefore,
	\[
	\G_{a,\mathrm E}-a
	=\bigl\{(\hat a-a)-(\hat a_{\mathrm{aux}}-a)\bigr\}\{1+o_p(1)\}
	=(\hat a-a)-(\hat a_{\mathrm{aux}}-a)+o_p(n^{-3/2}).
	\]
	Similarly,
	\[
	\G_{\lambda,\mathrm E}-\lambda
	=\lambda\frac{\hat\lambda}{\hat \lambda_{\mathrm{aux}}}-\lambda
	=\lambda\frac{\hat\lambda-\hat \lambda_{\mathrm{aux}}}{\hat \lambda_{\mathrm{aux}}}
	=\bigl\{(\hat\lambda-\lambda)-(\hat \lambda_{\mathrm{aux}}-\lambda)\bigr\}\frac{\lambda}{\hat \lambda_{\mathrm{aux}}}.
	\]
	Since $\hat \lambda_{\mathrm{aux}}\xrightarrow{P}\lambda$, it follows that ${\lambda}/{\hat \lambda_{\mathrm{aux}}}=1+o_p(1),$
	and hence
	\[
	\G_{\lambda,\mathrm E}-\lambda
	=\bigl\{(\hat\lambda-\lambda)-(\hat \lambda_{\mathrm{aux}}-\lambda)\bigr\}\{1+o_p(1)\}
	=(\hat\lambda-\lambda)-(\hat \lambda_{\mathrm{aux}}-\lambda)+o_p(n^{-1/2}).
	\]
	Collecting the two expansions above, we obtain
	\[
	\begin{pmatrix}
		n^{3/2}(\G_{a,\mathrm E}-a)\\
		n^{1/2}(\G_{\lambda,\mathrm E}-\lambda)
	\end{pmatrix}
	=
	\begin{pmatrix}
		n^{3/2}(\hat a-a)\\
		n^{1/2}(\hat\lambda-\lambda)
	\end{pmatrix}
	-
	\begin{pmatrix}
		n^{3/2}(\hat a_{\mathrm{aux}}-a)\\
		n^{1/2}(\hat \lambda_{\mathrm{aux}}-\lambda)
	\end{pmatrix}
	+o_p(1).
	\]
	Both vectors on the right converge to $N(\bm0,\Sigma_{\mathrm E}^{-1})$, and they are independent because the auxiliary sample is generated independently of the observed sample. Hence, by Slutsky's theorem,
	\[
	\begin{pmatrix}
		n^{3/2}(\G_{a,\mathrm E}-a)\\
		n^{1/2}(\G_{\lambda,\mathrm E}-\lambda)
	\end{pmatrix}
	\xrightarrow{d}
	N(\bm0,2\Sigma_{\mathrm E}^{-1}).
	\]

	\section{Proof of Theorem~\ref{thm:mle_gamgp}}\label{proof.mle_gamgp}

	Define
	\[
	\Phi_n(u)=\sum_{i=1}^n (n-2i+1)u^{\,i-1}X_i,\qquad u>0.
	\]
	Let $Y_i=a^{\,i-1}X_i,\ i=1,2,\ldots,n.$ 
	Then $Y_1,\ldots,Y_n \stackrel{\mathrm{i.i.d.}}{\sim} \Ga(\alpha,\beta)$, where $\alpha$ and $\beta$ are the shape and rate parameters, respectively. Thus, $E(Y_i)=\alpha/\beta$ and $\Var(Y_i)=\alpha/\beta^2$.
	Since $\sum_{i=1}^n(n-2i+1)=0$, we have
	\[
	\Phi_n(a)=\sum_{i=1}^n (n-2i+1)Y_i
	=\sum_{i=1}^n (n-2i+1)\Bigl(Y_i-\frac{\alpha}{\beta}\Bigr)=O_p(n^{3/2}).
	\]
	Moreover,
	\[
	\Phi_n'(a)=\frac1a\sum_{i=1}^n (n-2i+1)(i-1)Y_i.
	\]
	Let $c_{i,n}=(n-2i+1)(i-1)$, then 
	\[
	n^{-3} \Phi_n'(a)
	=
	\frac{\alpha}{a\beta n^{3}}
	\sum_{i=1}^n c_{i,n}
	+\frac{1}{a n^{3}}
	\sum_{i=1}^n c_{i,n}
	\left(Y_i-\frac{\alpha}{\beta}\right).
	\]
	Since $\sum_{i=1}^n c_{i,n} =-{(n^3-n)}/{6}$, the first part of \(n^{-3}\Phi_n'(a)\) converges to
	$-{\alpha}/{(6a\beta)}.$ For the second part, let $R_n={(a n^{3})^{-1}}\sum_{i=1}^n c_{i,n} \left(Y_i-{\alpha}/{\beta}\right).$ Then \(E(R_n)=0\). 
	Since \(\sum_{i=1}^n c_{i,n}^2=O(n^5)\), we have
	\[
	\Var(R_n)
	=\frac{\alpha}{a^2\beta^2 n^6}
	\sum_{i=1}^n c_{i,n}^2
	=O(n^{-1})\to 0.
	\]
	Therefore, \(R_n=o_p(1)\), and consequently $n^{-3}\Phi_n'(a) \xrightarrow{P} -{\alpha}/{(6a\beta)}.$

	We now prove the local uniform convergence of $n^{-3}\Phi_n'(u)$ in a shrinking
	neighborhood of $a$. For fixed $M>0$, define $I_{n,M}=\{u:|u-a|\le Mn^{-3/2}\}.$
	Similarly to Section \ref{proof.mle_exp} of the Supplementary Material, we have
	\[
	\max_{1\le i\le n}\sup_{u\in I_{n,M}}
	\left|
	\frac1u\left(\frac{u}{a}\right)^{i-1}-\frac1a
	\right|
	=O(n^{-1/2}).
	\]
	Therefore,
	\[
	\begin{aligned}
		\sup_{u\in I_{n,M}}|\Phi_n'(u)-\Phi_n'(a)|
		&\le
		\left[
		\max_{1\le i\le n}\sup_{u\in I_{n,M}}
		\left|
		\frac1u\left(\frac{u}{a}\right)^{i-1}-\frac1a
		\right|
		\right]\sum_{i=1}^n |c_{i,n}|Y_i =
		O(n^{-1/2})\sum_{i=1}^n |c_{i,n}|Y_i .
	\end{aligned}
	\]
	Since
	$E\left(\sum_{i=1}^n |c_{i,n}|Y_i\right)={\alpha}{\beta^{-1}}\sum_{i=1}^n |c_{i,n}|
	=O(n^3),$ we have $\sum_{i=1}^n |c_{i,n}|Y_i=O_p(n^3)$, and thus $\sup_{u\in I_{n,M}}n^{-3}|\Phi_n'(u)-\Phi_n'(a)|=o_p(1).$
	Together with $n^{-3}\Phi_n'(a)\xrightarrow{P}- \alpha /(6a\beta)$, the triangle
	inequality gives
	\[
	\begin{aligned}
		\sup_{u\in I_{n,M}}
		\left|n^{-3}\Phi_n'(u)+\frac{\alpha }{6a \beta}\right|
		&\le
		\sup_{u\in I_{n,M}}n^{-3}|\Phi_n'(u)-\Phi_n'(a)|
		+\left|n^{-3}\Phi_n'(a)+\frac{\alpha }{6a \beta}\right| =o_p(1).
	\end{aligned}
	\]
	Hence, $\sup_{u\in I_{n,M}} \left|n^{-3}\Phi_n'(u)+{\alpha }/{(6a \beta)}\right|
	\xrightarrow{P}0.$ Now let $u_n(t)=a+tn^{-3/2}$ for fixed $t$. By the mean value theorem, $\Phi_n(u_n(t))-\Phi_n(a)=tn^{-3/2}\Phi_n'(\tilde a_n),$ where $\tilde a_n$ lies between $a$ and $u_n(t)$. Since $\tilde a_n\in I_{n,|t|}$, the local uniform convergence implies $n^{-3}\Phi_n'(\tilde a_n)=-{\alpha }/{(6a \beta)}+o_p(1).$
	Therefore,
	\[
	n^{-3/2}\{\Phi_n(u_n(t))-\Phi_n(a)\}
	=t n^{-3}\Phi_n'(\tilde a_n)
	=-\frac{\alpha t }{6a\beta}+o_p(1).
	\]
	Let $Z_n=n^{-3/2}\Phi_n(a)$. Since $Z_n=O_p(1)$, it follows that
	\[
	n^{-3/2}\Phi_n(a \mp Mn^{-3/2})=Z_n \pm \frac{M \alpha}{6a\beta}+o_p(1).
	\]
	By the tightness of $Z_n$, for any $\varepsilon>0$, one can choose $M>0$
	such that
	\[
	\liminf_{n\to\infty}
	\Pr\left\{
	\Phi_n(a-Mn^{-3/2})>0,\ 
	\Phi_n(a+Mn^{-3/2})<0
	\right\}
	\ge 1-\varepsilon .
	\]
	Since $\Phi_n(u)$ is continuous in $u$, the intermediate value theorem implies that, on the above event, there exists $u_n^*\in[a-Mn^{-3/2},a+Mn^{-3/2}]$ such that $\Phi_n(u_n^*)=0$. Hence, for any $\varepsilon>0$, one can choose $M>0$ such that
	\[
	\liminf_{n\to\infty}
	\Pr\left\{
	\text{there exists a root of $\Phi_n(u)=0$ in }
	[a-Mn^{-3/2},a+Mn^{-3/2}]
	\right\}
	\ge 1-\varepsilon .
	\]
	Let $\hat a$ denote this local root corresponding to the ML estimator. Then, on this event, $|\hat a-a|\le Mn^{-3/2},$ and therefore
	$
	\liminf_{n\to\infty}
	\Pr\left\{n^{3/2}|\hat a-a|\le M\right\}
	\ge 1-\varepsilon .
	$
	This implies $n^{3/2}(\hat a-a)=O_p(1)$. In particular, $\hat a-a=o_p(1)$, and hence $\hat a\xrightarrow{P}a$.

	The estimators $\hat\alpha$ and $\hat\beta$ depend on $\hat a$ through the transformed sample
	\[
	\hat Y_i:=\hat a^{\,i-1}X_i
	=\left(\frac{\hat a}{a}\right)^{i-1}Y_i,\qquad i=1,2,\ldots,n.
	\]
	Since $n^{3/2}(\hat a-a)=O_p(1)$, we have
	$\max_{1\le i\le n}\left|\left({\hat a}/{a}\right)^{i-1}-1\right|
	=O_p(n^{-1/2})=o_p(1).$
	Therefore,
	\[
	\bar Y=\frac1n\sum_{i=1}^n \hat Y_i
	=\frac1n\sum_{i=1}^n Y_i+o_p(1)\xrightarrow{P}\frac{\alpha}{\beta}.
	\]
	Define $\tilde Y:=\Bigl(\prod_{i=1}^n \hat Y_i \Bigr)^{1/n}$. Similarly, we have
	\[
	\log \tilde Y
	=\frac1n\sum_{i=1}^n \log \hat Y_i
	=\frac1n\sum_{i=1}^n \log Y_i+\frac1n\sum_{i=1}^n (i-1)\log(\hat a/a)
	=\frac1n\sum_{i=1}^n \log Y_i+\frac{n-1}{2}\log(\hat a/a).
	\]
	Since $n^{3/2}(\hat a-a)=O_p(1)$, we have $\hat a-a=O_p(n^{-3/2})$. As $a>0$ is fixed,
	\[
	\frac{\hat a}{a}-1=\frac{\hat a-a}{a}=O_p(n^{-3/2})=o_p(1).
	\]
	Using the expansion $\log(1+x)=x+O(x^2)$ as $x\to0$, we obtain
	\[
	\frac{n-1}{2} \log\left(\frac{\hat a}{a}\right)
	=
	\frac{n-1}{2} \log\left(1+\frac{\hat a-a}{a}\right)
	=
	O_p(n^{-1/2}) = o_p(1).
	\]
	Hence, we have
	\[
	\log \tilde Y
	=\frac1n\sum_{i=1}^n \log Y_i+o_p(1)
	\xrightarrow{P}\psi(\alpha)-\log\beta.
	\]
	Therefore, $\log \bigl(\bar Y/\tilde Y\bigr)\xrightarrow{P}\log\alpha-\psi(\alpha).$ 
	From Lemma \ref{lem:T_alpha_limit_gam}, we know that $g(\alpha)=\log(\alpha)-\psi(\alpha)$ is strictly decreasing on $(0,\infty)$. Hence, $g$ is invertible. Then, by Equation~\eqref{eq:alpmle}, we have $g(\hat\alpha)=\log\!\bigl(\bar Y/\tilde Y\bigr),$ 
	so $\hat\alpha\xrightarrow{P}\alpha$. Finally, Equation~\eqref{eq:lammle} gives
	\[
	\hat\beta=\frac{n\hat\alpha}{\sum_{i=1}^n \hat Y_i}
	=\frac{\hat\alpha}{\bar Y}
	\xrightarrow{P}\beta.
	\]
	Thus $\hat{\boldsymbol{\theta}}_{\mathrm G}=(\hat a,\hat\alpha,\hat\beta)^T\xrightarrow{P}\boldsymbol{\theta}_{\mathrm G}=(a,\alpha,\beta)^T$.
	
	Having established the consistency and convergence rates of the ML estimator, we next derive its limiting distribution. The estimator $\hat a$ converges at the rate $n^{-3/2}$, whereas $\hat \alpha$ and $\hat \beta$ converge at the usual rate $n^{-1/2}$. This leads to the following nonstandard normalization $D_n=\diag(n^{3/2},n^{1/2},n^{1/2}).$ Let $S_n(\boldsymbol{\theta}_{\mathrm G})$ and $H_n(\boldsymbol{\theta}_{\mathrm G})$ denote the score vector and Hessian matrix of $\boldsymbol{\theta}_{\mathrm G}$. 
	
	At the true parameter, define
	$
	\xi_i=\beta Y_i-\alpha,\
	\zeta_i=\log Y_i-\{\psi(\alpha)-\log\beta\}.
	$
	Then $E(\xi_i)=E(\zeta_i)=0$, $\Var(\xi_i)=\alpha$, $\Var(\zeta_i)=\psi_1(\alpha)$, and $\Cov(\xi_i,\zeta_i)=1$. A direct calculation gives
	\[
	S_n(\boldsymbol{\theta}_{\mathrm G})
	=\sum_{i=1}^n
	\begin{pmatrix}
		-\dfrac{i-1}{a}\xi_i,
		&\zeta_i,
		&-\dfrac1\beta \xi_i
	\end{pmatrix}^T,
	\]
	and hence
	\[
	D_n^{-1}S_n(\boldsymbol{\theta}_{\mathrm G})
	=\sum_{i=1}^n
	\begin{pmatrix}
		-\dfrac{i-1}{a n^{3/2}}\xi_i,
		&\dfrac{1}{n^{1/2}}\zeta_i,
		&-\dfrac{1}{\beta n^{1/2}}\xi_i
	\end{pmatrix}^T.
	\]
	To apply the Lindeberg--Feller central limit theorem \cite[Proposition~2.27]{vdV1998}, we first verify two conditions. Define
	\[
	\bm W_{n,i}
	=
	\begin{pmatrix}
		-\dfrac{i-1}{a n^{3/2}}\xi_i,\
		\dfrac{1}{n^{1/2}}\zeta_i,\
		-\dfrac{1}{\beta n^{1/2}}\xi_i
	\end{pmatrix}^T,
	\qquad i=1,\ldots,n.
	\]
	Then $\bm W_{n,1},\ldots,\bm W_{n,n}$ are independent mean-zero random vectors, and
	$
	D_n^{-1}S_n(\boldsymbol{\theta}_{\mathrm G})
	=\sum_{i=1}^n \bm W_{n,i}.
	$
	Let
	$q_{n,i}={(i-1)^2}/{(a^2n^3)}+{1}/{n}+{1}/{(\beta^2n)}.$
	Since $(i-1)^2/n^3\le 1/n$, there exists a constant $C>0$ such that $q_{n,i}\le {C^2}/{n},\ i=1,\ldots,n.$
	Moreover,
	$
	\|\bm W_{n,i}\|^2
	=\left\{
	{(i-1)^2}/{(a^2n^3)}
	+{1}/{(\beta^2n)}
	\right\}\xi_i^2
	+\frac{1}{n}\zeta_i^2
	\le
	q_{n,i}(\xi_i^2+\zeta_i^2).
	$
	Hence, $\|\bm W_{n,i}\| \le {C n^{-1/2}}(|\xi_i|+|\zeta_i|).$
	Thus, for any $\eta>0$,
	$\{\|\bm W_{n,i}\|>\eta\} \subset \left\{|\xi_i|+|\zeta_i|>{\eta n^{1/2}}/{C}\right\}.$
	Since $E(\xi_1^2)<\infty$, $E(\zeta_1^2)<\infty$, and $\eta n^{1/2}/C\to\infty$, we have 
	\[
	\begin{aligned}
		\sum_{i=1}^n
		E\left[
		\|\bm W_{n,i}\|^2
		I\{\|\bm W_{n,i}\|>\eta\}
		\right] 
		& \le
		\sum_{i=1}^n
		\frac{C^2}{n}
		E\left[
		(\xi_i^2+\zeta_i^2)
		I\left\{
		|\xi_i|+|\zeta_i|>\frac{\eta n^{1/2}}{C}
		\right\}
		\right] \\
		& =
		C^2
		E\left[
		(\xi_1^2+\zeta_1^2)
		I\left\{
		|\xi_1|+|\zeta_1|>\frac{\eta n^{1/2}}{C}
		\right\}
		\right]
		\to 0.
	\end{aligned}
	\]
	The covariance matrix of $\bm W_{n,i}$ is given by
	\[
	\sum_{i=1}^n \Cov(\bm W_{n,i})
	=\sum_{i=1}^n
	\begin{pmatrix}
		\dfrac{(i-1)^2}{a^2n^3}\alpha &
		-\dfrac{i-1}{a n^2} &
		\dfrac{(i-1)\alpha}{a\beta n^2}\\[8pt]
		-\dfrac{i-1}{a n^2} &
		\dfrac{\psi_1(\alpha)}{n} &
		-\dfrac{1}{\beta n}\\[8pt]
		\dfrac{(i-1)\alpha}{a\beta n^2} &
		-\dfrac{1}{\beta n} &
		\dfrac{\alpha}{\beta^2 n}
	\end{pmatrix}.
	\]
	Using $n^{-3}\sum_{i=1}^n(i-1)^2\to 1/3$, $n^{-2}\sum_{i=1}^n(i-1)\to 1/2$, and $n^{-1}\sum_{i=1}^n1\to 1$, 
	we obtain
	\[
	\sum_{i=1}^n \Cov(\bm W_{n,i})
	\to
	\Sigma_{\mathrm G} =
	\begin{pmatrix}
		\dfrac{\alpha}{3a^2} &
		-\dfrac{1}{2a} &
		\dfrac{\alpha}{2a\beta}\\[8pt]
		-\dfrac{1}{2a} &
		\psi_1(\alpha) &
		-\dfrac{1}{\beta}\\[8pt]
		\dfrac{\alpha}{2a\beta} &
		-\dfrac{1}{\beta} &
		\dfrac{\alpha}{\beta^2}
	\end{pmatrix}.
	\]
	Therefore, by the Lindeberg--Feller central limit theorem, $D_n^{-1}S_n(\boldsymbol{\theta}_{\mathrm G}) =\sum_{i=1}^n\bm W_{n,i} \xrightarrow{d} N(\bm0,\Sigma_{\mathrm G}).$

	Next, consider the Hessian matrix $H_n(\boldsymbol{\theta}_{\mathrm G})$ as  
	\[
	H_n(\boldsymbol{\theta}_{\mathrm G})
	=
	\begin{pmatrix}
		H_{aa}(\boldsymbol{\theta}_{\mathrm G}) & H_{a\alpha}(\boldsymbol{\theta}_{\mathrm G}) & H_{a\beta}(\boldsymbol{\theta}_{\mathrm G})\\[8pt]
		H_{a\alpha}(\boldsymbol{\theta}_{\mathrm G}) &H_{\alpha\alpha}(\boldsymbol{\theta}_{\mathrm G})  & H_{\alpha\beta}(\boldsymbol{\theta}_{\mathrm G}) \\[8pt]
		H_{a\beta}(\boldsymbol{\theta}_{\mathrm G}) &H_{\alpha\beta}(\boldsymbol{\theta}_{\mathrm G}) &H_{\beta\beta}(\boldsymbol{\theta}_{\mathrm G})
	\end{pmatrix},
	\]
	where
	\begin{align*}
		&H_{aa}(\boldsymbol{\theta}_{\mathrm G})
		=
		-\frac{\alpha}{a^2}\sum_{i=1}^n(i-1)
		-\frac{\beta}{a^2}\sum_{i=1}^n(i-1)(i-2)Y_i,\\
		&H_{a\alpha}(\boldsymbol{\theta}_{\mathrm G})
		=
		\frac1a\sum_{i=1}^n(i-1),\\
		&H_{a\beta}(\boldsymbol{\theta}_{\mathrm G})
		=
		-\frac1a\sum_{i=1}^n(i-1)Y_i,\\
		&H_{\alpha\alpha}(\boldsymbol{\theta}_{\mathrm G})
		=-n\psi_1(\alpha),\\
		&H_{\alpha\beta}(\boldsymbol{\theta}_{\mathrm G})=\frac{n}{\beta},\\
		&H_{\beta\beta}(\boldsymbol{\theta}_{\mathrm G})=-\frac{n\alpha}{\beta^2}.
	\end{align*}
	After normalization by $D_n^{-1}$ on both sides, we obtain $-\,D_n^{-1}H_n(\boldsymbol{\theta}_{\mathrm G})D_n^{-1}\xrightarrow{P}\Sigma_{\mathrm G}.$ 
	Moreover, since $\hat{\boldsymbol{\theta}}_{\mathrm G}\xrightarrow{P}\boldsymbol{\theta}_{\mathrm G}$, we have $-D_n^{-1}H_n(\tilde{\boldsymbol{\theta}}_n)D_n^{-1}\xrightarrow{P}\Sigma_{\mathrm G}.$ 
	Finally, a first-order Taylor expansion of the score around the true parameter gives
	$\bm0
	=S_n(\hat{\boldsymbol{\theta}}_{\mathrm G})
	=S_n(\boldsymbol{\theta}_{\mathrm G})
	+H_n(\tilde{\boldsymbol{\theta}}_n)(\hat{\boldsymbol{\theta}}_{\mathrm G}-\boldsymbol{\theta}_{\mathrm G}),
	$
	which implies
	$
	D_n(\hat{\boldsymbol{\theta}}_{\mathrm G}-\boldsymbol{\theta}_{\mathrm G})
	=\Bigl[-D_n^{-1}H_n(\tilde{\boldsymbol{\theta}}_n)D_n^{-1}\Bigr]^{-1}
	D_n^{-1}S_n(\boldsymbol{\theta}_{\mathrm G}).
	$
	Therefore, by Slutsky's theorem,
	\[
	D_n(\hat{\boldsymbol{\theta}}_{\mathrm G}-\boldsymbol{\theta}_{\mathrm G})
	\xrightarrow{d}
	N(\bm0,\Sigma_{\mathrm G}^{-1}),
	\]
	where
	\[
	\Sigma_{\mathrm G}^{-1}=
	\begin{pmatrix}
		\dfrac{12a^2}{\alpha} & 0 & -\dfrac{6a\beta}{\alpha}\\[8pt]
		0 & \dfrac{\alpha}{\alpha\psi_1(\alpha)-1} & \dfrac{\beta}{\alpha\psi_1(\alpha)-1}\\[10pt]
		-\dfrac{6a\beta}{\alpha} & \dfrac{\beta}{\alpha\psi_1(\alpha)-1}
		& \dfrac{\beta^2\{4\alpha\psi_1(\alpha)-3\}}{\alpha\{\alpha\psi_1(\alpha)-1\}}
	\end{pmatrix}.
	\]

	\section{Proof of Lemma~\ref{lem:T_alpha_limit_gam}}\label{proof.T_alpha_gam}
	
	Recall that if $W\sim \mathrm{Ga}(\alpha,1)$, then
	$\E(W)=\alpha$, $\Var(W)=\alpha$, $\E(\log W)=\psi(\alpha)$, $\Var(\log W)=\psi_1(\alpha)$, and $\Cov(W,\log W)=1$, where $\psi(\alpha)$ and $\psi_1(\alpha)$ are the digamma and trigamma functions, respectively.
	Let $F_G(\cdot;\alpha)$ denote the cumulative distribution function of $\mathrm{Ga}(\alpha,1)$, and let $U_1,\ldots,U_n \stackrel{\mathrm{i.i.d.}}{\sim}\mathrm{Unif}(0,1)$. For each fixed $\alpha>0$, let $Q(u;\alpha)=F_G^{-1}(u;\alpha)$ for $0<u<1$.  By the probability integral transform,  $Q(U_i;\alpha)$ has the same distribution as $\mathrm{Ga}(\alpha,1)$. Define
	\[
	T_n(\alpha;\bmU)=\log A_n(\alpha;\bmU)-L_n(\alpha;\bmU),
	\]
	where $\bmU=(U_1,\ldots,U_n)$, 
	$A_n(\alpha;\bmU)= n^{-1} \sum_{i=1}^n Q(U_i;\alpha), \
	L_n(\alpha;\bmU)= n^{-1} \sum_{i=1}^n \log Q(U_i;\alpha).$
	Let $g(\alpha)=\log\alpha-\psi(\alpha)$. Since $Q(U_1;\alpha),\ldots,Q(U_n;\alpha)$ are i.i.d.\ with the same distribution as $\mathrm{Ga}(\alpha,1)$, the strong law of large numbers gives
	$A_n(\alpha;\bmU)\xrightarrow{\mathrm{a.s.}}\alpha,
	\ L_n(\alpha;\bmU)\xrightarrow{\mathrm{a.s.}}\psi(\alpha).$
	Hence, by continuity of $\log(\cdot)$ on $(0,\infty)$,
	$T_n(\alpha;\bmU)\xrightarrow{\mathrm{a.s.}}\log\alpha-\psi(\alpha)=g(\alpha),$
	which proves part (1).
	
	Next, let $\bmZ_n=\bigl(A_n(\alpha;\bmU),L_n(\alpha;\bmU)\bigr)^T$ and $\bmmu=\bigl(\alpha,\psi(\alpha)\bigr)^T$. Since $\bigl(Q(U_i;\alpha),\log Q(U_i;\alpha)\bigr)^T$, $i=1,\ldots,n$, are i.i.d.\ with mean $\bmmu$ and covariance matrix $\Sigma=\left(\begin{smallmatrix}
		\alpha & 1\\
		1 & \psi_1(\alpha)
	\end{smallmatrix}\right)$,
	the bivariate central limit theorem yields $\sqrt{n}\,(\bmZ_n-\bmmu)\xrightarrow{d}N(\bm0,\Sigma).$
	Let $h(z_1,z_2)=\log z_1-z_2$. Then $T_n(\alpha;\bmU)=h(\bmZ_n)$, $g(\alpha)=h(\bmmu)$, and $\nabla h(\bmmu)=(1/\alpha,-1)^T$. By the delta method, $\sqrt{n}\{T_n(\alpha;\bmU)-g(\alpha)\}
	\xrightarrow{d}
	N\!\left(0,\nabla h(\bmmu)^T\Sigma\nabla h(\bmmu)\right).$
	A direct calculation gives $\nabla h(\bmmu)^T\Sigma\nabla h(\bmmu) =\psi_1(\alpha)-1 / \alpha,$ and therefore
	\[
	\sqrt{n}\{T_n(\alpha;\bmU)-g(\alpha)\}
	\xrightarrow{d}
	N\!\left(0,\psi_1(\alpha)-\frac{1}{\alpha}\right),
	\]
	which proves part (2).
	
	Finally, since $g(\alpha)=\log\alpha-\psi(\alpha)$, we have $g'(\alpha)=1/\alpha-\psi_1(\alpha)$. Using the series representation $\psi_1(\alpha)=\sum_{j=0}^\infty 1 / (\alpha+j)^2,$ we obtain
	\[
	\psi_1(\alpha)>\int_0^\infty \frac{1}{(\alpha+x)^2}\,dx=\frac{1}{\alpha}.
	\]
	Hence $g'(\alpha)<0$ for all $\alpha>0$, so $g(\alpha)$ is strictly decreasing on $(0,\infty)$, which proves part (3).

	\section{Proof of Theorem~\ref{thm:alpha_star_unique_gam}}
	\label{proof.alpha_star_unique_gam}

	Since $U_1,\cdots,U_n$ are random samples from a continuous distribution $\mathrm{Unif}(0,1)$, its components are pairwise distinct almost surely. Fix such a realization and relabel if necessary so that $0<U_1<\cdots<U_n<1$. Then $\alpha\mapsto T_n(\alpha;\bmU)$ is continuous on $(0,\infty)$, since $Q(u;\alpha)$ is continuous in $\alpha$ for each fixed $u\in(0,1)$. Moreover, $T_n(\alpha;\bmU)$ has the same form as the statistic $\log(\bar x/\tilde x)$ in \cite{Iliopoulos2016} with $x_i=Q(U_i;\alpha)$; hence, by the same argument as in Lemmas A.2--A.3 therein, $T_n(\alpha;\bmU)$ is strictly decreasing on $(0,\infty)$.
	
	As $\alpha\to\infty$, the standard normal approximation for the gamma quantile gives $Q(U_i;\alpha)=\alpha+O(\alpha^{1/2})$ for each fixed $i$, so
	$\sum_{i=1}^n Q(U_i;\alpha)  / (n\alpha)\to1,
	\
	\left\{\prod_{i=1}^n Q(U_i;\alpha)\right\}^{1/n}  / \alpha\to1,$
	and hence $T_n(\alpha;\bmU)\to0$. Since the $Q(U_i;\alpha)$ are not all equal for any $\alpha>0$, the strict arithmetic--geometric mean inequality implies
	$\sum_{i=1}^n Q(U_i;\alpha) / n >
	\left\{\prod_{i=1}^n Q(U_i;\alpha)\right\}^{1/n},$
	so $T_n(\alpha;\bmU)>0$ for all $\alpha>0$. Therefore, $\lim_{\alpha\to\infty}T_n(\alpha;\bmU)=0^+.$
	
	As $\alpha\to0^+$, the lower-tail expansion $F_G(x;\alpha)=x^\alpha\Gamma(\alpha+1)^{-1}\{1+o(1)\}$
	implies $Q(U_i;\alpha)=\{U_i\Gamma(\alpha+1)\}^{1/\alpha}\{1+o(1)\}, \ i=1,\ldots,n.$ 
	Let $\ell_i=\log U_i$, $\bar\ell=n^{-1}\sum_{i=1}^n\ell_i$, and $\ell_{\max}=\max_{1\le i\le n}\ell_i$. Then
	\[
	\log Q(U_i;\alpha)=\frac{\ell_i}{\alpha}+\frac{\log\Gamma(\alpha+1)}{\alpha}+o(1),
	\]
	so
	\[
	\frac{1}{n}\sum_{i=1}^n \log Q(U_i;\alpha)
	=\frac{\bar\ell}{\alpha}+\frac{\log\Gamma(\alpha+1)}{\alpha}+o(1).
	\]
	We rewrite $\sum_{i=1}^n Q(U_i;\alpha)/n$ as
	\[
	\frac{1}{n}\sum_{i=1}^n Q(U_i;\alpha)
	=\frac{\Gamma(\alpha+1)^{1/\alpha}}{n}
	\sum_{i=1}^n \exp\!\left\{\frac{\ell_i}{\alpha}+o(1)\right\},
	\]
	and factor out the largest exponential term, which gives
	\[
	\log\!\left(\frac{1}{n}\sum_{i=1}^n Q(U_i;\alpha)\right)
	=\frac{\ell_{\max}}{\alpha}
	+\frac{\log\Gamma(\alpha+1)}{\alpha}
	+O(1).
	\]
	Hence
	\[
	T_n(\alpha;\bmU)
	=\log\!\left(\frac{1}{n}\sum_{i=1}^n Q(U_i;\alpha)\right)
	-\frac{1}{n}\sum_{i=1}^n \log Q(U_i;\alpha)
	=\frac{\ell_{\max}-\bar\ell}{\alpha}+O(1).
	\]
	Because $U_1,\ldots,U_n$ are pairwise distinct, and $\ell_{\max}>\bar\ell$, thus
	$\lim_{\alpha\to0^+}T_n(\alpha;\bmU)=+\infty.$
	
	Therefore, for almost every realization of $\bmU$, the function $\alpha\mapsto T_n(\alpha;\bmU)$ is continuous and strictly decreasing on $(0,\infty)$, with
	$\lim_{\alpha\to0^+}T_n(\alpha;\bmU)=+\infty, \ \lim_{\alpha\to\infty}T_n(\alpha;\bmU)=0^+.$
	Hence its range is $(0,\infty)$, and since $t_{\mathrm{obs}}>0$, Equation~\eqref{eq:alpha_star_def} has exactly one solution $\G_{\alpha,\mathrm G}\in(0,\infty)$.

	\section{Proof of Theorem~\ref{thm:AN_gamgp}}\label{proof.AN_gamgp}

	To analyze the GPQs, we introduce an auxiliary sample that is constructed from the pivotal variable $\bmU=(U_1,\ldots,U_n)$, where $U_1,\ldots,U_n \stackrel{\mathrm{i.i.d.}}{\sim} \mathrm{Unif}(0,1)$. Let  $\hat{\boldsymbol{\theta}}_{\mathrm G,\mathrm{aux}}=(\hat a_{\mathrm{aux}},\hat\alpha_{\mathrm{aux}},\hat\beta_{\mathrm{aux}})^T$ be the ML estimator computed from the auxiliary sample $X_{i,\mathrm{aux}}=Q(U_i;\alpha)/(\beta a^{\,i-1})$, $i=1,2,\ldots,n$. Since $Q(U_i;\alpha)\overset{d}{=}\beta Y_i$, the sample $X_{i,\mathrm{aux}}$ follows the same gamma GP model as the observed sample and is independent of it. Hence $D_n(\hat{\boldsymbol{\theta}}_{\mathrm G,\mathrm{aux}}-\boldsymbol{\theta}_{\mathrm G}) \xrightarrow{d} N(\bm0,\Sigma_{\mathrm G}^{-1}).$

	We first connect the GPQ with the observed-sample and auxiliary-sample ML estimators. Recall that  $\G_{\alpha,\mathrm G}$ is defined by $T_n(\G_{\alpha,\mathrm G};\bm U)=t_{\mathrm{obs}}.$ For the observed sample, the likelihood equation gives $g(\hat\alpha)=t_{\mathrm{obs}}$, where $g(\alpha)=\log \alpha-\psi(\alpha).$ Define the auxiliary-sample estimator $\hat\alpha_{\mathrm{aux}}$ by $g(\hat\alpha_{\mathrm{aux}})=T_n(\alpha;\bm U).$ Then $T_n(\G_{\alpha,\mathrm G};\bm U)-T_n(\alpha;\bm U)=g(\hat\alpha)-g(\hat\alpha_{\mathrm{aux}}).$
	We next show that $\G_{\alpha,\mathrm G}$ is consistent. By Lemma~1, for each
	fixed $t>0$, $T_n(t;\bm U)\xrightarrow{P}g(t).$ Since $\hat\alpha\xrightarrow{P}\alpha$ and $g(\cdot)$ is continuous, we have $
	g(\hat\alpha)\xrightarrow{P}g(\alpha).$ For any fixed $0<\varepsilon<\alpha$, the strict monotonicity of $g$ implies $g(\alpha+\varepsilon)<g(\alpha)<g(\alpha-\varepsilon).$ Together with $T_n(\alpha\pm\varepsilon;\bm U)\xrightarrow{P} g(\alpha\pm\varepsilon)$, we obtain $\Pr\left\{T_n(\alpha+\varepsilon;\bm U) <g(\hat\alpha)<T_n(\alpha-\varepsilon;\bm U)\right\}\to 1.$ 
	Since $t_{\mathrm{obs}}$ is the realized value of $g(\hat\alpha)$, in the asymptotic argument, we write the defining equation as $T_n(\G_{\alpha,\mathrm G};\bm U)=g(\hat\alpha)$. Since $T_n(t;\bm U)$ is strictly decreasing in $t$ with probability tending to one, the above event implies $\alpha-\varepsilon<\G_{\alpha,\mathrm G}  <\alpha+\varepsilon.$ Therefore, $\G_{\alpha,\mathrm G}\xrightarrow{P}\alpha.$

	Together with $\hat\alpha\xrightarrow{P}\alpha$ and $\hat\alpha_{\mathrm{aux}}\xrightarrow{P}\alpha$, this consistency result ensures that the intermediate points in the following mean-value expansions lie in a neighborhood of $\alpha$ with probability tending to one. Hence, we can compare $\G_{\alpha,\mathrm G}-\alpha$ with $\hat\alpha-\hat\alpha_{\mathrm{aux}}$ as follows. By the mean value theorem, for some $\bar\alpha_{n,1}$ between $\G_{\alpha,\mathrm G}$ and $\alpha$, and some $\bar\alpha_{n,2}$ between $\hat\alpha$ and $\hat\alpha_{\mathrm{aux}}$,
	\[
	\frac{\partial T_n(\bar\alpha_{n,1};\bm U)}{\partial \alpha}
	(\G_{\alpha,\mathrm G}-\alpha)
	=
	g'(\bar\alpha_{n,2})(\hat\alpha-\hat\alpha_{\mathrm{aux}}).
	\]
	Therefore,
	\[
	\G_{\alpha,\mathrm G}-\alpha
	=\frac{g'(\bar\alpha_{n,2})}{\partial T_n(\bar\alpha_{n,1};\bmU)/\partial \alpha}
	(\hat\alpha-\hat\alpha_{\mathrm{aux}}).
	\]
	Since $\G_{\alpha,\mathrm G}$, $\hat\alpha$, and $\hat\alpha_{\mathrm{aux}}$ are all consistent for $\alpha$, the intermediate points satisfy $\bar\alpha_{n,1},\bar\alpha_{n,2}\xrightarrow{P}\alpha$. Moreover, by the differentiability of the gamma quantile and the local uniform law of large numbers, for some $\delta>0$,
	$
	\sup_{|t-\alpha|\le \delta}
	\left|
	{\partial T_n(t;\bm U)}/{\partial t}-g'(t)
	\right|
	\xrightarrow{P}0.
	$
	Therefore,
	$
	{\partial T_n(\bar\alpha_{n,1};\bm U)}/{\partial \alpha}
	\xrightarrow{P}g'(\alpha),
	\
	g'(\bar\alpha_{n,2})\xrightarrow{P}g'(\alpha).
	$
	Since $g'(\alpha)=1/\alpha-\psi_1(\alpha)<0$, it follows that
	$
	{g'(\bar\alpha_{n,2})}/
	{[\partial T_n(\bar\alpha_{n,1};\bm U)/\partial \alpha]}
	\xrightarrow{P}1.
	$
	Hence,
	\[
	\G_{\alpha,\mathrm G}-\alpha
	=
	(\hat\alpha-\hat\alpha_{\mathrm{aux}})+o_p(n^{-1/2})
	=
	(\hat\alpha-\alpha)-(\hat\alpha_{\mathrm{aux}}-\alpha)+o_p(n^{-1/2}).
	\]

	For $\G_{a,\mathrm G}$, consider the auxiliary analogue of the likelihood equation for $a$, obtained by replacing $\alpha$ with $\G_{\alpha,\mathrm G}$. Let $\bmV^*=(V_1^*,\ldots,V_n^*)$ be the resulting auxiliary sample with $V_i^* \sim \chi^2(2\G_{\alpha,\mathrm G})$. By construction, the defining equation of $\phi_{a,\mathrm G}(\bmV^*)$ coincides with this auxiliary likelihood equation. Hence, $\phi_{a,\mathrm G}(\bmV^*)=\hat a_{\mathrm{aux}}/a+o_p(n^{-3/2}).$
	Since $\G_{a,\mathrm G}=\hat a/\phi_{a,\mathrm G}(\bmV^*)$, a first-order expansion yields $\G_{a,\mathrm G}-a=(\hat a-a)-(\hat a_{\mathrm{aux}}-a)+o_p(n^{-3/2}).$ Similarly, Equation~\eqref{eq:beta_gamgp} is the auxiliary counterpart of the likelihood equation for $\beta$, and hence $\G_{\beta,\mathrm G}-\beta=(\hat\beta-\beta)-(\hat\beta_{\mathrm{aux}}-\beta)+o_p(n^{-1/2}).$ Therefore, 
	$D_n(\G_{\boldsymbol{\theta}_{\mathrm G}}-\boldsymbol{\theta}_{\mathrm G})
	=D_n(\hat{\boldsymbol{\theta}}_{\mathrm G}-\boldsymbol{\theta}_{\mathrm G})
	-D_n(\hat{\boldsymbol{\theta}}_{\mathrm G,\mathrm{aux}}-\boldsymbol{\theta}_{\mathrm G})
	+o_p(1).$
	Since the first two terms on the right converge to independent $N(\bm0,\Sigma_{\mathrm G}^{-1})$ limits, Slutsky's theorem implies $D_n(\G_{\boldsymbol{\theta}_{\mathrm G}}-\boldsymbol{\theta}_{\mathrm G}) \xrightarrow{d} N(\bm0,\,2\Sigma_{\mathrm G}^{-1}),$
	which is exactly the stated result.

	\section{Proof of Theorem \ref{thm:mle_weigp}}\label{proof.mle_weigp}
	
	Using the same argument as in the gamma GP case, there exists a local ML estimator $\hat{\boldsymbol\theta}_{\mathrm W}=(\hat a,\hat m,\hat\eta)^T$ satisfying
	$
	\hat a-a=O_p(n^{-3/2}),\
	\hat m-m=O_p(n^{-1/2}),\
	\hat\eta-\eta=O_p(n^{-1/2}).
	$
	In particular, $\hat{\boldsymbol\theta}_{\mathrm W}\xrightarrow{P}\boldsymbol\theta_{\mathrm W}$.
	We therefore use the nonstandard normalization $D_n=\diag(n^{3/2},n^{1/2},n^{1/2}).$
	Define
	\[
	Y_i:=a^{i-1}X_i,\qquad
	Z_i:=\left(\frac{Y_i}{\eta}\right)^m,\qquad
	L_i:=\log Z_i,\qquad i=1,2,\ldots,n.
	\]
	Then $Y_1,\ldots,Y_n \stackrel{\mathrm{i.i.d.}}{\sim} \mathrm{WB}(m,\eta)$, and hence $Z_1,\ldots,Z_n \stackrel{\mathrm{i.i.d.}}{\sim} \mathrm{Exp}(1)$. For convenience, let $A_i=1+(1-Z_i)L_i.$
	Since $E(Z_i)=1$, $E(L_i)=-c_E$, and $E(Z_iL_i)=1-c_E$, we have
	$E(1-Z_i)=0,\ E(A_i)=0$, where $c_E$ is Euler's constant.  Moreover, $\Var(1-Z_i)=1,\ \Var(A_i)={\pi^2}/{6}+(1-c_E)^2, \ \Cov(1-Z_i,A_i)=1-c_E,\ \Cov(1-Z_i,Z_i-1)=-1.
	$
	The score vector $S_n(\boldsymbol{\theta}_{\mathrm W})$ is given by
	\[
	S_n(\boldsymbol{\theta}_{\mathrm W})
	=\sum_{i=1}^n
	\begin{pmatrix}
		\dfrac{m(i-1)}{a}(1-Z_i),
		&\dfrac{1}{m}A_i,
		&\dfrac{m}{\eta}(Z_i-1)
	\end{pmatrix}^T,
	\]
	and we have
	\[
	D_n^{-1}S_n(\boldsymbol{\theta}_{\mathrm W})
	=\sum_{i=1}^n
	\begin{pmatrix}
		\dfrac{m(i-1)}{a n^{3/2}}(1-Z_i),
		&\dfrac{1}{m n^{1/2}}A_i,
		&\dfrac{m}{\eta n^{1/2}}(Z_i-1)
	\end{pmatrix}^T.
	\]
	To apply the Lindeberg--Feller central limit theorem \cite[Proposition~2.27]{vdV1998}, we first verify two conditions. Define
	\[
	\bm\xi_{n,i}
	=\left(
	\dfrac{m(i-1)}{a n^{3/2}}(1-Z_i),\,
	\dfrac{1}{m n^{1/2}}A_i,\,
	\dfrac{m}{\eta n^{1/2}}(Z_i-1)
	\right)^T,
	\qquad i=1,\ldots,n .
	\]
	Then $\bm\xi_{n,1},\ldots,\bm\xi_{n,n}$ are independent mean-zero random vectors, and
	$D_n^{-1}S_n(\boldsymbol{\theta}_{\mathrm W}) =\sum_{i=1}^n\bm\xi_{n,i}.$
	Let $q_{n,i}={m^2(i-1)^2}/{(a^2n^3)}+{1}/{(m^2n)}+{m^2}/{(\eta^2n)}.$ Also define $B_i^2=(1-Z_i)^2+A_i^2+(Z_i-1)^2 .$
	Then 
	$\|\bm\xi_{n,i}\|^2 \le q_{n,i}B_i^2. $
	Since
	\[
	\max_{1\le i\le n}q_{n,i}
	\le
	\frac{m^2}{a^2n}
	+
	\frac{1}{m^2n}
	+
	\frac{m^2}{\eta^2n}
	\le
	\frac{C^2}{n}
	\]
	for some constant $C>0$, we have
	$
	\|\bm\xi_{n,i}\|
	\le
	\frac{C}{ n^{1/2}}B_i .
	$
	Thus, for any $\delta>0$,
	$
	\{\|\bm\xi_{n,i}\|>\delta\}
	\subset
	\left\{
	B_i>\frac{\delta n^{1/2}}{C}
	\right\}.
	$
	Therefore,
	\begin{align*}
		\sum_{i=1}^n
		E\left[
		\|\bm\xi_{n,i}\|^2
		I\{\|\bm\xi_{n,i}\|>\delta\}
		\right]  
		&\le
		\sum_{i=1}^n
		q_{n,i}
		E\left[
		B_i^2
		I\left\{
		B_i>\frac{\delta n^{1/2}}{C}
		\right\}
		\right]  \\
		&=
		\left(\sum_{i=1}^n q_{n,i}\right)
		E\left[
		B_1^2
		I\left\{
		B_1>\frac{\delta n^{1/2}}{C}
		\right\}
		\right].
	\end{align*}
	Moreover,
	\[
	\sum_{i=1}^n q_{n,i}
	=\frac{m^2}{a^2n^3}\sum_{i=1}^n(i-1)^2
	+\frac{1}{m^2n}\sum_{i=1}^n1
	+\frac{m^2}{\eta^2n}\sum_{i=1}^n1
	=O(1).
	\]
	Since $Z_i\sim\mathrm{Exp}(1)$ and
	$A_i=1+(1-Z_i)\log Z_i$, we have $E(B_1^2)<\infty$. Hence,
	$
	E\left[
	B_1^2
	I\left\{
	B_1>\frac{\delta n^{1/2}}{C}
	\right\}
	\right]\to0 .
	$
	It follows that
	$
	\sum_{i=1}^n
	E\left[
	\|\bm\xi_{n,i}\|^2
	I\{\|\bm\xi_{n,i}\|>\delta\}
	\right]\to0.
	$
	It remains to verify the covariance condition. From
	$\Var(1-Z_i)=1,\
	\Var(A_i)={\pi^2}/{6}+(1-c_E)^2,\
	\Var(Z_i-1)=1,$
	and
	$\Cov(1-Z_i,A_i)=1-c_E,\
	\Cov(1-Z_i,Z_i-1)=-1,\
	\Cov(A_i,Z_i-1)=-(1-c_E),$
	we have
	\[
	\sum_{i=1}^n\Cov(\bm\xi_{n,i})
	=\sum_{i=1}^n
	\begin{pmatrix}
		\dfrac{m^2(i-1)^2}{a^2n^3} &
		\dfrac{(1-c_E)(i-1)}{a n^2} &
		-\dfrac{m^2(i-1)}{a\eta n^2}\\[8pt]
		\dfrac{(1-c_E)(i-1)}{a n^2} &
		\dfrac{\pi^2/6+(1-c_E)^2}{m^2 n} &
		-\dfrac{1-c_E}{\eta n}\\[8pt]
		-\dfrac{m^2(i-1)}{a\eta n^2} &
		-\dfrac{1-c_E}{\eta n} &
		\dfrac{m^2}{\eta^2 n}
	\end{pmatrix}.
	\]
	Since $n^{-3}\sum_{i=1}^n(i-1)^2\to 1/3$, $n^{-2}\sum_{i=1}^n(i-1)\to 1/2$, and $n^{-1}\sum_{i=1}^n1\to 1$, 
	we obtain
	\[
	\sum_{i=1}^n\Cov(\bm\xi_{n,i})
	\to
	\Sigma_{\mathrm W} =
	\begin{pmatrix}
		\dfrac{m^2}{3a^2} & \dfrac{1-c_E}{2a} & -\dfrac{m^2}{2a\eta}\\[8pt]
		\dfrac{1-c_E}{2a} & \dfrac{\pi^2/6+(1-c_E)^2}{m^2} & -\dfrac{1-c_E}{\eta}\\[8pt]
		-\dfrac{m^2}{2a\eta} & -\dfrac{1-c_E}{\eta} & \dfrac{m^2}{\eta^2}
	\end{pmatrix}.
	\]
	Therefore,
	$
	D_n^{-1}S_n(\boldsymbol{\theta}_{\mathrm W})
	\xrightarrow{d}
	N(\bm0,\Sigma_{\mathrm W}).
	$

	Next, consider the Hessian matrix $H_n(\boldsymbol{\theta}_{\mathrm W})$ as
	\[
	H_n(\boldsymbol{\theta}_{\mathrm W})
	=
	\begin{pmatrix}
		H_{aa}(\boldsymbol{\theta}_{\mathrm W}) & H_{a m}(\boldsymbol{\theta}_{\mathrm W}) & H_{a\eta}(\boldsymbol{\theta}_{\mathrm W})\\[8pt]
		H_{a m}(\boldsymbol{\theta}_{\mathrm W}) &H_{m m}(\boldsymbol{\theta}_{\mathrm W})  & H_{m \eta}(\boldsymbol{\theta}_{\mathrm W}) \\[8pt]
		H_{a \eta}(\boldsymbol{\theta}_{\mathrm W}) &H_{m \eta}(\boldsymbol{\theta}_{\mathrm W}) &H_{\eta \eta}(\boldsymbol{\theta}_{\mathrm W})
	\end{pmatrix},
	\]
	where 
	\begin{align*}
		&H_{aa}(\boldsymbol{\theta}_{\mathrm W})
		=
		-\frac{m}{a^2}\sum_{i=1}^n(i-1)(1-Z_i)
		-\frac{m^2}{a^2}\sum_{i=1}^n(i-1)^2Z_i,\\
		&H_{am}(\boldsymbol{\theta}_{\mathrm W})
		=
		\frac1a\sum_{i=1}^n(i-1)(1-Z_i)
		-\frac1a\sum_{i=1}^n(i-1)Z_iL_i,\\
		&H_{a\eta}(\boldsymbol{\theta}_{\mathrm W})
		=
		\frac{m^2}{a\eta}\sum_{i=1}^n(i-1)Z_i,\\
		&H_{mm}(\boldsymbol{\theta}_{\mathrm W})
		=
		-\frac{n}{m^2}
		-\frac1{m^2}\sum_{i=1}^n Z_iL_i^2,\\
		&H_{m\eta}(\boldsymbol{\theta}_{\mathrm W})
		=
		\frac1\eta\sum_{i=1}^n (Z_i-1)+\frac1\eta\sum_{i=1}^n Z_iL_i,\\
		&H_{\eta\eta}(\boldsymbol{\theta}_{\mathrm W})
		=
		-\frac{m}{\eta^2}\sum_{i=1}^n(Z_i-1)
		-\frac{m^2}{\eta^2}\sum_{i=1}^n Z_i.
	\end{align*}
	Using the law of large numbers together with $E(Z_i)=1$, $E(Z_iL_i)=1-c_E$, and $E(Z_iL_i^2)={\pi^2}/{6}-1+(1-c_E)^2,$ we obtain $-\,D_n^{-1}H_n(\boldsymbol{\theta}_{\mathrm W})D_n^{-1}\xrightarrow{P}\Sigma_{\mathrm W}.$ Moreover, since $\hat{\boldsymbol{\theta}}_{\mathrm W}\xrightarrow{P}\boldsymbol{\theta}_{\mathrm W}$, the intermediate point
	$\tilde{\boldsymbol{\theta}}_n$ between $\hat{\boldsymbol{\theta}}_{\mathrm W}$ and $\boldsymbol{\theta}_{\mathrm W}$ also satisfies  $\tilde{\boldsymbol{\theta}}_n\xrightarrow{P}\boldsymbol{\theta}_{\mathrm W}$, and hence $-\,D_n^{-1}H_n(\tilde{\boldsymbol{\theta}}_n)D_n^{-1}\xrightarrow{P}\Sigma_{\mathrm W}.$
	
	Finally, a first-order Taylor expansion of the score around the true parameter gives
	$
	\bm0
	=S_n(\hat{\boldsymbol{\theta}}_{\mathrm W})
	=S_n(\boldsymbol{\theta}_{\mathrm W})
	+H_n(\tilde{\boldsymbol{\theta}}_n)(\hat{\boldsymbol{\theta}}_{\mathrm W}-\boldsymbol{\theta}_{\mathrm W}),
	$
	which implies
	$
	D_n(\hat{\boldsymbol{\theta}}_{\mathrm W}-\boldsymbol{\theta}_{\mathrm W})
	=\Bigl[-D_n^{-1}H_n(\tilde{\boldsymbol{\theta}}_n)D_n^{-1}\Bigr]^{-1}
	D_n^{-1}S_n(\boldsymbol{\theta}_{\mathrm W}).
	$
	Therefore, by Slutsky's theorem,
	\[
	D_n(\hat{\boldsymbol{\theta}}_{\mathrm W}-\boldsymbol{\theta}_{\mathrm W})
	\xrightarrow{d}
	N(\bm0,\Sigma_{\mathrm W}^{-1}),
	\]
	where
	\[
	\Sigma_{\mathrm W}^{-1}
	=\begin{pmatrix}
		\dfrac{12a^2}{m^2} & 0 & \dfrac{6a\eta}{m^2}\\[8pt]
		0 & \dfrac{6m^2}{\pi^2} & \dfrac{6(1-c_E)\eta}{\pi^2}\\[10pt]
		\dfrac{6a\eta}{m^2} & \dfrac{6(1-c_E)\eta}{\pi^2}
		& \dfrac{\eta^2}{m^2}\!\left(4+\dfrac{6(1-c_E)^2}{\pi^2}\right)
	\end{pmatrix}.
	\]

	\section{Proof of Theorem \ref{thm:AN_weigp}}\label{proof.AN_weigp}

	Let $\G_{\boldsymbol{\theta}_{\mathrm W}}=(\G_{a,\mathrm W},\G_{m,\mathrm W},\G_{\eta,\mathrm W})^T$ denote a generalized fiducial draw from $f_F(\boldsymbol{\theta}_{\mathrm W};\bmx)$. Since $J(\bmx;\boldsymbol{\theta}_{\mathrm W}) \propto (am\eta)^{-1}$, then the log fiducial density is
	\[
	\ell_F(\boldsymbol{\theta}_{\mathrm W})
	=\ell_n(\boldsymbol{\theta}_{\mathrm W};\bmx)-\log a-\log m-\log\eta+\text{const},
	\]
	where $\ell_n(\boldsymbol{\theta}_{\mathrm W};\bmx)$ is the log-likelihood function for $\boldsymbol{\theta}_{\mathrm W}$ based on the observed sample $\bmx$. To study the local behavior of the fiducial draw around the ML estimator $\hat{\boldsymbol{\theta}}_{\mathrm W}$, let $\boldsymbol{\theta}_{\mathrm W}=\hat{\boldsymbol{\theta}}_{\mathrm W}+D_n^{-1}\bq,$ where $\bq=(q_1,q_2,q_3)^T$ denotes the local coordinate vector.
	Then
	\small{
		\[
		\ell_F(\hat{\boldsymbol{\theta}}_{\mathrm W}+D_n^{-1}\bq)-\ell_F(\hat{\boldsymbol{\theta}}_{\mathrm W})
		=
		\ell_n(\hat{\boldsymbol{\theta}}_{\mathrm W}+D_n^{-1}\bq)-\ell_n(\hat{\boldsymbol{\theta}}_{\mathrm W})
		-\log\!\left(1+\frac{q_1}{\hat a n^{3/2}}\right)
		-\log\!\left(1+\frac{q_2}{\hat m n^{1/2}}\right)
		-\log\!\left(1+\frac{q_3}{\hat\eta n^{1/2}}\right).
		\]
	}
	Since $S_n(\hat{\boldsymbol{\theta}}_{\mathrm W})=\bm0$, a second-order Taylor expansion gives
	\[
	\ell_n(\hat{\boldsymbol{\theta}}_{\mathrm W}+D_n^{-1}\bq)-\ell_n(\hat{\boldsymbol{\theta}}_{\mathrm W})
	=-\frac12 \bq^T\Sigma_{\mathrm W}\bq+o_p(1),
	\]
	uniformly on compact sets. On the other hand,
	\[
	\log\!\left(1+\frac{q_1}{\hat a n^{3/2}}\right)
	+\log\!\left(1+\frac{q_2}{\hat m n^{1/2}}\right)
	+\log\!\left(1+\frac{q_3}{\hat\eta n^{1/2}}\right)
	=o_p(1),
	\]
	uniformly on compact sets. Therefore,
	\[
	\ell_F(\hat{\boldsymbol{\theta}}_{\mathrm W}+D_n^{-1}\bq)-\ell_F(\hat{\boldsymbol{\theta}}_{\mathrm W})
	=
	-\frac12 \bq^T\Sigma_{\mathrm W}\bq+o_p(1).
	\]
	Hence the localized fiducial density converges to the density of $N(\bm0,\Sigma_{\mathrm W}^{-1})$. Equivalently, the conditional distribution of  $D_n(\G_{\boldsymbol{\theta}_{\mathrm W}}-\hat{\boldsymbol{\theta}}_{\mathrm W})$ given $\bmx$ converges weakly in probability to $N(\bm0,\Sigma_{\mathrm W}^{-1})$.
	Finally,
	$
	D_n(\G_{\boldsymbol{\theta}_{\mathrm W}}-\boldsymbol{\theta}_{\mathrm W})
	=D_n(\G_{\boldsymbol{\theta}_{\mathrm W}}-\hat{\boldsymbol{\theta}}_{\mathrm W})
	+D_n(\hat{\boldsymbol{\theta}}_{\mathrm W}-\boldsymbol{\theta}_{\mathrm W}).
	$
	The first term on the right converges conditionally to $N(\bm0,\Sigma_{\mathrm W}^{-1})$, whereas the second term converges to $N(\bm0,\Sigma_{\mathrm W}^{-1})$ by the result above. Therefore,
	$
	D_n(\G_{\boldsymbol{\theta}_{\mathrm W}}-\boldsymbol{\theta}_{\mathrm W})
	\xrightarrow{d}
	N(\bm0,\,2\Sigma_{\mathrm W}^{-1}).
	$

	\section{Algorithms for confidence intervals and model discrimination}

	\subsection{Generalized confidence intervals under the exponential GP}\label{app.alg:gci_exp}
	
	This subsection presents a Monte Carlo procedure for constructing generalized confidence intervals for the reliability and hazard rate under the exponential GP.
	
	\begin{algorithm}[H]
		\begin{spacing}{1}
			\caption{{Generalized confidence intervals for $R_{i, \mathrm E}(\tau)$ and $h_{i, \mathrm E}(\tau)$ under the exponential GP.}}
			\label{supp.alg:gci_exp}
			\renewcommand{\algorithmicrequire}{\textbf{Input:}}
			\renewcommand{\algorithmicensure}{\textbf{Output:}}
			\begin{algorithmic}[1]
				\REQUIRE Observed data $\bmx$, failure index $i$, and time $\tau$.
				\ENSURE The $(1-\gamma)$ GCIs for $R_{i, \mathrm E}(\tau)$ and $h_{i, \mathrm E}(\tau)$.
				\STATE Compute the ML estimate ${\hat{\boldsymbol{\theta}}}_{\mathrm E} = (\hat a, \hat \lambda)^T$ from Equations~\eqref{eq:a}--\eqref{eq:lambda}.
				\FOR{$b = 1$ to $B$}
				\STATE Generate $Z_1,\ldots,Z_n \sim \mathrm{Exp}(1)$ independently.
				\STATE Compute $\phi_{a,\mathrm E}(\bmZ)$ as the root of Equation~\eqref{eq:g1exp}, and compute $\G_{a,\mathrm E}^{(b)}$ from Equation~\eqref{eq:a_gpqexp}.            
				\STATE Compute $\phi_{\lambda,\mathrm E}(\bmZ)$ from Equation~\eqref{eq:g2exp} and then obtain $\G_{\lambda,\mathrm E}^{(b)}$ from Equation~\eqref{eq:lam_gpqexp}.
				\STATE Compute $\G_{R_{i,\mathrm E}(\tau)}^{(b)}$ and $\G_{h_{i,\mathrm E}(\tau)}^{(b)}$ from Equation~\eqref{eq:rihi_expgp}.
				\ENDFOR

				\STATE Sort the simulated values $\{\G_{R_{i,\mathrm E}(\tau)}^{(b)}\}_{b=1}^{B}$ and $\{\G_{h_{i,\mathrm E}(\tau)}^{(b)}\}_{b=1}^{B}$ in ascending order.
				
				\STATE Obtain the empirical $\gamma/2$ and $1-\gamma/2$ quantiles of
				the simulated values.
				
				\STATE Report the intervals formed by the corresponding empirical quantiles as the $100(1-\gamma)\%$ GCIs for $R_{i,\mathrm E}(\tau)$ and $h_{i,\mathrm E}(\tau)$, respectively.

				\vspace{0.5\baselineskip}
			\end{algorithmic}
		\end{spacing}
	\end{algorithm}

	\subsection{Generalized confidence intervals under the gamma GP}\label{app.alg:gci_gamma}

	This subsection presents a Monte Carlo procedure for constructing generalized confidence intervals for the reliability and hazard rate under the gamma GP.
	
	\begin{algorithm}[H]
		\begin{spacing}{1}
			\caption{{Generalized confidence intervals for $R_{i,\mathrm G}(\tau)$ and $h_{i,\mathrm G}(\tau)$ under the gamma GP.}}
			\label{alg:gci_gamma}
			\renewcommand{\algorithmicrequire}{\textbf{Input:}}
			\renewcommand{\algorithmicensure}{\textbf{Output:}}
			\begin{algorithmic}[1]
				\REQUIRE Observed data $\bmx$, failure index $i$, and time $\tau$.
				\ENSURE The $(1-\gamma)$ GCIs for $R_{i,\mathrm G}(\tau)$ and $h_{i,\mathrm G}(\tau)$.
				
				\STATE Compute the ML estimate $\hat{\boldsymbol{\theta}}_{\mathrm G}$ from Equations~\eqref{eq:amle}--\eqref{eq:lammle}.
				\STATE Compute $t_{\mathrm{obs}}$ from Equation~\eqref{eq:tobs_gamgp}.
				
				\FOR{$b=1$ to $B$}
				\STATE Generate $U_1,\ldots,U_n \sim \mathrm{Unif}(0,1)$ independently, and compute $T_n(\alpha;\bmU)$ from Equation~\eqref{eq:T_alpha}.
				
				\STATE Compute $\G_{\alpha,\mathrm G}^{(b)}$ from  Equation~\eqref{eq:alpha_star_def}.
				
				\STATE Generate $V_i^* \sim \chi^2\!\bigl(2\G_{\alpha,\mathrm G}^{(b)}\bigr)$ independently for $i=1,\ldots,n$.
				
				\STATE Compute $\phi_{a,\mathrm G}(\bmV^*)$ as the root of Equation~\eqref{eq:g3_gamgp}, and then compute $\G_{a,\mathrm G}^{(b)}$ from Equation~\eqref{eq:a_gamgp}.
				
				\STATE Compute $\G_{\beta,\mathrm G}^{(b)}$ from Equation~\eqref{eq:beta_gamgp}.
				
				\STATE Compute $\G_{R_{i,\mathrm G}(\tau)}^{(b)}$ and $\G_{h_{i,\mathrm G}(\tau)}^{(b)}$ from Equation~\eqref{eq:rihi_gamgp}.
				\ENDFOR
				
				\STATE Sort the simulated values $\{\G_{R_{i,\mathrm G}(\tau)}^{(b)}\}_{b=1}^{B}$ and $\{\G_{h_{i,\mathrm G}(\tau)}^{(b)}\}_{b=1}^{B}$ in ascending order.
				
				\STATE Obtain the empirical $\gamma/2$ and $1-\gamma/2$ quantiles of these simulated values.
				
				\STATE Report the intervals formed by the corresponding empirical quantiles as the $100(1-\gamma)\%$ GCIs for $R_{i,\mathrm G}(\tau)$ and $h_{i,\mathrm G}(\tau)$.

				\vspace{0.5\baselineskip}
			\end{algorithmic}
		\end{spacing}
	\end{algorithm}
	
	\subsection{Generalized confidence intervals under the Weibull GP}\label{app.alg:gfd_weibull_gp}

	This subsection presents a Monte Carlo procedure for constructing generalized confidence intervals for the reliability and hazard rate under the Weibull GP.
	
	\begin{algorithm}[H]
		\begin{spacing}{1}
			\caption{{Generalized confidence intervals for $R_{i, \mathrm W}(\tau)$ and $h_{i, \mathrm W}(\tau)$ under the Weibull GP.}}
			\label{supp.alg:gfd_weibull_gp}
			\renewcommand{\algorithmicrequire}{\textbf{Input:}}
			\renewcommand{\algorithmicensure}{\textbf{Output:}}
			\begin{algorithmic}[1]
				\REQUIRE Observed data $\bmx$, failure index $i$, and time $\tau$.
				\ENSURE The $(1-\gamma)$ GCIs for $R_{i, \mathrm W}(\tau)$ and $h_{i, \mathrm W}(\tau)$.
				
				\STATE Compute the ML estimate $\hat{\boldsymbol{\theta}}_{\mathrm W}$ from $\bmx$ and use it as the initial value $(\G_{a,\mathrm W}^{(0)},\G_{m,\mathrm W}^{(0)},\G_{\eta,\mathrm W}^{(0)})$.

				\FOR{$b=1$ to $B$}
				\STATE Generate $\xi \sim \mathrm{IGa}\!\left(n,\sum_{i=1}^n\left[(\G_{a,\mathrm W}^{(b-1)})^{\,i-1}x_i\right]^{\G_{m,\mathrm W}^{(b-1)}}\right)$ and update
				$\G_{\eta,\mathrm W}^{(b)}=\xi^{1/\G_{m,\mathrm W}^{(b-1)}}$.
				\STATE Generate $\G_{a,\mathrm W}^{(b)}$ from $\pi(a\mid \G_{m,\mathrm W}^{(b-1)},\G_{\eta,\mathrm W}^{(b)},\bmx)$ by a Metropolis--Hastings step.
				\STATE Generate $\G_{m,\mathrm W}^{(b)}$ from $\pi(m\mid \G_{a,\mathrm W}^{(b)},\G_{\eta,\mathrm W}^{(b)},\bmx)$ by a Metropolis--Hastings step.
				\STATE Compute $\G_{R_{i,\mathrm W}(\tau)}^{(b)}$ and $\G_{h_{i,\mathrm W}(\tau)}^{(b)}$ from Equation~\eqref{eq:rihi_weigp}.
				\ENDFOR
				
				\STATE Discard the first $L$ fiducial draws and sort the retained draws
				$\{\G_{R_{i,\mathrm W}(\tau)}^{(b)}\}_{b=L+1}^{B}$ and $\{\G_{h_{i,\mathrm W}(\tau)}^{(b)}\}_{b=L+1}^{B}$ in ascending order.
				
				\STATE Obtain the empirical $\gamma/2$ and $1-\gamma/2$ quantiles of the retained fiducial draws.
				
				\STATE Report the intervals formed by the corresponding empirical quantiles as  the $100(1-\gamma)\%$ GCIs for $R_{i,\mathrm W}(\tau)$ and $h_{i,\mathrm W} (\tau)$.

				\vspace{0.5\baselineskip}
			\end{algorithmic}
		\end{spacing}
	\end{algorithm}

	\subsection{Generalized-inference discrimination of GP baseline distributions}\label{app.alg:gpq_discrimination}
	
	This subsection presents a generalized-inference procedure for discriminating among GP baseline distributions.
	
	\begin{algorithm}[H]
		\begin{spacing}{1}
			\caption{{Generalized-inference discrimination procedure for GP baseline distributions.}}
			\label{supp.alg:gpq_discrimination}
			\renewcommand{\algorithmicrequire}{\textbf{Input:}}
			\renewcommand{\algorithmicensure}{\textbf{Output:}}
			\begin{algorithmic}[1]
				\REQUIRE Observed GP data $\bmx$, significance level $\gamma$.
				\ENSURE Decision on $H_0$ versus $H_1$.
				
				\STATE Compute the observed statistic $\mathcal S_0=\mathcal S(\bmx)$ from Equation~\eqref{eq:Tstat}.
				
				\FOR{$b=1$ to $B$}
				\STATE Generate draws $\G_{a_0}^{(b)}$ and $\G_{\boldsymbol{\vartheta}_0}^{(b)}$ from the inferential distribution under $H_0$, namely the distribution of GPQs for the exponential or gamma GP, or the generalized fiducial distribution for the Weibull GP.
				\STATE Generate pseudo baseline observations $y_1^{*(b)},\ldots,y_n^{*(b)}\sim f_0(\cdot;\G_{\boldsymbol{\vartheta}_0}^{(b)})$.
				\STATE Transform them back to the GP scale by setting
				\[
				x_i^{*(b)}=\frac{y_i^{*(b)}}{(\G_{a_0}^{(b)})^{\,i-1}}, \qquad i=1,\ldots,n,
				\]
				and denote the resulting pseudo dataset by $\bmx^{*(b)} = \{ x_i^{*(b)},i=1,2,\dots,n\}$.
				\STATE Refit both candidate baseline models to $\bmx^{*(b)}$ and compute
				$\mathcal S^{(b)}=\mathcal S(\bmx^{*(b)})$.
				\ENDFOR
				
				\STATE Let $\mathcal S_c$ be the empirical $\gamma$-quantile of $\{\mathcal S^{(1)},\ldots,\mathcal S^{(B)}\}$.
				\STATE Reject $H_0$ in favor of $H_1$ if $\mathcal S_0<\mathcal S_c$; otherwise, fail to reject $H_0$.
				\vspace{0.5\baselineskip}
			\end{algorithmic}
		\end{spacing}
	\end{algorithm}

	\section{Additional simulation results for exponential and gamma baselines}\label{app.simulation}

	This section reports additional Monte Carlo results for the exponential and gamma GP models. As in the main text, we evaluate interval performance using coverage probability (CP) and average interval length (AL). For the cycle-specific reliability measures, one-sided bounds are considered, namely lower bounds for $R_i$ and upper bounds for $h_i$. The proposed GCIs are compared with the bootstrap confidence intervals (BPCIs) and Bayesian credible intervals (BCIs) of \citet{ChenLiLam2010}. All results are based on $10{,}000$ Monte Carlo replications, with $B=10{,}000$ used in the corresponding algorithms. Throughout, we consider $a\in\{0.5,0.9,1,1.1,1.5\}$ and $n\in\{10,20,30\}$.

	\subsection{Exponential GP}
	
\begin{figure}[H]
		\centering
		\includegraphics[width=1\linewidth]{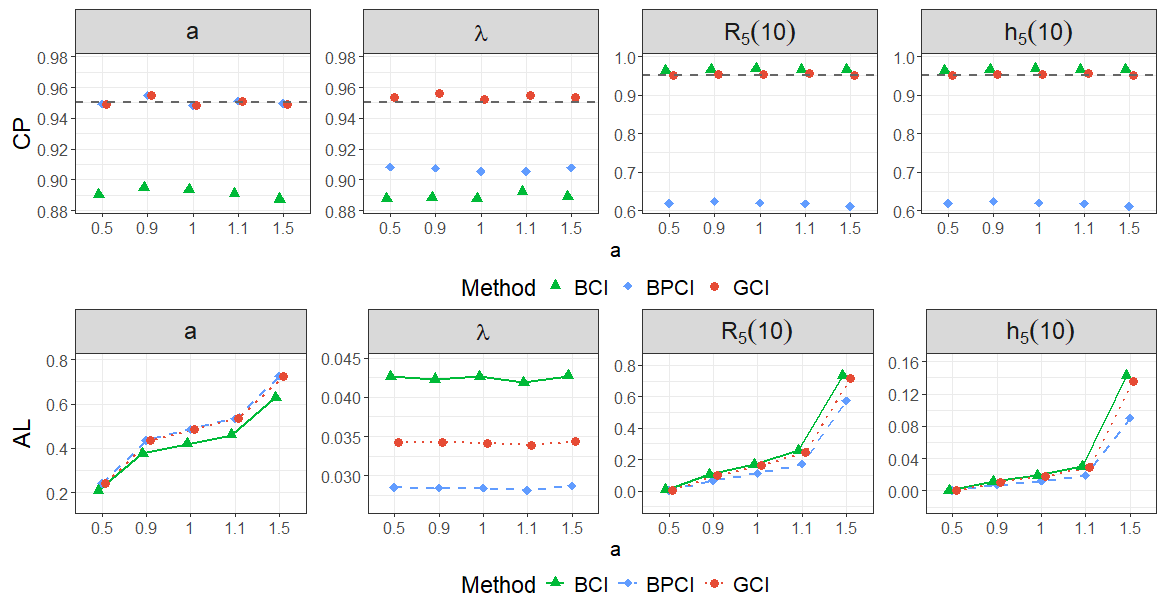}
		\caption{CPs and ALs of 95\% intervals for the exponential GP with $n=10$.}
		\label{fig:exp_n10}
	\end{figure}

\begin{figure}[H]
		\centering
		\includegraphics[width=1\linewidth]{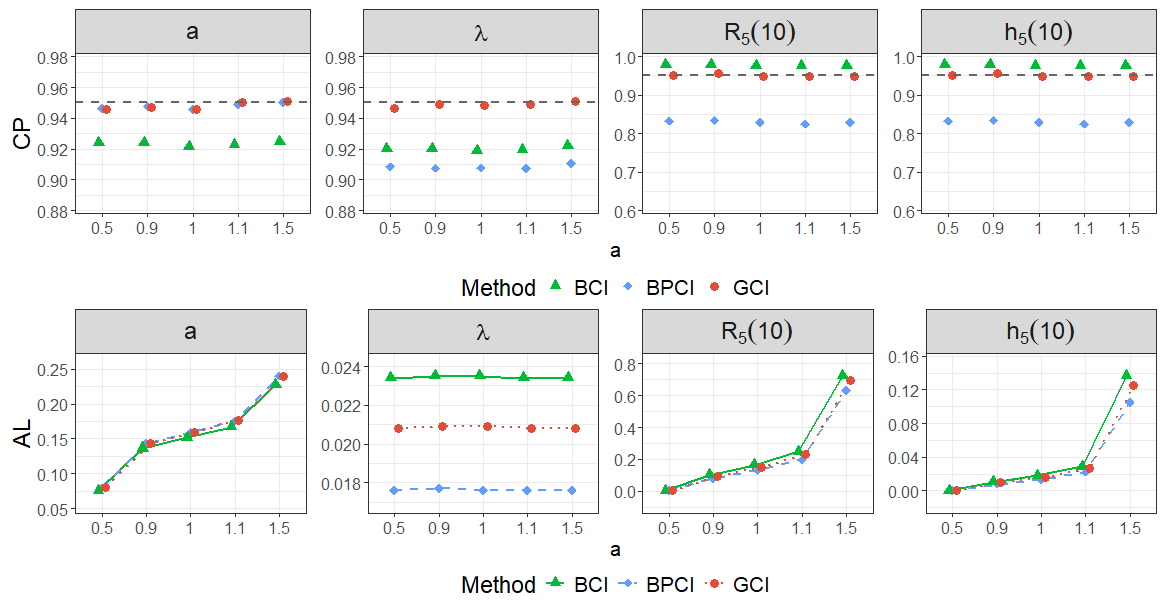}
		\caption{CPs and ALs of 95\% intervals for the exponential GP with $n=20$.}
		\label{fig:exp_n20}
	\end{figure}

\begin{figure}[H]
		\centering
		\includegraphics[width=1\linewidth]{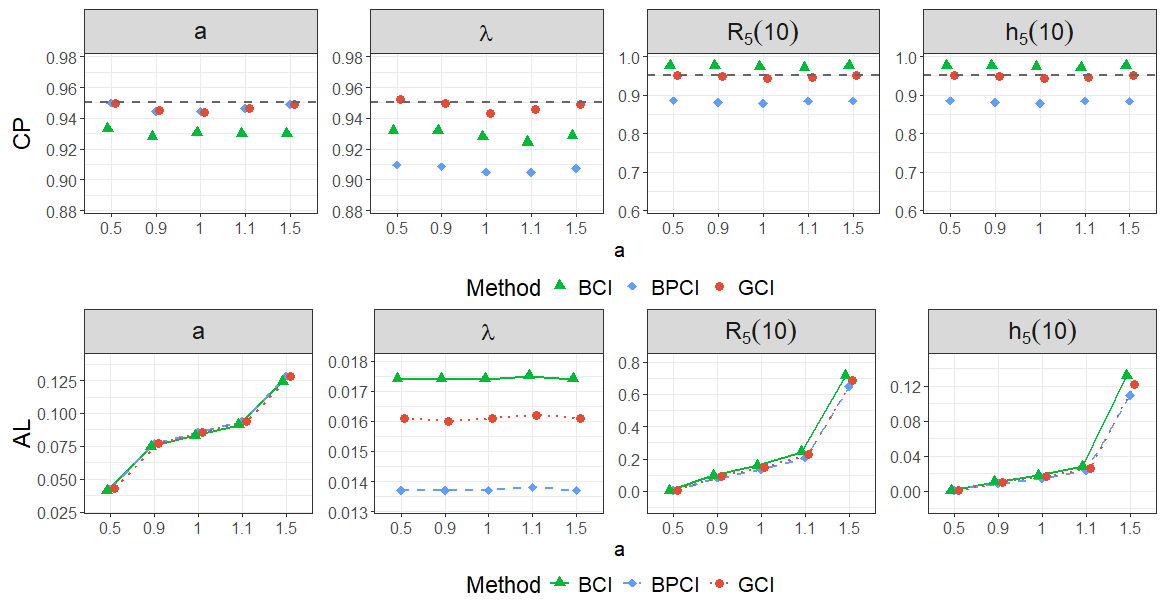}
		\caption{CPs and ALs of 95\% intervals for the exponential GP with $n=30$.}
		\label{fig:exp_n30}
	\end{figure}

	In this subsection, we consider the exponential GP with $\lambda=0.01$. The prior distributions used for constructing the BCIs are $a\sim \mathrm{Unif}(0.1,3)$ and $\lambda\sim \mathrm{Ga}(1,0.001)$. The results are shown in Figures~\ref{fig:exp_n10}-\ref{fig:exp_n30}. 
	The main finding is that the proposed GCIs achieve CPs close to the nominal level $0.95$ across all sample sizes for both the model parameters $(a,\lambda)$ and the cycle-specific reliability measures $R_{5}(10)$ and $h_{5}(10)$. By comparison, the BCIs exhibit noticeable undercoverage for $a$ and $\lambda$, but tend to be conservative for $R_{5}(10)$ and $h_{5}(10)$, with CPs above $0.95$ in many settings. The BPCIs give satisfactory coverage for $a$, but still show undercoverage for $\lambda$, $R_{5}(10)$ and $h_{5}(10)$, especially when the sample size is small. 
	In terms of AL, the ALs for $(a,\lambda)$ generally decrease as $n$ increases. The BCIs often produce noticeably longer intervals for $\lambda$ than either the BPCIs or the GCIs, whereas the BPCIs tend to give the shortest intervals. For $R_{5}(10)$ and $h_{5}(10)$, the corresponding ALs are broadly comparable across the three methods and tend to increase with $a$, especially when $a>1$. These results suggest that, under the exponential GP, the proposed GCIs provide the most reliable coverage performance among the three methods.

	\subsection{Gamma GP}
	
\begin{figure}[H]
		\centering
		\includegraphics[width=1\linewidth]{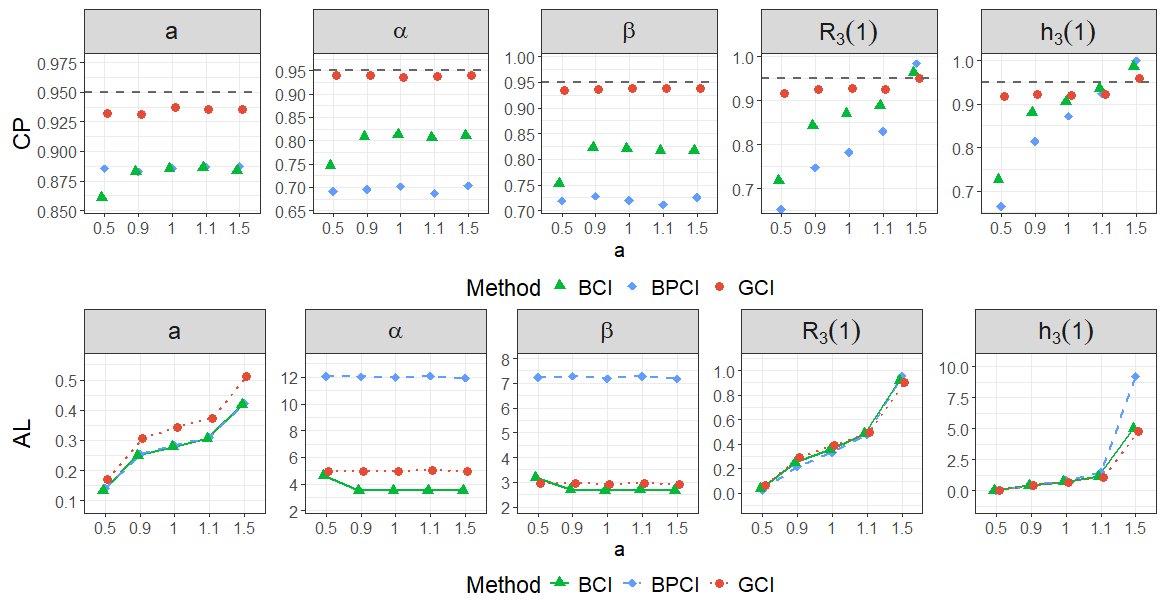}
		\caption{CPs and ALs of 95\% confidence intervals for the gamma GP with $n=10$.}
		\label{fig:ga_n10}
	\end{figure}

\begin{figure}[H]
		\centering
		\includegraphics[width=1\linewidth]{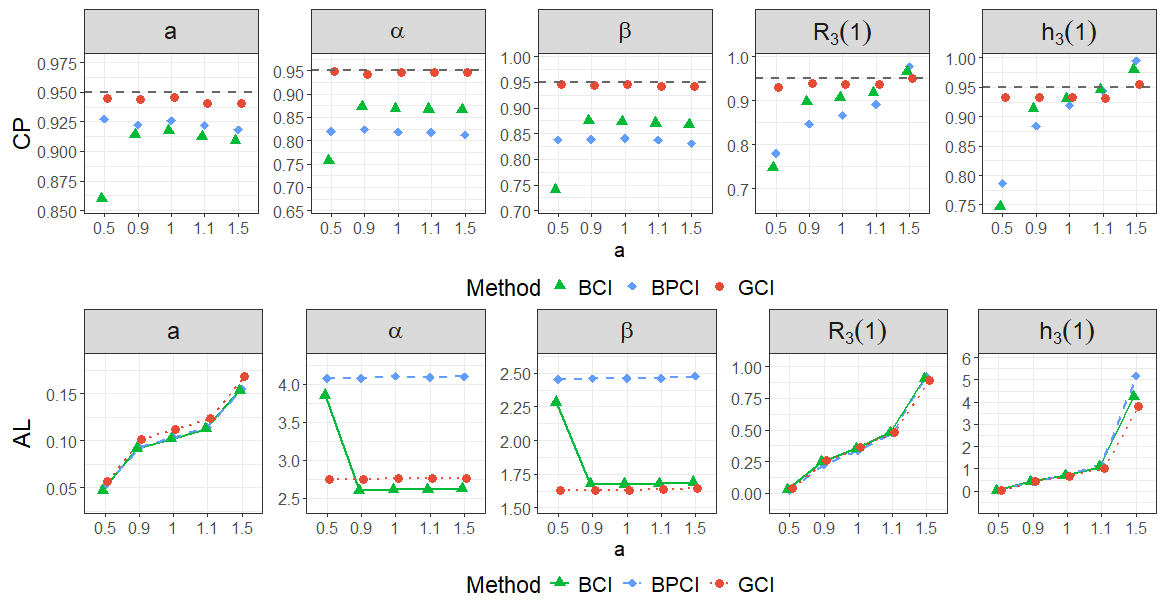}
		\caption{CPs and ALs of 95\% confidence intervals for the gamma GP with $n=20$.}
		\label{fig:ga_n20}
	\end{figure}

\begin{figure}[H]
		\centering
		\includegraphics[width=1\linewidth]{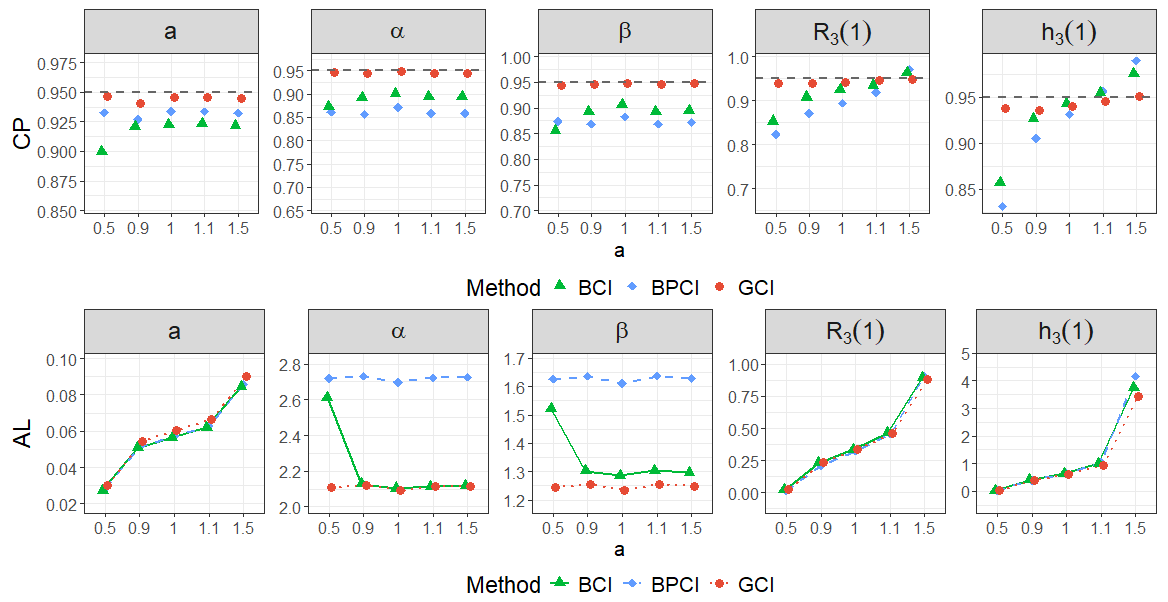}
		\caption{CPs and ALs of 95\% confidence intervals for the gamma GP with $n=30$.}
		\label{fig:ga_n30}
	\end{figure}

	In this subsection, we consider the gamma GP with $\alpha = 2$ and $\beta = 0.8$. The prior distributions used for constructing the BCIs are $a \sim \mathrm{Unif}(0.001, 3)$, $\alpha \sim \mathrm{Unif}(0.001, 6)$, and $\beta \sim \mathrm{Ga}(1, 0.001)$. The corresponding results are shown in Figures~\ref{fig:ga_n10}--\ref{fig:ga_n30}. 
	The main finding is that the proposed GCIs achieve CPs close to the nominal level $0.95$ across all sample sizes for both the model parameters $(a,\alpha,\beta)$ and the cycle-specific reliability measures $R_{3}(1)$ and $h_{3}(1)$. By comparison, both the BCIs and the BPCIs show substantial undercoverage for $a,\alpha$ and $\beta$, and this problem is especially pronounced when the sample size is small. For the derived measures $R_{3}(1)$ and $h_{3}(1)$, the CPs of the corresponding BCIs and BPCIs improve as $n$ increases, but they remain noticeably less stable across different values of $a$. In terms of AL, the ALs for the model parameters generally decrease as $n$ increases. The BPCIs produce substantially wider intervals for $\alpha$ and $\beta$ than either the BCIs or the GCIs, yet this additional width does not translate into better coverage. For $R_{3}(1)$ and $h_{3}(1)$, the corresponding ALs tend to increase with $a$, especially when $a>1$.

	Combining the Weibull results in the main text with the additional exponential and gamma results reported here, the proposed GCIs consistently achieve CPs close to the nominal level of 95\% for both the model parameters and the derived reliability measures over different values of the GP ratio parameter and different sample sizes, showing stable finite-sample performance. In contrast, the BPCIs often show noticeable undercoverage, especially for nonlinear reliability measures. 
	The BCIs tend to be conservative for the derived reliability measures under the exponential and Weibull GP models, but show less stable coverage for the corresponding measures under the gamma GP model, suggesting that their performance is sensitive to the choice of baseline distribution.
	Although the coverage of all methods generally improves as the sample size increases, the GCIs still provide the most stable and reliable coverage for small to moderate samples. This is particularly useful in practical maintenance decision making, where uncertainty in parameter and reliability estimates can directly affect inspection and replacement decisions.
	
	{\setlength{\bibsep}{0pt plus 0.3ex}
		\bibliographystyle{apalike-ejor}
		\bibliography{ref_gp}}
\end{document}